\documentclass[12pt]{elsarticle}

\usepackage{amssymb}
\usepackage{amsmath}

\journal{arXiv.org}

\begin{document}

\begin{frontmatter}

\title{XTQ: A Declarative Functional XML Query Language\tnoteref{sponsor,ack}}
\tnotetext[sponsor]{This research is partially supported by National Science
Foundation of China and Open Foundation of State Key Lab of Software Engineering.}
\tnotetext[ack]{The authors wish to thank the anonymous referees for their
valuable comments and suggestions, which greatly improved the technical content and the
presentation of the paper. Also thanks to Tieyun Qian and Ming Zhong for their valuable advice and efforts in revising the submission.}
\author[label1]{Xuhui Li\corref{cor1}}

\ead{lixuhui@whu.edu.cn}

\author[label3]{Mengchi Liu}

\ead{mengchi@scs.carleton.ca}

\author[label1,label4]{Shanfeng Zhu}
\ead{zhusf@fudan.edu.cn}

\author[label2]{Arif Ghafoor}
\ead{ghafoor@purdue.edu}

%

\cortext[cor1]{Corresponding Author}

\address[label1]{State Key Lab. of Software Engineering, Wuhan University, Wuhan, China}
\address[label3]{School of Computer Science, Carleton University, Ottawa, Canada}
\address[label4]{School of Computer Science, Fudan University, Shanghai, China}
\address[label2]{School of Electrical and Computer Engineering, Purdue University, West Lafayette, U.S.A.}
\begin{abstract}
Various query languages have been proposed to extract and restructure information in XML documents. These languages, usually claiming to be declarative, mainly consider the conjunctive relationships among data elements. In order to present the operations where the hierarchical and the disjunctive relationships need to be considered, such as restructuring hierarchy and handling heterogeneity, the programs in these languages often exhibit a procedural style and thus the declarativeness in them is not so prominent as in conventional query languages like SQL.

In this paper, we propose a declarative pattern-based functional XML query language named XML Tree Query (XTQ). XTQ adopts expressive composite patterns to present data extraction, meanwhile establishing the conjunctive, the disjunctive and the hierarchical relationships among data elements. It uses the matching terms, a composite structure of the variables bound to the matched data elements, to present a global sketch of the extracted data, and develops a deductive restructuring mechanism of matching terms to indicate data transformation, especially for restructuring hierarchy and handling heterogeneity. Based on matching terms, XTQ employs a coherent approach to function declaration and invocation to consistently extract and construct composite data structure, which integrates features of conventional functional languages and pattern-based query languages. Additionally, XTQ also supports data filtering on composite data structure such as hierarchical data, which is seldom deliberately considered in other studies. We demonstrate with various examples that XTQ can declaratively present complex XML queries which are common in practice.

\end{abstract}


\end{frontmatter}

\section{Introduction}
With the wide spread of XML documents over the Internet, XML query has become an important and interesting topic in utilizing and managing XML data. An XML query specifies certain retrieval operations, i.e., the extraction, filtering, transformation and  construction operations on XML data and is often presented in a certain XML query language. In the past decade, some XML query languages have been proposed, such as XPath/XQuery~\cite{xpath, xquery}, Xcerpt~\cite{xcerpt}, XTreeQuery~\cite{xtree}, XDuce~\cite{xduce}, CDuce~\cite{cduce} and TQL~\cite{tql}, among which XPath/XQuery have already become the standards advocated by W3C. Nevertheless, due to the increasing interests in XML documents management, people are still engaged in proposing new query languages~\cite{ksql} for XML or adding new features to existing languages like XQuery~\cite{xquery3}.

An XML document often consists of the data elements with various tags in a hierarchical structure. The elements with different tags can have similar structure and thus show the homogeneous semantics from a certain viewpoint, meanwhile the elements with the same tags can often have heterogeneous structure. This hierarchical-semistructured feature brings forth four kinds of requests to XML query. Firstly, similar to the ``select'' clause of a SQL query, an XML query often combines the interrelated data elements from the source documents to generate the output document; secondly, an XML query needs to parse the source document hierarchy and to build the target document hierarchy; thirdly, an XML query needs to discriminate the heterogeneous data and treat them respectively; and fourthly, an XML query needs to extract and reorganize the nested sets of (syntactically or semantically) homogeneous data elements embedded in the document hierarchy. These four kinds of requests can be intertwined orthogonally in a practical query, and thus a query language should have an expressive and coherent framework to present them.

Query languages often claim to be ``declarative'', as SQL did in relational query, aiming at enabling users to focus on presenting query requests without considering the procedures to carry them out. To achieve the declarativeness, a query language should be expressive enough to efficiently present the common query requests and should have a proper mechanism to infer the operations for implementing the query tasks. However, the declarativeness of existing XML query languages is not as prominent as in SQL when the query involves multiple kinds of the query requests, especially in handling heterogeneity and reorganizing nested sets. For example, given a sample document ``bib.xml'' storing book information which is composed of the elements of the authors, the editors, the title, the price and the publisher. Consider the following query: for the books whose title contains one or more authors' first name or last name, fetch these authors and all the editors, and for each of them list his/her name and the titles of the books that he/she plays a role (as author or editor) in. This query can be implemented in XQuery as Q1 shows.

\begin{footnotesize}\begin{verbatim}
Q1.
<results>{
    let $tempresults :=
            for $b := doc("bib.xml")/bib/book,
                $t = $b/title
            where (some $a in $b/author
                   satisfies contains($t,$a/first) or contains($t,$a/last))
            return
                <book>
                   {$t}
                   { for $a1 in $b/author
                     where  contains($t,$a1/first) or contains($t,$a1/last)
                     return {$a1} }
                   {$b/editor}
                </book>
    let $p := $tempresults/book(/author|/editor)
    for $last in distinct-values($tempresult/last),
        $first in distinct-values ($p[last=$last]/first)
    orderby $last, $first
    return
      <result>
         <person>
            <last>{ $last }</last> <first>{ $first }</first>
         </person>
         {  for $b1 in $$tempresults/book,
                $ae in $b1(/author | /editor)
                      [last=$last and first=$first]
            where exist($ae)
            return
               <book>
               { $b1/title
                 if (exist($b1/author[.=$ae])) then <role>author</role>
                 else ()
                 if (exist($b1/editor[.=$ae])) then <role>editor</role>
                 else () }
               </book>  }
      </result>   }
</results>
\end{verbatim}\end{footnotesize}

\noindent Q1 firstly uses a subquery to fetch the books and store them in a temporary document, and then uses the \emph{for}-clauses to get the distinct values of the person names in the temporary document; after that, in the subquery of the return clause it looks for the books for each person, and finally it uses conditional statements to decide his/her role. This program is little declarative for two reasons: a) it presents the query request with the detailed procedures, which requires the users be quite familiar with the procedure programming; and b) it uses temporary document and has redundant data retrieval operations on the people information, which increases the difficulty for the query processor to analyze and optimize the program.

With the exploration and the comparison to the existing XML query languages, we find that there is the lack of an expressive query expression and the associated mechanisms to process the independent data elements. Such a lack often leads to the insufficiency of the declarativeness in practical queries. Existing query languages usually conjunctively combine the interrelated data elements as a tuple, however, in XML queries it is also common for the semantically independent data elements to be combined and be processed under the same document context. The independent elements can occur exclusively, like the ``author'' or ``editor'' roles in the output document in Q1, or they can occur simultaneously but be subjectively treated as ``independent'' ones by the user, like the ``author'' and the ``editor'' elements in the source document. The former case is common in handling heterogeneous data, and the latter case is common in processing the sets where the elements are treated as independent to each other.

In this paper, we propose a new  XML query language called XTQ, standing for XML Tree Query, to overcome the lack. XTQ is a first-order functional query language which adopts expressive patterns with coherent mechanisms to achieve declarativeness in presenting the aformentioned query requests. The language has the following contributions:

a) XTQ introduces a novel structure to constitute the patterns for data extraction and construction. The patterns are tree-shape ones with the conjunctive and the disjunctive composite operators, for extracting data and presenting data structure from a logic perspective. In comparison with the existing studies which uses the tree-shape structural patterns and conjunctively combines the data elements, XTQ also employees the ``disjunctive'' combination of the subordinate patterns to indicates the data elements matching the sub-patterns to be treated as independent ones. This pattern structure can coherently specify the various kinds of combination of data elements and thus facilitate presenting common query requests.

b) XTQ employees a flexible and deductive mechanism  to declaratively present data transformation requests. It introduces a term expression named \emph{matching term} to present the composite structure of the data elements matching the patterns, and adopts a rewriting system for restructuring the matching terms to indicate the transformation requests such as reorganizing the nested sets.
In comparison with the rigid data types or schemas in existing studies, the matching terms can flexibly and coherently represent the data structure throughout the process from data extraction to construction. In comparison with other functional query languages which often specify the details of the restructuring and
derives the data structure (i.e., type) from the function body, XTQ uses a restructured matching term to declaratively specify the structure of the output data and leaves resolving the restructuring process to the parser.

c) XTQ allows data filtering on composite data structure by composing conditions in a bottom-up way along the inner hierarchy of matching terms. Existing languages often use predicates on tuples for data filtering, inheriting the mechanisms from conventional database, thus procedural steps are often required to filter data elements in disjunctive or hierarchical structure. The data filtering mechanism in XTQ introduces a structural condition composition and specifies a sound semantic model for data filtering. This mechanism can flexibly and consistently filter each portion of the restructured data hierarchy without decomposing it with separate clauses.

d) XTQ provides a uniform query style for function declaration and invocation. XTQ functions deploy a pattern matching mechanism for parsing the composite   argument structure, which is consistent with the one for document extraction. In addition, it uses a virtual function value to present the function invocation in the query clause,  which enables the functions be seamlessly embedded in to the query and thus enhances the expressiveness and the flexibility of  function invocation.


This paper substantially improves the work we once published in \cite{xtqdexa} which is a rudiment of the XTQ language. In the current version, we redesign the extraction patterns and the term restructuring mechanisms to make the language more expressive and clearer. We also introduce novel mechanisms of function declaration and invocation so that the language is more suitable for presenting data query than existing functional languages. Besides, we explicitly specify the semantics of the language and propose and prove the important properties of the language in this paper.

The remainder of the paper is organized as follows. Section 2 introduces the related works of XML query languages. Section 3 introduces the XTQ language on the major issues concerning document extraction and data transformation. Section 4 introduces the data filtering mechanism, especially discussing the filtering of composite data structure of XTQ. Section 5 introduces the enhanced issues  specific for functions in XTQ. Section 6 concludes the paper by making a comparison between XTQ and typical XML query languages on common features which are important to declarativeness.

\section{Related Works}

In catering to the hierarchical form and the descriptive tags of XML document, XML query languages usually adopt a pattern-based style for data operations in one way or another. In fact, the languages are distinct from each other in the design and the usage of the patterns, which makes it reasonable to compare them from the angle of presenting data operations with the patterns.

\textbf{Data extraction.} The patterns for data extraction play a fundamental role in XML query program, because they indicate the way of carrying out data extraction and directly specify the composite structure of the extracted data elements. Generally they can be classified into three kinds: the ones based on nested navigational expressions, the ones based on hierarchically organized rules, and the ones based on regular expression structure.

Navigational expression originates from object query and is adopted by early semi-structured data query languages such as Lorel~\cite{lorel} and XQL~\cite{xql}. It is also adopted by XPath, and thus is used in XQuery and XSLT\cite{xslt} as the basic means of data extraction. XPath deploys the powerful but complicated ``step'' functions as navigation operators to freely navigate along arbitrary routes between nodes. It also uses the syntax sugars ``/'' and ``//'' to navigate children and descendants nodes.  Navigation expressions often only extract homogeneous data elements. To conjunctively compose multiple data elements extracted by different XPath expressions, XQuery program uses the nested \emph{for}-clause and \emph{let}-clause in its FLWOR statement, as shown in Q1. However, this would result in two problems. Firstly, XQuery adopts a plain tuple as the only composite structure for the extracted data elements, so it is hard to present hierarchical and disjunctive relationships and needs more work to present the query requests involving them. Secondly,  the nested \emph{for}-clauses or \emph{let}-clauses indicate the procedures of data extraction. This brings extra costs to query processor for generating efficient query plan. To make the path expression more expressive, the logic language XPathLog~\cite{xpathlog} extends XPath by binding multiple variables to the query terms in predicate, which records the ``route'' of navigation. However, it still uses the conjunctive tuple of variables to represent data structure. Disjunction is also concerned in the extension of XPath expressions such as in \cite{xpathdisj}, but it is used as a union of the sub-patterns and no further mechanism is developed to use it to discriminately processing the independent data elements.

Hierarchical rule-based patterns are widely adopted in XML query languages as the sketch of the data of interest. Early studies such as UnQL~\cite{unql} and XML-QL~\cite{xmlql} use conjunctive pattern compositions and nested sub-queries to form a pattern hierarchy.  More considerate pattern-based languages like  XML-RL~\cite{xmlrl}, XTreeQuery, Xcerpt and TQL are then developed using tree-shape patterns to hierarchically organize the extract data. Additionally, some studies also investigate the pattern-based equivalent forms of subsets of XQuery \cite{next, material}. The obvious advantage of the tree-shape pattern against navigational expressions is that the pattern style is in accordance with the data hierarchy and thus it is more intuitive. Some languages like XTreeQuery and Xcerpt can extract the plain sets of data elements in matching the tree-shape patterns.

Regular expression structures, such as DTD schemas, of certain XML documents enable query languages to adopt more subtle patterns, usually based on the specific types, for data extraction. XDuce~\cite{xduce} is the first language which introduces the regular expression into the type constructors and forms a first-order functional language with the static types specific to XML documents. Then CDuce~\cite{cduce} extends the type system with the general types in simple functional languages, the Boolean connective type constructors and an interesting theory on semantic subtyping~\cite{semsub}.  CQL~\cite{cql} uses a SQL-like select-from statement embedded with the XML specific types and the pattern matching mechanism of CDuce, which is relatively easy for query users. On the contrary, Xtatic~\cite{xtatic} embeds the XDuce types into a C$^{\#}$-like object-oriented language to make the common programming language able to directly manipulate XML values. These languages employee the flexible type-based pattern construction in ML-like functional languages and the regular expression patterns can efficiently and accurately extract the preferred elements in the regular structure. Furthermore, the extracted data elements are also well typed so that the coherence of the data structure during the query is guaranteed. Some other studies also try to introduce the data types representing XML schemas into the conventional functional languages like Haskell~\cite{haskellxml}.
In addition, some query languages \cite{ksql} without a type system also adopt regular expression in their pattern structure and thus can be used to efficiently query the documents with regular structures. Even though regular expression patterns are powerful in data extraction, the languages can hardly work on the documents without the inner regular structure.

\textbf{Data construction.} Most of the XML query languages, e.g., XML-QL, XQuery, Xcerpt, XTreeQuery, TQL, XDuce, CDuce and CQL, use tree-shape patterns for data construction so that the output hierarchy can be intuitively presented. XQuery inherits the construction pattern  from XML-QL which can be treated as a function on the tuple of variables, but the lack of processing the nested sets compels the program to resort to the sub-queries for hierarchically building nested sets. TQL is similar with XQuery in presenting data construction, and it is more expressive since it supports disjunction patterns for construction. Xcerpt and XTreeQuery can utilize the simple hierarchical relationship in data construction, and thus the program is compact in building data hierarchy. However, Xcerpt and XTreeQuery don't consider disjunctive relationships so that it would incur multiple steps of queries to construct results where disjunctive data elements are to be separated or combined.

The typed languages like XDuce and CDuce usually intertwine the construction patterns with the extraction patterns in the function body. CQL separately present the construction pattern in the select clause, which is more intuitive for query users. Besides the common data transformation and construction features in other languages, these languages can also support the disjunctive combination of the patterns which handles the data elements exclusively occurred, and also support the specific nested sets, i.e., sequence of elements in regular structure.

\textbf{Data transformation.} Data transformation in XML query mainly focus on reorganizing the nested sets of the extracted data elements for constructing results. The languages like XQuery and XML-QL do not support the composite data structure except plain tuple and thus involve no data transformation but rearranging the locations of the data elements in the output document hierarchy. The logic languages like XTreeQuery and Xcerpt can implicitly transform a tuple of a plain set and the remainder, e.g., \textsl{(a, \{b\})} into a set of the tuples \textsl{\{(a, b)\}} by converting the original conjunctive relationship to the one between \textsl{a} and \textsl{b}. Due to the insufficient expressiveness in composite data structure, these languages often resort to sub-queries to carry out complex transformation such as grouping the set of tuples by distinct values.

The typed functional languages like XDuce, CDuce and CQL can process the nested sequence of elements with the pattern matching statements in the top-down way, and they also provide a flattening operation of the nested sequence of elements. Even though, they should also use the nested sub-queries to present the complex transformation involving value grouping.

\textbf{Data filtering.} The importance of data filtering is often underestimated in existing XML query languages. However, it actually affects the declarativeness of the language when a composite data structure is to be filtered. Query languages often use a ``\textsl{where}'' clause containing predicate conditions to present data filtering requests. The composite structure of the data elements are seldom concerned  in the conditions since they are often treated as conjunctive ones as in XQuery. If a complicated filtering request is to be put on a composite data structure, say, a nested set or a union of sets, it would inevitably incur multiple subqueries to specify a reasonable filtering procedure. Especially, the languages such as XDuce, CDuce and TQL which support the disjunctive combination of patterns also use the sub-queries combined disjunctively to specify the data filtering on the disjunctive patterns.

Additionally, many studies on XML query explore various ways to make the query languages more expressive and practical in the certain aspects. First of all, some studies enhance the XML query from the angle of information retrieval. For example, XIRQL~\cite{xirql} extends XPath with typical retrieval features such as index term weighting, specificity-oriented search and structural vagueness, and some recent studies like \cite{keyword, xmlkeyword} show interests in XML keyword search. Secondly, some studies try to improve the interaction of presenting queries through a certain user-friendly interface. For example, XQBE~\cite{xqbe} is a visual query language for expressing a large subset of XQuery in a visual form; and NaLIX~\cite{nalix} is a generic interactive
natural language query interface to translate an arbitrary English language sentence into an XQuery expression that can be evaluated against an XML database. Thirdly, some studies extend the query languages to query the unspecified structure of XML documents. For example, in \cite{meaningful} a new notion of Meaningful Query Focus is designed to enhance the XQuery language for finding related nodes within a schema-free XML document; in \cite{fuzzyxpath} a Fuzzy-Set-based querying framework called FuzzyXPath is designed to enhance the XPath language with fuzzy conditions for the definition of
flexible constraints on stored data. Fourthly, as temporal and spatial XML documents are quite general in practical applications, some query languages or methods are introduced to specify the temporal or the spatial XML document queries. For example, there are some extension of XPath~\cite{sxpath, txpath} or XQuery~\cite{tempxml}, and we also propose a temporal extension of the original version of XTQ~\cite{tempxtq}.
These enhancements in the specific aspects enable the query languages to present the certain kinds of requests, however, they provide no solution to the deficiencies mentioned above.

Except for retrieval operations of XML query, the update operation of XML documents is also universally studied in the past decade. However, it is beyond the scope of this paper and thus we don't discuss this topic here.

\textbf{Problems in existing languages.} For the four kinds of XML query requests, handling the heterogeneity and reorganizing the nested sets are seldom easy for existing query languages to present. Although some languages support disjunctive combination of query patterns, they still need to use sub-queries to present the complex queries intertwining those requests. As far as we know, no XML query language has proposed a coherent model to specify the extraction and the transformation on the interrelated and independent data elements and further develop the sound data filtering and construction mechanisms base on them. This limitation hinders the efficient and declarative presentation of many complex query requests.

\section{Core XTQ Language}
\subsection{Synopsis of XTQ}
XTQ language is a pattern-based functional language. It adopts the \textbf{Query-Where-Construct} (QWC) statement which presents data extraction, data filtering and data construction separately in the three clauses.  The synopsis of its abstract syntax is listed in Figure 1, and listed below is an example of XTQ query which implements the query request described in Section 1.

\begin{footnotesize}\begin{verbatim}
Q2. query doc("bib.xml")//book(/title=>$t,(/author=>$a||/editor=>$e)
                                               (/last=>$l, /first=>$f))
    where contains($t, ($l hid $a)) or contains($t, ($f hid $a))
    construct results=>{result=>(person=>(last=>$l, first=>$f)%,
                                 {book=>(title=>$t%,
                                   {(role=>"author" hid $a) |
                                    (role=>"editor" hid $e) })} )}
                        orderby (($l hid $f)% asc, ($f hid $l)% asc)
\end{verbatim}\end{footnotesize}

\begin{figure}
\begin{center}
\begin{small}
\begin{tabular}{|p{0.13\textwidth} p{0.02\textwidth} p{0.75\textwidth}|}
  \hline
  $<$qwcstmt$>$ &::=& `\textbf{query}' $<$queryexpr$>$  `\textbf{construct}' $<$consptn$>$ $|$\\
  && `\textbf{query}' $<$queryexpr$>$ `\textbf{where}' $<$condition$>$ \\&& `\textbf{construct}' $<$consptn$>$\\
  $<$fundecl$>$ &::=& `\textbf{declare}' $<$funsig$>$ `\textbf{as}' \textbf{(}$<$qwcstmt$>$\textbf{)} \\
  $<$funsig$>$ &::=& $<$funname$>$ `('$<$valtype$>$`)' `:' $<$valtype$>$ \\
  $<$queryexpr$>$ &::=& $<$datasrc$>$ $<$extrptn$>$ $|$ $<$compqexpr$>$\\
 $<$datasrc$>$ &:=& $<$docsrc$>$ $|$ $<$valsrc$>$\\
  \hline
\end{tabular}
\caption{Syntax synopsis of XTQ}
\end{small}
\end{center}
\end{figure}

The clauses in the QWC statement employs several patterns to present data operations. The \textsl{query} clause uses (composite) query expressions consisting of a data source, e.g., ``\textsl{doc(`bib.xml')}'' and an extraction pattern, e.g., ``\textsl{//book(...)}''. The composite value consisting of the data elements matching the extraction pattern is represented by a special expression named \textsl{matching term}, a structural composition of variables, which plays a central role in the whole query. A matching term can be restructured to another one following certain restructuring rules, which indicates transforming data bound to the variables to be of the new structure.

In the \textsl{where} clause and the \textsl{construct} clause, a data construction request is presented with a special kind of function invocations named \textsl{construction pattern}, e.g., ``\textsl{person$\Rightarrow$(last$\Rightarrow$\$l, first$\Rightarrow$\$f)\%}''. A construction pattern has a (restructured) matching term as its argument, indicating to generate a document fragment from the composite value corresponding to the term. The \textsl{where} clause deploys a (compound) condition, consisting of one or more predicate function invocations, e.g., ``\textsl{contains(\$t, (\$l hid \$a))}'',  on elements of the original matching terms, to structurally indicate filtering the restructured data. The filtered data is then used in the \textsl{construct} clause to build the query result.

XTQ uses first-order functions to present sub-queries. A standard function in XTQ is declared as a \textsl{QWC} statement with the function signature indicating the input value type and output value type of the function. The functions can be directly invoked in construction patterns or be indirectly invoked in query expressions.

The features of the language are illustrated in detail with the definitions of these patterns and the mechanisms of the data operations in the following subsections or sections.

%
%
%
%

\subsection{Data Model of XTQ}
XTQ deploys a hierarchical data model named Set-based XML Data (SXD)
model to represent the XML document and indicate the data structure
in manipulation. We assume the existence of a set \textsl{C} of
constants used as the attribute and the element values, a set \textsl{N}
of names used as the attribute and the element names, and a set \textsl{L} of location constants used as the location values of an element or an attribute in the document. The location set \textsl{L} is accompanied with the binary relations ``/'' and  ``$\ll$'' which respectively indicates that the two elements be the parent-children elements or be the preceding-succeeding siblings. Especially, we treat the attributes as a special kind of elements whose names have a  prefix ``@'', and treat a content value of an element as a special kind of element whose name is ``$\sim$''.

The SXD model is constituted by the set-based document fragments which are sets of valid elements.
\newdefinition{definition}{Definition}
\begin{definition}
A document fragment set \textsl{D} and an element set \textsl{E} are mutually and inductively defined as follows:\\
1. \textsl{E} $\subset$ \textsl{N$\times$C$\times$L} $\cup$ \textsl{N$\times$D$\times$L};\\
2. \textsl{D} $\subset$ \textsl{2$^{E}$};\\
3. For \textsl{e$\in$E} that \textsl{e}=\textsl{(n,d,l)} and \textsl{d}=\textsl{\{e$_{i}$\}}, if  \textsl{e$_{i}$}=\textsl{(n$_{i}$,c$_{i}$,l$_{i}$)} or \textsl{e$_{i}$}=\textsl{(n$_{i}$,d$_{i}$,l$_{i}$)} then \textsl{l/l$_{i}$} holds;\\
4. For \textsl{d$\in$D}, \textsl{d}=$\emptyset$, or there is a permutation \emph{p} of the elements in \emph{d} that p = (e$_{1}$, \dots, e$_{n}$) where that \textsl{e$_{i}$}=\textsl{(n$_{i}$,c$_{i}$,l$_{i}$)} or \textsl{(n$_{i}$,d$_{i}$,l$_{i}$) (i$\in$\{1..n\})} and \textsl{l$_{j}$$\ll$l$_{j+1}$(j$\in$\{1..n-1\})} hold.
\end{definition}

SXD is designed following the idea that the conjunctive relationships between the elements are established only if they are necessary. That is, in the model the elements are independent to each other if users don't explicitly consider the latent relationships, e.g., parent-children, siblings, among them. Comparing to the typed-based models which treat elements as being interrelated with one another, SXD model puts a fairly loose restriction on the document structure.


For convenience, we often use \textsl{d} to denote the content of an element which can be either a fragment or a constant, and denote the element \textsl{(n, d, l)} as ``\textsl{n$\Rightarrow_{l}$d}'' or simply ``\textsl{n$\Rightarrow$d}'' when the location \textsl{l} can be inferred from the context. Since the elements of a set have successive locations, we can use the list   \textsl{d$_{l}$}=\textsl{\{$|$e$_{1}$;\dots;e$_{n}$$|$\}} to represent the successive permutation of the element set \textsl{d} based on the order derived from the location relation ``$\ll$''. In specifying the composition of the document fragments in the successive way, we treat \textsl{\{$|$d$_{1}$;\dots;d$_{n}$$|$\};\{$|$d'$_{1}$;\dots;d'$_{m}$$|$\}} = \textsl{\{$|$d$_{1}$;\dots;d$_{n}$; d'$_{1}$;\dots;d'$_{m}$$|$\}} and \textsl{\{$|$\{$|$d$_{1}$;\dots;d$_{n}$$|$\}$|$\}} = \textsl{\{$|$d$_{1}$;\dots;d$_{n}$$|$\}}. The successive representation of the SXD model can omit the location information and facilitate structurally presenting the document.  For example, a sample fragment in a ``bib.xml'' document can be represented as

%
\begin{footnotesize}\begin{verbatim}
 {| book=> {|
      @year=>"1996"; 
      title=>"Selected Poems And Four Plays of William Butler Yeats";
      author=>{|last=>"Yeats"; first=>"Willam"; 
                email=>"wbyeats@amazon.com"|};
      author=>{|last=>"Artificial"; first=>"Data"|};          
      editor=>{|last=>"Rosenthal"; first=>"M.L.";
                email=>"mlrosenthal@amazon.com"|};
      price=>"15.87" |};      
    book=> {|
      @year=>"1989"; title=>"The Selected Poems of Paul Blackburn";
      author=>{|last=>"Blackburn"; first=>"Paul";
                email=>"pbburn@amazon.com"|};
      author=>{|last=>"Aritifical"; first=>"Info."|};
      editor=>{|last=>"Rosenthal"; first=>"M.L.";
                email=>"mlrosenthal@amazon.com"|};
      editor=>{|last=>"Jarolim"; first=>"Edie"; 
                email=>"ejarolim@amazon.com"|};
      price=>"55.00" |};  ... |}                       (sampledoc)
\end{verbatim}\end{footnotesize}
\noindent where a \textsl{book} element contains several
sub-elements such as \textsl{title}, \textsl{author},
\textsl{editor} and an attribute \textsl{year} (some data such as the email elements are artificial). This sample fragment works as a running example in the paper.
For brevity, we respectively
denote the content of the \textsl{book}, \textsl{title},
\textsl{author} and \textsl{editor} elements as
\textsl{d$_{b1}$}, \textsl{d$_{t1}$}, \textsl{d$_{a11}$}, \textsl{d$_{a12}$}, \textsl{d$_{e1}$},\textsl{d$_{b2}$}, \textsl{d$_{t2}$}, \textsl{d$_{a21}$}, \textsl{d$_{a22}$}, \textsl{d$_{e21}$} and \textsl{d$_{e22}$}, and we also denote the location of those elements as \textsl{l$_{n}$} whose subscripts \textsl{n} are the same as the ones of the element content \textsl{d$_{n}$}.

Since a document fragment is solely determined by the content and the location of its elements, there is naturally a equivalence between two fragments \textsl{d} and \textsl{d'} if they share the same content and there is a isomorphism between the locations of the elements in \textsl{d} and \textsl{d'}.

\begin{definition}
The document fragments \textsl{d} and \textsl{d'} are structurally equivalent with each other, denoted as  \textsl{d$\equiv$d'}, iff:\\
1. \textsl{d}=\textsl{c} and \textsl{d'}=\textsl{c};\\
2. \textsl{d$_{l}$} = \textsl{\{$|$e$_{1}$;\dots;e$_{n}$$|$\}} and \textsl{d'$_{l}$}=\textsl{\{$|$e'$_{1}$;\dots;e'$_{n}$$|$\}} where \textsl{n$_{i}$}=\textsl{n'$_{i}$} and \textsl{d$_{i}$}$\equiv$\textsl{d'$_{i}$} for each \textsl{e$_{i}$} = \textsl{n$_{i}$$\Rightarrow$d$_{i}$} and \textsl{e'$_{i}$} = \textsl{n'$_{i}$$\Rightarrow$d'$_{i}$}.
\end{definition}


\subsection{Data Extraction}

A query in XTQ begins with data extraction which is specified by query expressions indicating matching an extraction pattern with a data source. The syntax of data extraction is specified in Figure 2. A data source is a SXD document specified as a ``doc()'' function invocation like ``\textsl{doc(`bib.xml')}'', or a value being of a certain type defined in XTQ. An extraction pattern extracts data elements from the data source and establishes a composite data structure as the extraction result. The query expressions can be composed using the operators ``,'' and ``$||$''.
In this subsection, we only discuss the issue of extracting data from documents, i.e., using the \textsl{doc()} function, with the document extraction patterns and the general composite query expressions. The value extraction pattern which is often used in the query clause involving the function specification and invocation would be introduced in Section 5.

The semantics of  data extraction is listed in  Figure 3. Here we use \textsl{n} to denote the element names, use \textsl{e} to denote the elements, use \textsl{d} to denote the document fragments or constant values, use \textsl{p} to denote the extraction patterns,  use \textsl{b} to denote the bindings which are the results of data extraction.
The semantics of data extraction is specified in the form of ``\textsl{d $|$ p $\rightarrow$ b}'', indicating that matching the pattern \textsl{p} with the fragment \textsl{d} should result in the binding \textsl{b}.

\begin{figure}
\begin{center}
\begin{small}
\begin{tabular}{|p{0.15\textwidth} p{0.02\textwidth} p{0.73\textwidth}|}
  \hline
    $<$queryexpr$>$ &::=& $<$datasrc$>$ $<$extrptn$>$ $|$
    $<$compqexpr$>$\\
    $<$compqexpr$>$ &::=& 
     `\textbf{(}'$<$queryexpr$>`$\textbf{)}' $|$$<$queryexpr$>$`\textbf{,}'$<$queryexpr$>$ $|$  \\
 && $<$queryexpr$>$`\textbf{$||$}'$<$queryexpr$>$ \\
   $<$extrptn$>$ &::=&  $<$docexptn$>$ $|$ `['$<$valexptn$>$`]' $|$ $<$extrptn$>$`,'$<$extrptn$>$ $|$ \\
    &&  $<$extrptn$>$ `$||$' $<$extrptn$>$ $|$ `('$<$extrptn$>$`)'\\
    $<$docexptn$>$ &::=&  $<$elemptn$>$ $|$ $<$attrptn$>$ $|$ $<$elemptn$>$ $<$docexptn$>$ $|$ \\
    && `/'$<$atomptn$>$`\textbf{;}'$<$docexptn$>$ $|$ $<$docexptn$>$`,'$<$docexptn$>$ $|$ \\
    &&  $<$docexptn$>$ `$||$' $<$docexptn$>$ $|$ `('$<$docexptn$>$`)'\\
   $<$elemptn$>$ &::=&  `/'$<$atomptn$>$ $|$ `//'$<$atomptn$>$\\ 
   $<$attrptn$>$ &::=& `@'$<$atomptn$>$\\
   $<$atomptn$>$ &::=& $<$name$>$`\textbf{$\Rightarrow$}'$<$const$>$ $|$ $<$name$>$`\textbf{$\Rightarrow$}'$<$var$>$ $|$ \\
    && $<$var$>$`\textbf{$\Rightarrow$}'$<$const$>$ $|$ $<$var$>$`\textbf{$\Rightarrow$}'$<$var$>$ \\
\hline
\end{tabular}
\caption{Syntax of extraction patterns}
\end{small}
\end{center}
\end{figure}

\begin{figure}[h]
\begin{center}
\begin{footnotesize}
\begin{tabular}{|p{\textwidth}|}
  \hline
\textbf{Atomic pattern matching} \\
$\dfrac{}{n\Rightarrow_{l}d\hspace{2pt}  | \hspace{2pt} n\Rightarrow\$x \rightarrow \$x\mapsto(d,l)} $ (cont-match) \hspace{10pt}
$\dfrac{n \neq n'}{n{\Rightarrow_{l}}d \hspace{2pt} | \hspace{2pt}n'\Rightarrow\$x \rightarrow \$x{\mapsto}null}$  (cont-mis) \\
\vspace{1pt}
$\dfrac{}{n\Rightarrow_{l}d \hspace{2pt}  | \hspace{2pt}  \$x{\Rightarrow}d \rightarrow \$x\mapsto(n,l)}$  (name-match) \hspace{10pt}
$\dfrac{d \neq d'} {n\Rightarrow_{l}d\hspace{2pt} |\hspace{2pt} \$x{\Rightarrow}d' \rightarrow \$x{\mapsto}null} $  (name-mis) \\
\vspace{1pt}
$\dfrac{}{n\Rightarrow_{l}d \hspace{2pt}|\hspace{2pt} \$x\Rightarrow\$y \rightarrow (\$x\mapsto(n,l),\$y\mapsto(d,l))}$ (name-cont-match) \\
\hline
\textbf{Element and attribute pattern matching} \\
$\dfrac{ d \hspace{2pt}|\hspace{2pt} p_{e} \rightarrow b,  d' \hspace{2pt}|\hspace{2pt} p_{e} \rightarrow b' }
{ d \cup d'\hspace{2pt} |\hspace{2pt} p_{e} \rightarrow  \#(b \cup b')}$ (ep-comp) $^{\dag}$ \hspace{10pt} $\dfrac{n\Rightarrow_{l}c \hspace{2pt}|\hspace{2pt} p_{a} \rightarrow  b} {\{ @n\Rightarrow_{l}c\} \hspace{2pt}|\hspace{2pt} @p_{a} \rightarrow  \{b\}}$  (attr-match)\\
\vspace{1pt}
$\dfrac{e \hspace{2pt}|\hspace{2pt} p_{a} \rightarrow  b }{  \{ e \} | /p_{a} \rightarrow  \{b\} }$   (pc-match)
\hspace{5pt}
$\dfrac{ d \hspace{2pt}|\hspace{2pt} //p_{a} \rightarrow \{b_{i} \hspace{2pt}|\hspace{2pt} i\in\{1..n\}\}, n\Rightarrow_{l}d \hspace{2pt}|\hspace{2pt} p_{a} \rightarrow b_{0}  }
{ \{n\Rightarrow_{l}d\}\hspace{2pt} |\hspace{2pt} //p_{a} \hspace{2pt}  \rightarrow \hspace{2pt} \#(\{b_{i}| i\in\{0..n\}\})}$  (ad-match)
\\
\hline
\textbf{Composite pattern matching} \\
$\dfrac{ d \hspace{2pt}|\hspace{2pt} p_{t} \rightarrow b,  d' \hspace{2pt}|\hspace{2pt} p_{t} \rightarrow b' }
{ d \cup d'\hspace{2pt} |\hspace{2pt} p_{t} \rightarrow  \#(b \cup b')}$ (tree-comp)  $\dfrac{ d \hspace{2pt}|\hspace{2pt} p \rightarrow b,  \hspace{2pt} n\Rightarrow_{l}d \hspace{2pt}|\hspace{2pt} p_{a} \rightarrow b',  \neg null(b')}{
\{ n\Rightarrow_{l}d \} \hspace{2pt}|\hspace{2pt} /p_{a}(p) \rightarrow \{(b',b)\}}$  (pc-tree-match) $^{\ddag}$ \\
\vspace{1pt}
$\dfrac{ d \hspace{2pt}|\hspace{2pt} //p_{a}(p) \rightarrow  \{b_{i} | i\in\{1..n\}\},
\{ n\Rightarrow_{l}d \} \hspace{2pt}|\hspace{2pt} /p_{a}(p) \rightarrow  \{b_{0}\} }
{ \{n\Rightarrow_{l}d\}\hspace{2pt} |\hspace{2pt} //p_{a}(p) \rightarrow \#(\{b_{i} | i\in\{0..n\}\}) }$(ad-tree-match)\\
\vspace{1pt}
$\dfrac{d \hspace{2pt}|\hspace{2pt} /p_{a} \rightarrow \{b_{i} | i\in\{1..n\}, \neg null(b_{i})\}, d \hspace{2pt}|\hspace{2pt} p \rightarrow b}{ d \hspace{2pt}|\hspace{2pt} /p_{a};p \rightarrow \{(b_{i},b)| i\in\{1..n\}\}}$ (fold-comp)
$\dfrac{d \hspace{2pt}|\hspace{2pt} /p_{a}\rightarrow\{b\},  null(b)}{ d \hspace{2pt}|\hspace{2pt} /p_{a};p \rightarrow \{b\}}$ (fold-null)\\
\vspace{1pt}
$\dfrac{ d \hspace{2pt}|\hspace{2pt} p \rightarrow b, d \hspace{2pt}|\hspace{2pt} p' \rightarrow b'}{  d \hspace{2pt}| \hspace{2pt}(p, p') \rightarrow (b, b') }$     (conj-comp) \hspace{10pt}
$\dfrac{ d \hspace{2pt}|\hspace{2pt} p \rightarrow b, d \hspace{2pt}|\hspace{2pt} p' \rightarrow b'}{  d \hspace{2pt}|\hspace{2pt} (p \hspace{2pt}||\hspace{2pt} p') \rightarrow (b\hspace{2pt}||\hspace{2pt}b')}$   (disj-comp)
\\
\vspace{0.5pt}
$\dag$ \#(b) denotes the result of reducing the redundant null bindings from the set binding b.\\
$\ddag$ null(b) indicates that the binding b be null.\\
\hline
\end{tabular}
\end{footnotesize}
\caption{Semantic rules of data extraction}
\end{center}
\end{figure}

As Figure 3 indicates, a binding is  specified as \\
\indent \textsl{b ::= x$\mapsto$(d,l) $|$ x$\mapsto$`null' $|$ (b$_{1}$,\dots, b$_{n}$) $|$ b$_{1}$$||$\dots$||$b$_{n}$ $|$ \{b$_{1}$,\dots, b$_{n}$\}}\\
where \textsl{x} denotes a variable, \textsl{l} denotes a location value and ``\textsl{null}'' is a specifical value indicating the mismatching. The composite binding has three kinds of  structures, the conjunctive one like \textsl{(b$_{1}$,\dots, b$_{n}$)}, the disjunctive one like \textsl{b$_{1}$$||$\dots$||$b$_{n}$} and the set one like \textsl{\{b$_{1}$,\dots, b$_{n}$\}}.

As Figure 2 shows, the fundamental expressions of an extraction pattern are the atomic patterns evolved from the element pair ``\textsl{n$\Rightarrow$d}'' or ``\textsl{n$\Rightarrow$c}'' by substituting one or both sides with variable(s) , e.g., ``\textsl{title$\Rightarrow$\$t}''
or ``\textsl{\$n$\Rightarrow$\$c}''. As shown in the atomic pattern matching rules in Figure 3, when an element ``\textsl{n$\Rightarrow_{l}$d}'' is matched against an atomic pattern, the variable(s) would be bound to the value(s) in the corresponding place(s) and the location \textsl{l} as the result of data extraction, if the non-variable part of the pattern matches the corresponding values in the element. Otherwise, the variable would be bound to a ``null'' value.  For example, matching the atomic pattern ``\textsl{title$\Rightarrow$\$t}'' with the element ``\textsl{title$\Rightarrow_{l_{t1}}$d$_{t1}$}''  would result in a variable-value binding ``\textsl{\$t $\mapsto$(d$_{t1}$,l$_{t1}$)}'', while another element ``\textsl{author$\Rightarrow_{l_{a11}}$d$_{a11}$}'' would result in  ``\textsl{\$t$\mapsto$null}''.

An \textbf{element pattern} or an \textbf{attribute pattern} is composed of a prefix and an atomic pattern. The location prefix  ``/'' or
``//''  respectively indicates to match the children elements or the descendant elements. In the element pattern matching rules, the \textsl{ep-comp} rule shows that matching an element or attribute pattern \textsl{p$_{e}$} with an element set \textsl{d} should be inferred from matching the pattern with the union of the singleton element sets. The \textsl{pc-match} rule shows that matching an children element pattern ``\textsl{/p$_{a}$}'' with a singleton element set \textsl{\{e\}} would result in a singleton set containing the binding for matching the atomic pattern ``\textsl{p$_{a}$}'' with the element \textsl{e}. The \textsl{ad-match} rule shows that matching the descendant element pattern ``\textsl{//p$_{a}$}'' with the singleton element set \textsl{\{e\}} would recursively combine the bindings of matching \textsl{/p$_{a}$} with all the descendant elements of \textsl{e}. The \textsl{at-match} rule works similarly as the \textsl{pc-match} rule. Especially, a notation ``\#'' is introduced in the post-processing of the set binding of an element pattern or tree pattern which might contain redundant null bindings due to the mismatching. For a set binding \textsl{b}, if it only contains the null bindings, then \textsl{\#(b)} is a singleton set which contains a null binding; otherwise, \textsl{\#(b)} is the result of eliminating all the null binding from \textsl{b}.
For example, matching the element pattern ``\textsl{//author$\Rightarrow$\$a}'' with  \textsl{sampledoc} would result in a binding set \textsl{\{\$a$\mapsto$(d$_{a11}$,l$_{a11}$), \$a$\mapsto$(d$_{a12}$,l$_{a12}$), \$a$\mapsto$(d$_{a21}$,l$_{a21}$),...\}}.

XTQ uses the hierarchical extraction patterns to generate nested sets of data elements. A common case is the tree pattern ``\textsl{p$_{r}$ p$_{b}$}'' composed of an element pattern \textsl{p$_{r}$} and a general extraction pattern \textsl{p$_{b}$}. It is used to match a hierarchical document fragment where the patterns in the branch pattern \textsl{p$_{b}$} are to be matched with the content of each element matching the root pattern \textsl{p$_{r}$}. As the \textsl{tree-comp}, the \textsl{pc-tree-match} and the \textsl{ad-tree-comp} rules show, the matching of a tree pattern works like the one of an element pattern. For example, matching the tree pattern ``\textsl{//book$\Rightarrow$\$b/author$\Rightarrow$\$a}'' with  \textsl{sampledoc} would result in a nested set binding like ``\textsl{\{(\$b$\mapsto$(d$_{b1}$,l$_{b1}$),\{\$a$\mapsto$(d$_{a11}$,l$_{a11}$), \$a$\mapsto$(d$_{a12}$,l$_{a12}$)\}), \dots\}}''.

Another kind of hierarchical extraction patterns is the folding patterns in the form of ``\textsl{/p$_{a}$;p}'', which is often used to generate a nested set from a plain set of sibling elements. As the \textsl{fold-comp} rule shows, for each binding \textsl{b$_{i}$} of matching the children element pattern \textsl{/p$_{a}$}, the binding \textsl{b} of matching the pattern \textsl{p} would be duplicated and conjunctively associated with \textsl{b$_{i}$}. With a proper filtering on \textsl{(b$_{i}$,b)}, the duplications of \textsl{b} would be various with respect to \textsl{b$_{i}$}. For example, for the folding pattern ``\textsl{//books$\Rightarrow$\$b(/author$\Rightarrow$\$a;/editor $\Rightarrow$\$e)}'' we can restrict that the emails of \textsl{\$e} have the same domain as the one of \textsl{\$a}, thus generating a nested set binding of \textsl{\{(\$a,\{\$e\})\}} where each binding of \textsl{\{\$e\}} is specific to the binding of \textsl{\$a} it is associated with.

The extraction pattern can be composed conjunctively or disjunctively using the infix operator ``,'' or ``$||$'', indicating that the data elements matching the subordinate patterns be treated as being interrelated or independent. The bindings of the subordinate patterns are correspondingly combined with the operator ``,'' or ``$||$'', as the \textsl{conj-comp} and the \textsl{disj-comp} rules show. The composition of the bindings with the operators ``,'' or ``$||$'' are associative, e.g., \textsl{(b$_{1}$, (b$_{2}$, b$_{3}$)) = (b$_{1}$, b$_{2}$, b$_{3}$)} holds.

The conjunction and the disjunction patterns exhibit the logic feature of the extraction patterns in XTQ, that is, the extraction patterns can be treated as predicates on the document fragments, and the predicates can be conjunctively or disjunctively combined in a logical way rather than a structural way. For example, for each book which has less than 3 authors to find the authors that have a certain last name, we can use the pattern ``\textsl{//book$\Rightarrow$\$b(/author$\Rightarrow$\$a1/last$\Rightarrow$\$l, /author$\Rightarrow$\$a2)}'' that for each book element generates two different but conjunctively combined set binding of authors. By restricting the binding of \textsl{\$l} be the certain last name and restricting the cardinality of the set binding \textsl{\{\$a2\}} be less than 3, the filtering result of \textsl{\{\$a1\}} would be the authors required.

The disjunction pattern is used to treat data elements as being logically independent rather than be solely used to match exclusively occurred elements in some typed query languages. The result of a disjunction pattern is a disjoint union of the results of the subpatterns. The goal of introducing disjunction pattern is to handle heterogeneity.  On one hand, the elements in the disjoint union can be regarded as in the same kind against the outside conjunctive context, thus heterogenous data can be extracted and treated homogeneously. For example, in the sample fragment the the content of the authors and the editors are heterogeneous but can be treated homogeneously with respect to the title. On the other hand, the elements can be enumerated and treated separately, thus homogeneous data can be extracted and treated heterogeneously. For example, the pattern ``\textsl{(/author$\Rightarrow$\$a $||$ /author$\Rightarrow$\$e)}'' indicates that the  content of the
authors would be extracted and bound to \textsl{\$a} and
\textsl{\$e} independently (but not exclusively), and \textsl{\$a} and \textsl{\$e} can be
filtered and used in different ways.

Especially, different element or tree patterns composed disjunctively can contain same variables, e.g., the variables \textsl{\$l} and \textsl{\$f} in the following extraction pattern

\begin{footnotesize}\begin{verbatim}
E1. //book=>$b(/title=>$t, (/author=>$a(/last=>$l,/first=>$f) ||
                            /editor=>$e(/last=>$l,/first=>$f) ) )
\end{verbatim}\end{footnotesize}
Additionally, disjunction patterns having a common branch pattern can adopt a syntax sugar to use the pattern as common suffix, e.g., \textsl{(/author$\Rightarrow$\$a $||$ /editor$\Rightarrow$\$e)(/last$\Rightarrow$\$l,/first$\Rightarrow$\$f)}''.

The pattern E1 indicates to find all the book elements and for each book find the year, the title and the authors or the editors with their last and first names. Here the authors and the editors are treated as being independent to each other. Following the semantic rules, the result of matching E1 with \textsl{sampledoc} is the binding like: \\
\begin{small}
\textsl{
\{(\$b$\mapsto$(d$_{b1}$,l$_{b1}$), \\
\indent( \{\$t$\mapsto$(d$_{t1}$,l$_{t1}$)\},\\
\indent   ( \{(\$a$\mapsto$(d$_{a11}$,l$_{a11}$),(\{\$l$\mapsto$("Yeats",l$_{l11}$)\},\{\$f$\mapsto$("Willam" ,l$_{f11}$)\})), \\
\indent            (\$a$\mapsto$(d$_{a12}$,l$_{a12}$),(\{\$l$\mapsto$("Artificial",l$_{l12}$)\},\{\$f$\mapsto$("Data",l$_{f12}$)\}))\} $||$   \\
\indent         \{(\$e$\mapsto$(d$_{e1}$,l$_{e1}$),(\{\$l$\mapsto$("Rosenthal",l$_{l13}$)\},\{\$f$\mapsto$("M.L.",l$_{f13}$)\}))\} ) ) ),  \\
  (\$b$\mapsto$(d$_{b2}$,l$_{b2}$),  \\
 \indent     ( \{\$t$\mapsto$(d$_{t2}$,l$_{t2}$)\},  \\
\indent        ( \{(\$a$\mapsto$(d$_{a21}$,l$_{a21}$),(\{\$l$\mapsto$("Blackburn",l$_{l21}$)\},\{\$f$\mapsto$("Paul",l$_{f21}$)\})),  \\
\indent            (\$a$\mapsto$(d$_{a22}$,l$_{a22}$),(\{\$l$\mapsto$("Aritifical",l$_{l22}$)\},\{\$f$\mapsto$("Info.",l$_{f22}$)\}))\} $||$  \\
\indent          \{(\$e$\mapsto$(d$_{e21}$,l$_{e21}$),(\{\$l$\mapsto$("Rosenthal",l$_{l23}$)\},\{\$f$\mapsto$("M.L.",l$_{f23}$)\})),,  \\
\indent            (\$a$\mapsto$(d$_{e22}$,l$_{e22}$),(\{\$l$\mapsto$("Jarolim",l$_{l24}$)\},\{\$f$\mapsto$("Edie",l$_{f24}$)\}))\}  ) ) )\\
  \dots \}    }    (sample-binding)
\end{small}
\\For brevity, we use \textsl{b$_{x,n}$} to denote the variable binding \textsl{x$\mapsto$(d$_{n}$,l$_{n}$)} in \textsl{sample-binding}. For example, \textsl{b$_{\$b,b1}$} denotes ``\textsl{\$b$\mapsto$(d$_{b1}$,l$_{b1}$)}''.

A composite query expression indicates to compose the bindings from the subordinate query expressions with the operators ``,'' and ``$||$'' as the result of the query clause.  For example, in the following query clauses, the former one indicates to extract the book information from different files to combine them as homogenous data, the latter one indicates to extract the book information and the journal information form different files and join them to find the latent relationships.

\begin{small}\begin{verbatim}
query (doc("bib1.xml")||doc("bib2.xml"))//book=>$b/title=>$t
query doc("bib.xml")//book=>$b/author=>$a1,
      doc("journal.xml")//paper=>$p/author=>$a2
\end{verbatim}\end{small}

\subsection{Matching Terms and Term Restructuring}
\subsubsection{Matching Terms}

To explicitly specify the structure of the data elements extracted and transformed in the query, XTQ introduces an expression named \textsl{matching term}.
\begin{definition}
A \textbf{matching term} is  inductively defined as:\\
\indent\indent \textsl{t\hspace{5pt} ::=\hspace{5pt}x \hspace{5pt}$|$ \hspace{5pt}(t,t)\hspace{5pt} $|$ \hspace{5pt}(t$||$t) \hspace{5pt}$|$ \hspace{5pt}(t$|$t) \hspace{5pt}$|$ \hspace{5pt}\{t\}$_{t}$ \hspace{5pt}$|$\hspace{5pt} t\%\hspace{5pt}$|$\hspace{5pt}$\epsilon$}
\end{definition}
Here we use \textsl{t} to range over the matching terms, and use \textsl{x} to range over the variables.
As the definition shows, a matching term is often composed of the variables and the conjunctive operator ``,'', the disjunctive operators ``$||$'' and ``$|$', the set operator ``\{ \}'', and the content suffix \%. For brevity, we call the term \textsl{$\epsilon$} a ``\textbf{unit}'' term,  \textsl{(t,t')} a ``\textbf{tuple}'' term,  \textsl{t$||$t'} a ``\textbf{disjoint union}'' term or ``\textbf{d-union}'' in short,  \textsl{t$|$t'} an ``\textbf{option}'' term, \textsl{\{t\}$_{t'}$} a ``\textbf{set}'' term, and  \textsl{t\%} a ``\textbf{content}'' term. Especially, for the set \textsl{\{t\}$_{r}$} the term \textsl{r} is called the \textbf{index term} of the set. The unit term $\epsilon$ is an auxiliary term working as an identity element for tuple terms.

\begin{definition}
A binding \textsl{b} is in accordance with a matching term \textsl{t}, denoted as \textsl{b:t},  inductively defined as follows:\\
1) if \textsl{b} = \textsl{x$\mapsto$(s,l)}, then \textsl{b:x};\\
2) if \textsl{b} = \textsl{\{b$_{i}$:t$|$i$\in$\{1..n\}\}}, then \textsl{b:\{t\}$_{r}$} iff:\\
\indent i) for each \textsl{b$_{i}$:t(i$\in$\{1..n\})}, there is \textsl{b$_{ri}$ possibly occurring in b$_{i}$} such that \textsl{b$_{ri}$:r} and \textsl{$\forall$b'$_{ri}$:r possibly occurring in b$_{i}$. b$_{ri}$=b'$_{ri}$};\\
\indent ii) for any \textsl{b$_{i}$, b$_{j}$:t(i,j$\in$\{1..n\} and i$\neq$j)}, $\forall$ \textsl{b$_{ri}$:r possibly occurring in b$_{i}$} and \textsl{b$_{rj}$:r possibly occurring in b$_{j}$}. \textsl{b$_{ri}$$\neq$b$_{rj}$}. The unique binding b$_{ri}$:r for each b$_{i}$ is called the index sub-binding of b$_{i}$\\
3) if \textsl{b} = \textsl{(b$_{1}$,b$_{2}$), b$_{1}$:t$_{1}$} and \textsl{b$_{2}$:t$_{2}$, then b:(t$_{1}$,t$_{2}$)};\\
4) if \textsl{b} = \textsl{(b$_{1}$$||$b$_{2}$), b$_{1}$:t$_{1}$} and \textsl{b$_{2}$:t$_{2}$}, then \textsl{b:(t$_{1}$$||$t$_{2}$)};\\
5) if \textsl{b:t}, then \textsl{b:(t$|$t')} if \textsl{t$|$t'} is valid;\\
6) if \textsl{b:t} and \textsl{b} is a null binding, then \textsl{b:(t,t')} if \textsl{(t,t')} is valid;\\
7) if \textsl{b:t}, then \textsl{\%(b):t\%} where \textsl{\%(b)} is a function on a binding \textsl{b} which substitutes all the locations bound to the variables in \textsl{b} to the null location \textsl{l$_{\bot}$};
8) if \textsl{b} = $\varepsilon$, then \textsl{b:$\epsilon$}.
\end{definition}
For the binding \textsl{b} and the term \textsl{t} that \textsl{b:t} holds, we also say that \textsl{b is a binding of t} and \textsl{t is a term of b}. The definition of \textsl{b' possibly occurring in b} would be described in the next subsection. For the binding \textsl{b} = \textsl{(b$_{1}$,...,b$_{n}$)}, each \textsl{b$_{i}$} possibly occurs in \textsl{b}.

A binding \textsl{b} can also be called a ``tuple'', ``disjoint union'', ``set'' or a ``content'' accordingly.  In the Definition 3, 1)-4) specify the deduction of the matching term of a binding based on its composite structure;  5) and 6) specify the compatible extension of the matching term to cater for the requirement of the alignment of the matching terms of the various bindings in data transformation; and 7) specifies the matching term of a special kind of bindings which only contains the document content information of a binding consisting of both content and location information. The matching terms \textsl{t$|$t'} and \textsl{t\%} are used in data transformation.
The validity of a matching term would be defined in the next subsection.

For an extraction pattern \textsl{p} we can directly infer a matching term indicating the structure of the extracted data.
\begin{definition}
The matching term of the document extraction pattern or atomic pattern \textsl{p} denoted by the function \textsl{mt(p)} is deductively defined as below:\\
1) if \textsl{p} = \textsl{n$\Rightarrow$x} or \textsl{x$\Rightarrow$c}, \textsl{mt(p)} = \textsl{x};\\
2) if \textsl{p} = \textsl{x$\Rightarrow$y}, \textsl{mt(p)} = \textsl{(x,y)};\\
3) if \textsl{p} = \textsl{/p'} or \textsl{//p'} or\textsl{ @p'}, \textsl{mt(p)} = \textsl{\{mt(p')\}$_{mt(p')}$};\\
4) if \textsl{p} = \textsl{p$_{1}$ p$_{2}$} and \textsl{mt(p$_{1}$)} = \textsl{\{t\}$_{t'}$}, \textsl{mt(p)} = \textsl{\{(t,mt(p$_{2}$))\}$_{t'}$};\\
5) if \textsl{p} = \textsl{/p$_{1}$;p$_{2}$}, \textsl{mt(p)} = \textsl{\{(mt(p$_{1}$), mt(p$_{2}$))\}$_{mt(p_{1})}$};\\
6) if \textsl{p} = \textsl{(p$_{1}$,p$_{2}$)}, \textsl{mt(p)} = \textsl{(mt(p$_{1}$),mt(p$_{2}$))};\\
7) if \textsl{p} = \textsl{p$_{1}$$||$p$_{2}$}, \textsl{mt(p)} = \textsl{(mt(p$_{1}$)$||$mt(p$_{2}$))}.
\end{definition}
For example, the extraction pattern E1 has the matching term

\noindent\textsl{ T1. \{(\$b,(\{\$t\}$_{\$t}$,\{(\$a,(\{\$l\}$_{\$l}$,\{\$f\}$_{\$f}$))\}$_{\$a}$$||$\{(\$e,(\{\$l\}$_{\$l}$,\{\$f\}$_{\$f}$))\}$_{\$e}$))\}$_{\$b}$}.

Especially, the matching term of a document extraction pattern is also called an \textsl{original term}. Naturally we have the following proposition.
\newtheorem{prop}{Proposition}
\begin{prop} \label{bindterm}
For a binding \textsl{b} of a document extraction pattern \textsl{p}, \textsl{b:mt(p)} holds.
\end{prop}



\subsubsection{Restructuring Matching Terms}

Original extracted data are often required to be transformed to meet practical requests of data construction. Therefore, XTQ employs a deductive restructuring mechanism for the matching terms to specify the transformation of the bindings accordingly. Restructured terms indicate the data structure to which the original data is expected to be transformed, and the deductive restructuring mechanism which is a term rewriting system can provide a means to infer the route of  data transformation. They work together to make the requests of data transformation being presented declaratively. The restructuring rules of the matching terms are listed in Figure 4.

\begin{small}
\begin{figure}[h]
\begin{center}
\begin{tabular}{|p{\textwidth}|}
\hline
(t$_{1}$,\dots, t$_{i}$, t$_{i+1}$,\dots, t$_{n}$)$\leadsto$ (t$_{1}$,\dots, t$_{i+1}$, t$_{i}$,\dots, t$_{n}$)   (tpl-comm)\\
(t$_{1}$,\dots, t$_{j}$, t$_{j+1}$,\dots, t$_{n}$)$\leftrightsquigarrow$ (t$_{1}$,\dots, t$_{j}$, (t$_{j+1}$,\dots, t$_{n}$))    (tpl-assoc) \\
(t$_{1}$$||$\dots$||$t$_{i}$$||$t$_{i+1}$$||$\dots$||$t$_{n}$)$\leadsto$ (t$_{1}$$||$\dots$||$t$_{i+1}$$||$t$_{i}$$||$\dots$||$t$_{n}$)       (dun-comm)\\
(t$_{1}$$||$\dots$||$ t$_{j}$$||$t$_{j+1}$$||$\dots$||$t$_{n}$)$\leftrightsquigarrow$ (t$_{1}$$||$\dots$||$t$_{j}$$||$(t$_{j+1}$$||$\dots$||$t$_{n}$))     (dun-assoc)\\
(t$_{1}$$|$\dots$|$t$_{i}$$|$t$_{i+1}$$|$\dots$|$t$_{n}$)$\leadsto$ (t$_{1}$$|$\dots$|$t$_{i+1}$$|$t$_{i}$$|$\dots$|$t$_{n}$)       (opt-comm)\\
(t$_{1}$$|$\dots$|$ t$_{j}$$|$t$_{j+1}$$|$\dots$|$t$_{n}$)$\leftrightsquigarrow$ (t$_{1}$$|$\dots$|$t$_{j}$$|$(t$_{j+1}$$|$\dots$|$t$_{n}$))     (opt-assoc)\\

t$\leftrightsquigarrow$(t,$\epsilon$) (tpl-extd) \hspace{20pt}  t $\leadsto$ (t, t) (tpl-dupl) \\
(t$_{1}$, \{t$_{2}$\}$_{r}$) $\leadsto$ \{(t$_{1}$, t$_{2}$)\}$_{r}$   if t$_{1}$ and r share no common variables, t$_{1}$ is not a content term and \{t$_{2}$\}$_{r}$ is not a folded set.$^{\dag}$ (set-distr) \\
\{\{t\}$_{r_{1}}$\}$_{r_{2}}$  $\leadsto$ \{t\}$_{(r_{2},r_{1})}$ (set-flatten) \\ 
\{(t$_{1}$, t$_{2}$)\}$_{r}$ $\leadsto$ \{(\{(t$_{1}$, t$_{2}$)\}$_{r}$, t$_{1}$\%)\}$_{t_{1}\%}$ (set-fold)   \\
(\{t$_{1}$\}$_{r_{1}}$ $||$\{t$_{2}$\}$_{r_{2}}$)  $\leadsto$ \{t$_{1}$ $|$ t$_{2}$\}$_{r_{1}|r_{2}}$ (dun-set-merge) \\
t $|$ t  $\leadsto$  t  (opt-idem) \hspace{10pt}  (\{t\}$_{r}$ $|$\{t'\}$_{r}$)  $\leadsto$ \{t $|$ t'\}$_{r}$ (opt-set-alter) \\
(t, (t' $|$ t'') ) $\leftrightsquigarrow$  ((t, t') $|$ (t, t''))  (opt-tpl-alter)\\
$\dag$ a folded set is a term which results form the set-fold rule.\\
\hline
\end{tabular}
\caption{Restructuring rules of matching terms}
\end{center}
\end{figure}
\end{small}
The rewriting rules in the form of ``\textsl{t$\leadsto$t'} '' specify the relation of the valid restructuring between the matching terms. We use $\stackrel{*}{\leadsto}$ to denote the reflexive-transitive closure of $\leadsto$, indicating the relation between the matching terms such that one can be eventually restructured to the other.
\begin{definition}
A matching term \textsl{t} is \textbf{valid} to a document extraction pattern \textsl{p} iff  \textsl{mt(p)$\stackrel{*}{\leadsto}$t}.
\end{definition}

For example, for the extraction pattern \emph{(/author$\Rightarrow$\$a, /title$\Rightarrow$\$t)}, the matching term \emph{\{(\{(\$a,\$t)\}$_{\$a}$,\$t)\}$_{\$t)}$} is valid; however, \emph{\{(\{\$a,\$t\}$_{\$t}$,\$t)\}$_{\$a}$} is invalid. In fact, a matching term containing the portion like \emph{(\$x,\{\$x\}$_{\$x}$)} is always invalid. We say a matching term is valid if it is valid to a certain extraction pattern.

In Figure 4, the commutative and the associative rules, i.e., \textsl{tpl-comm}, \textsl{tpl-assoc}, \textsl{dun-comm}, \textsl{dun-assoc}, \textsl{opt-comm} and \textsl{opt-assoc} and the extension rule \textsl{tpl-extd}, are auxiliary rules to rearrange the sub-terms' locations or to add padding empty terms for further restructuring. The terms restructured following these rules are treated as structurally equivalent and we use a structural equivalence relation ``\textsl{t$\equiv$t'} '' to denote it.

We introduce the following subordination relations between the matching terms.

\begin{definition}
For two matching terms \textsl{t} and \textsl{t'},
 \textsl{t} is a \textsl{subterm} of \textsl{t'}, denoted by \textsl{t $<$ t'},
if \textsl{t'} $\equiv$ \textsl{(t$_{1}$, t$_{2}$, \dots, t$_{n}$)}, \textsl{t'} $\equiv$ \textsl{(t$_{1}$ $||$
t$_{2}$ $||$ \dots $||$ t$_{n}$)} or \textsl{t'} $\equiv$ \textsl{(t$_{1}$ $|$ t$_{2}$ $|$ \dots $|$ t$_{n}$)} and \textsl{t} = \textsl{t$_{i}$} for some \textsl{i$\in$\{1..n\}},  or \textsl{t'} = \textsl{\{t\}}.
%
\end{definition}
We also use the symbol ``$\leqslant$'' to denote the  reflexive-transitive closure of ``$<$'' between matching terms. For example, we have \{\$f\}$<$(\$a,(\{\$l\},\{\$f\})) and \$f$\leqslant$ \{(\$a,(\{\$l\},\{\$f\}))\}.

The set-related rules in Figure 4 play a central role in term restructuring. They are used to flatten the nested sets, merge the disjoint union of sets and distribute the set elements to a conjunctively coupled term. There is an obvious property of restructuring the set terms.

\begin{prop}
For a valid set term \textsl{\{t\}$_{r}$} where r is not a content term, \textsl{r} is composed of the variables and the ``,'' and ``$|$'' operators.
\end{prop}

It is not required that the rewriting system for restructuring the matching terms be confluent since an original term can be restructured to two different terms which would not converge. We are only interested in that, given an original term \textsl{t} and a restructured term \textsl{t'}, whether there is a restructuring route from \textsl{t} to \textsl{t'} following the rules in Figure 4. In fact, the restructuring system has the following property.

\begin{prop}
For an original term \textsl{t} and a term \textsl{t'},  whether \textsl{t$\stackrel{*}{\leadsto}$t'} is decidable.
\end{prop}

For example, the matching term T1 can be restructured to the  term

\noindent\textsl{T2. \{((\$l,\$f)\%,\{(\$t\%,\{((\$a$|$\$e),(\$b,\$t,\$l,\$f))\}$_{(\$b,\$t,((\$a|\$e),(\$l,\$f)))}$)\}$_{\$t\%}$)\}$_{(\$l,\$f)\%}$}\\
following the restructuring route as listed in Figure 5(In the route we omit the application of the commutative and the associative rules).

\begin{figure}[h]
\begin{center}
\begin{footnotesize}
\begin{tabular}{|p{0.02\textwidth} p{0.70\textwidth} p{0.25\textwidth}|}
\hline
 &\{(\$b,(\{\$t\}$_{\$t}$,\{(\$a,(\{\$l\}$_{\$l}$,\{\$f\}$_{\$f}$))\}$_{\$a}$$||$\{(\$e,(\{\$l\}$_{\$l}$,\{\$f\}$_{\$f}$))\}$_{\$e}$))\}$_{\$b}$ &\\
$\leadsto$ & \multicolumn{2}{p{.95\textwidth}|}
{\{(\$b,(\{\$t\}$_{\$t}$,\{(\$a,(\{\$l\}$_{\$l}$,\{\$f\}$_{\$f}$)) $|$
(\$e,(\{\$l\}$_{\$l}$,\{\$f\}$_{\$f}$))\}$_{\$a|\$e}$))\}$_{\$b}$  \hspace{20pt}(dun-merge)} \\
 $\leadsto$ & \{(\$b,(\{\$t\}$_{\$t}$,\{((\$a$|$\$e),(\{\$l\}$_{\$l}$,\{\$f\}$_{\$f}$))\}$_{\$a|\$e}$))\}$_{\$b}$ &(opt-tpl-alter)\\
 $\leadsto$ & \{(\$b,(\{\$t\}$_{\$t}$,\{((\$a$|$\$e),\{(\$l,\{\$f\}$_{\$f}$)\}$_{\$l}$)\}$_{\$a|\$e}$))\}$_{\$b}$ &(set-distr)\\
 $\leadsto$ & \{(\$b,(\{\$t\}$_{\$t}$,\{((\$a$|$\$e),\{\{(\$l,\$f)\}$_{\$f}$\}$_{\$l}$)\}$_{\$a|\$e}$))\}$_{\$b}$ &(set-distr)\\
 $\leadsto$ & \{(\$b,(\{\$t\}$_{\$t}$,\{((\$a$|$\$e),\{(\$l,\$f)\}$_{(\$l,\$f)}$)\}$_{\$a|\$e}$))\}$_{\$b}$ &(set-flatten)\\
 $\stackrel{*}{\leadsto}$ & \{(\$b,\{\$t\}$_{\$t}$,\{((\$a$|$\$e),\$l,\$f)\}$_{((\$a|\$e),(\$l,\$f))}$)\}$_{\$b}$ &(set-distr,set-flatten)\\
 $\stackrel{*}{\leadsto}$ & \{(\$b,\$t,(\$a$|$\$e),\$l,\$f)\}$_{(\$b,\$t,((\$a|\$e),(\$l,\$f)))}$ &(set-distr,set-flatten)\\
 $\stackrel{*}{\leadsto}$ & \{((\$l,\$f)\%,\{(\$b,\$t,(\$a$|$\$e),\$l,\$f)\}$_{(\$b,\$t,((\$a|\$e),(\$l,\$f)))}$)\}$_{(\$l,\$f)\%}$ &(set-fold)\\
 $\stackrel{*}{\leadsto}$ & \multicolumn{2}{p{.95\textwidth}|}{ \{((\$l,\$f)\%,\{(\$t\%,\{((\$a$|$\$e),(\$b,\$t,\$l,\$f))\}$_{(\$b,\$t,((\$a|\$e),(\$l,\$f)))}$)\}$_{\$t\%}$)\}$_{(\$l,\$f)\%}$\hspace{5pt}(set-fold)}\\
 \hline
 \end{tabular}
\caption{Example of term restructuring}
\end{footnotesize}
\end{center}
\end{figure}

A restructuring rule ``\textsl{t$\leadsto$t'} '' indicates that a binding of the term \textsl{t} should be transformed to a binding of the term \textsl{t'}. These rules are used in accompany with the transformation rules of the bindings, which are specified in Figure 6. Especially, the restructuring rules \textsl{opt-idem}, \textsl{opt-set-alter} and \textsl{opt-tpl-alter} do not actually incur the transformation.

\begin{figure}[h]
\begin{center}
\begin{small}
\begin{tabular}{|p{\textwidth}|}
\hline
b $\curvearrowright$ b (reflex) \hspace{20pt} b $\curvearrowright$ (b, $\varepsilon$) (tpl-extd)\\
(b$_{1}$,\dots, b$_{i}$, b$_{i+1}$,\dots, b$_{n}$) $\curvearrowright$ (b$_{1}$,\dots, b$_{i+1}$, b$_{i}$,\dots, b$_{n}$)   (tpl-comm)\\
(b$_{1}$,\dots, b$_{j}$, b$_{j+1}$,\dots, b$_{n}$)$\frac{\curvearrowleft}{\curvearrowright}$ (b$_{1}$,\dots, b$_{j}$, (b$_{j+1}$,\dots, b$_{n}$))    (tpl-assoc) \\
(b$_{1}$$||$\dots$||$b$_{i}$$||$b$_{i+1}$$||$\dots$||$t$_{n}$)$\curvearrowright$ (b$_{1}$$||$\dots$||$b$_{i+1}$$||$b$_{i}$$||$\dots$||$b$_{n}$)       (dun-comm)\\
(b$_{1}$$||$\dots$||$ b$_{j}$$||$b$_{j+1}$$||$\dots$||$b$_{n}$)$\leftrightsquigarrow$ (b$_{1}$$||$\dots$||$b$_{j}$$||$(b$_{j+1}$$||$\dots$||$b$_{n}$))     (dun-assoc)\\
b $\curvearrowright$ (b, b) (tpl-dupl) \hspace{10pt}
(b,\{b$_{i}$$|$ i$\in$\{1..n\}\}) $\curvearrowright$ \{(b,b$_{i}$)$|$ i$\in$\{1..n\}\}  (set-distr)\\
\{c$_{i}$=\{b$_{i,j}$ $|$ j$\in$\{1..n$_{i}$\} $|$ i$\in$\{1..n\}\} $\curvearrowright$ \{b$_{i,j}$ $|$ i$\in$\{1..n\}, j$\in$\{1..n$_{i}$\} \} (set-flatten) \\
\{b$_{i}$$|$ i$\in$\{1..n\}\}$||$\{b'$_{j}$$|$ j$\in$\{1..m\}\} $\curvearrowright$ \{b$_{i}$$|$ i$\in$\{1..n\}\}$\cup$\{b'$_{j}$$|$ j$\in$\{1..m\}\}  (dun-merge) \\
B=\{(b$_{i}$, b'$_{i}$) $|$ i$\in$\{1..n\}\} $\curvearrowright$ \\
\hspace{40pt}\{(c, \{(b,b')$\in$B $|$ \%(b)=c\}) $|$ c$\in$\{\%b$_{i}$$|$i$\in$\{1..n\}\}\} (set-fold)\\
$\dfrac{b \curvearrowright b'}{ (b,b'') \curvearrowright (b',b'')}$ (tpl-comp)\hspace{10pt}
$\dfrac{b \curvearrowright b'}{ b || b'' \curvearrowright b'||b''}$ (dun-comp)\\
$\dfrac{ b_{i}\curvearrowright b'_{i} (i\in\{1..n\})}{\{b_{i}|i\in\{1..n\}\} \curvearrowright \{b'_{i}|i\in\{1..n\}\}}$ (set-comp) \\
\hline
\end{tabular}
\caption{Transformation rules of bindings}
\end{small}
\end{center}
\end{figure}

As the rules in both the figures show, the transformation of the bindings can be classified into three categories. The first one is the auxiliary rules to rearrange the terms and the bindings accordingly so as to facilitate presenting the further restructuring, such as the commutative and the associative rules for the tuples and the d-unions. We also use ``\textsl{b$\equiv$b'} ''to denote the structural equivalence between the bindings. The second one is the duplication of the bindings in a tuple, i.e., the \textsl{tpl-dupl} rule, which is often used for duplicating the bindings for different uses in data construction. The third one mainly focuses on restructuring the nested sets, such as the \emph{set-flatten} rule which flattens a nested set into a plain set, the \textsl{set-fold} rule which groups the set of the tuples by the values of a certain part of the tuple, and the \textsl{set-distr} rule which transforms the tuple containing a set to a distributed set of tuples. 

Similar to Definition 7, we introduce the following definitions to describe the subordination relation between bindings.
\begin{definition}
For two bindings \textsl{b} and \textsl{b'}, \textsl{b} is an \textsl{element binding} of \textsl{b'}, denoted by \textsl{b$<$b'}, if \textsl{b'} = \textsl{(b$_{1}$, b$_{2}$, \dots, b$_{n}$)}, \textsl{b'} = \textsl{(b$_{1}$ $||$
b$_{2}$ $||$ \dots $||$ b$_{n}$)}, or \textsl{b'}=\textsl{\{b$_{i}$$|$i$\in$\{1..n\}\}}  and \textsl{b} = \textsl{b$_{i}$} for some \textsl{i$\in$\{1..n\}};
%
\end{definition}
We use the symbol ``$\leqslant$'' to denote the reflexive-transitive closure of the relation ``$<$'' between the bindings.

Further, in practice the following subordination relations for terms and bindings are often used, as  Definition 9 shows.

\begin{definition}
1. For the two terms \textsl{t} and \textsl{u}, \textsl{u} \textbf{possibly occurs} in \textsl{t}, denoted by \textsl{u$\unlhd$t} if there is a term \textsl{t'} that \textsl{t$\stackrel{*}{\leadsto}$t'} and \textsl{u$\leqslant$t'}.\\
2. For the two bindings \textsl{b} and \textsl{c}, \textsl{c} \textbf{possibly occurs} in \textsl{b}, denoted by \textsl{c$\unlhd$b} if there is a binding \textsl{b'} that \textsl{b$\curvearrowright$b'} and \textsl{c$\leqslant$b'}.\\
\end{definition}

We have the following propositions for the soundness of the term restructuring and data transformation.
\begin{prop}
For \textsl{b:t} and \textsl{t$\stackrel{*}{\leadsto}$t'}, there is \textsl{b':t'} that \textsl{b$\curvearrowright$b'}.
\end{prop}
\begin{prop}
For \textsl{b:t} and \textsl{t$\stackrel{*}{\leadsto}$t'}, if \textsl{b$\curvearrowright$b':t'} and \textsl{b$\curvearrowright$b'':t'}, then \textsl{b'=b''}.
\end{prop}

Following the restructuring route in Figure 5, the sample-binding in Section 3.3  would be transformed to the following bindings:\\
\begin{footnotesize}
\textsl{
\{\indent b$_{r1}$ = (b$_{\$b,b1}$, b$_{\$t,t1}$, b$_{\$a,a11}$, \$l$\mapsto$("Yeats",l$_{l11}$), \$f$\mapsto$("Willam" ,l$_{f11}$)),\\
\indent b$_{r2}$ =(b$_{\$b,b1}$, b$_{\$t,t1}$, b$_{\$a,a12}$, \$l$\mapsto$("Artificial",l$_{l12}$), \$f$\mapsto$("Data",l$_{f12}$)),\\
\indent b$_{r3}$ =(b$_{\$b,b1}$, b$_{\$t,t1}$, b$_{\$e,e1}$, \$l$\mapsto$("Rosenthal",l$_{l13}$), \$f$\mapsto$("M.L.",l$_{f13}$)),\\
\indent b$_{r4}$ =(b$_{\$b,b2}$, b$_{\$t,t2}$, b$_{\$a,a21}$, \$l$\mapsto$("Blackburn",l$_{l21}$), \$f$\mapsto$("Paul",l$_{f21}$)),\\
\indent b$_{r5}$ =(b$_{\$b,b2}$, b$_{\$t,t2}$, b$_{\$a,a22}$, \$l$\mapsto$("Aritifical",l$_{l22}$)\},\{\$f$\mapsto$("Info.",l$_{f22}$)),\\
\indent b$_{r6}$ =(b$_{\$b,b2}$, b$_{\$t,t2}$, b$_{\$e,e21}$, \$l$\mapsto$("Rosenthal",l$_{l23}$), \$f$\mapsto$("M.L.",l$_{f23}$)),\\
\indent b$_{r7}$ =(b$_{\$b,b2}$, b$_{\$t,t2}$, b$_{\$e,e22}$, \$l$\mapsto$("Jarolim",l$_{l24}$), \$f$\mapsto$("Edie",l$_{f24}$)), \dots \} \\
: \{(\$b,\$t,(\$a$|$\$e),\$l,\$f)\}$_{(\$b,\$t,((\$a|\$e),(\$l,\$f)))}$ (restr-binding)\\
\\
\{ (\$l$\mapsto$("Yeats",l$_{\bot}$), \$f$\mapsto$("Willam" ,l$_{\bot}$), \\
\{(\$t$\mapsto$(d$_{t1}$,l$_{\bot}$), \{(b$_{\$a,a11}$, (b$_{\$b,b1}$, b$_{\$t,t1}$, \$l$\mapsto$("Yeats",l$_{l11}$), \$f$\mapsto$("Willam" ,l$_{f11}$)) )\})\}),\\
(\$l$\mapsto$("Rosenthal",l$_{\bot}$), \$f$\mapsto$("M.L." ,l$_{\bot}$), \\
\{(\$t$\mapsto$(d$_{t1}$,l$_{\bot}$),
\{(b$_{\$e,e1}$, (b$_{\$b,b1}$, b$_{\$t,t1}$, \$l$\mapsto$("Rosenthal",l$_{l13}$), \$f$\mapsto$("M.L." ,l$_{f13}$)) )\}),\\
(\$t$\mapsto$(d$_{t2}$,l$_{\bot}$), 
\{(b$_{\$e,e21}$, (b$_{\$b,b2}$, b$_{\$t,t2}$,  \$l$\mapsto$("Rosenthal",l$_{l23}$), \$f$\mapsto$("M.L." ,l$_{f23}$)) )\})\} ), \dots \} 
 : \{((\$l,\$f)\%,\{(\$t\%,\{((\$a$|$\$e),(\$b,\$t,\$l,\$f))\}$_{(\$b,\$t,((\$a|\$e),(\$l,\$f)))}$)\}$_{\$t\%}$)\}$_{(\$l,\$f)\%}$ \\
}\end{footnotesize}
\subsection{Data Construction}
\subsubsection{Construction Pattern}
XTQ uses construction pattern, a function invocation on a matching term, to specify data construction request
for generating document fragments or values with the extracted data. A construction pattern  can be specified as a normal function invocation like \textsl{f(t)} or, more often, by embedding the argument matching term into a composite ``\textsl{name$\Rightarrow$content}'' structure with some valid constant values.
The syntax of construction pattern is specified in Figure 7.

\begin{figure}
\begin{center}
\begin{small}
\begin{tabular}{|p{0.15\textwidth} p{0.05\textwidth} p{0.75\textwidth}|}
  \hline
$<$nameptn$>$ &::=& $<$name$>$ $|$ $<$var$>$ $|$ $<$funinvoke$>$ $|$\\
&&@$<$name$>$ $|$ `@'$<$var$>$ $|$ `@'$<$funinvoke$>$\\
$<$ctntptn$>$ &::=& $<$const$>$ $|$ $<$var$>$ $|$ $<$funinvoke$>$\\
$<$consptn$>$ &::=& $<$ctntptn$>$ $|$ $<$nameptn$>$ `$\Rightarrow$' $<$consptn$>$ $|$ \\
&& $<$label$>$ `\&' $<$consptn$>$ $|$ `('$<$tupleptn$>$`)' $|$ \\
&& `('$<$dunionptn$>$`)' $|$ `('$<$optionptn$>$`)' $|$ $<$setptn$>$\\
$<$tupleptn$>$ &::=& $<$consptn$>$ `,' $<$consptn$>$ $|$ $<$tupleptn$>$ `,' $<$consptn$>$ $|$\\
&& $<$consptn$>$ `\textbf{hid}' $<$term$>$\\
$<$dunionptn$>$ &::=& $<$consptn$>$`$||$'$<$consptn$>$ $|$ $<$dunionptn$>$`$||$'$<$consptn$>$ $|$ \\
&& $<$consptn$>$ `\textbf{elim}' $<$term$>$\\
$<$optionptn$>$ &::=& $<$consptn$>$`$|$'$<$consptn$>$ $|$ $<$optionptn$>$`$|$'$<$consptn$>$\\
$<$setptn$>$ &::=& `\{'$<$consptn$>$`\}' $|$ `\{'$<$consptn$>$`\}' `\textbf{groupby}' $<$term$>$ $|$ \\
&&`\{'$<$consptn$>$`\}' `\textbf{groupby}' $<$term$>$ `\textbf{orderby}' ($<$orderlist$>$)\\
&& $<$consptn$>$ `::' $<$setptn$>$\\
$<$orderlist$>$ &::=& $<$orderitem$>$ $|$ $<$orderitem$>$`,' $<$orderlist$>$ \\
$<$orderitem$>$ &::=& $<$orderunit$>$ `\textbf{asc}' $|$ $<$orderunit$>$ `\textbf{desc}' \\
$<$orderunit$>$ &::=& $<$term$>$ $|$ $<$funinvoke$>$  $|$ $<$orderunit$>`|'<$orderunit$>$\\
\hline
\end{tabular}
\caption{Syntax of construction patterns}
\end{small}
\end{center}
\end{figure}

A construction pattern should always have a valid matching term as its argument.

\begin{definition}
The argument matching term of the construction pattern \textsl{p} denoted by the function \textsl{at(p)} is deductively defined as below:\\
1) if \textsl{p} = \textsl{t}, \textsl{at(p)} = \textsl{t};\\
2) if \textsl{p} = \textsl{lab\&p'} or  \textsl{f(p')}, \textsl{at(p)} = \textsl{at(p')};\\
3) if \textsl{p} = \textsl{p$_{1}$ $\Rightarrow$ p$_{2}$}, \textsl{(p$_{1}$,p$_{2}$)} or \textsl{p$_{1}$ hid p$_{2}$}, \textsl{at(p)} = \textsl{(at(p$_{1}$),at(p$_{2}$))};\\
4) if \textsl{p} = \textsl{p$_{1}$$||$p$_{2}$} or \textsl{p$_{1}$ elim p$_{2}$}, \textsl{at(p)} = \textsl{(at(p$_{1}$)$||$at(p$_{2}$))};\\
5) if \textsl{p} = \textsl{p$_{1}$$|$p$_{2}$}, \textsl{at(p)} = \textsl{(at(p$_{1}$)$|$at(p$_{2}$))};\\
6) if \textsl{p} = \textsl{\{p'\} groupby r orderby (o$_{1}$,..., o$_{n}$)}, \textsl{at(p)} = \textsl{\{(at(p')\}$_{r}$};\\
7) if \textsl{p} = \textsl{p$_{1}$::p$_{2}$}, \textsl{at(p)} = \textsl{(at(p$_{1}$),at(p$_{2}$))}.\\
\end{definition}

In data construction, the original terms are restructured to the valid argument matching terms indicated by the construction pattern. However, it is common that a valid matching term might contain the sub-terms that are not to be used in constructing the values. XTQ uses the \textbf{hid} suffix or the \textbf{elim} suffix to specify the sub-term(s) of a tuple or a d-union that should be omitted in construction. Furthermore, the \textbf{groupby} suffix of a set construction pattern indicates the index term of the argument set term. However, it is often redundant and clumsy to list all the index terms and the terms to be hidden or eliminated. Fortunately, the \textsl{hid}, the \textsl{elim} and the \textsl{groupby} suffixes can often be simplified or reduced based on the following properties of the matching terms and the restructuring mechanism.

\begin{definition}
For the terms \textsl{t}, \textsl{\{t$_{1}$\}$_{r_{1}}$} and \textsl{\{t$_{2}$\}$_{r_{2}}$} that  \textsl{\{t$_{1}$\}$_{r_{1}}$$\leqslant$t} and \textsl{\{t$_{2}$\}$_{r_{2}}$$\leqslant$t}, the index term \textsl{r$_{1}$} is \textbf{inner than} \textsl{r$_{2}$} in \textsl{t}, denoted as \textsl{r$_{1}$$\prec$r$_{2}$}, if \textsl{\{t$_{1}$\}$_{r_{1}}$ $\leqslant$ t$_{2}$}.
\end{definition}

\begin{prop}For an original term t and a matching term \textsl{t'} that \textsl{t$\stackrel{*}{\leadsto}$t'}:
1. if \textsl{\{\{t$_{0}$\}$_{r_{0}}$\}$_{r_{1}}$$\leqslant$t'} and there are the index terms \textsl{r$_{a}$}, \textsl{r$_{b}$}, \textsl{r$_{c}$} in \textsl{t} that \textsl{r$_{a}$$\prec$r$_{b}$$\prec$r$_{c}$}, \textsl{r$_{a}$$\leqslant$r$_{0}$} and \textsl{r$_{c}$$\leqslant$r$_{1}$}, then either \textsl{r$_{b}$$\leqslant$r$_{0}$} or \textsl{r$_{b}$$\leqslant$r$_{1}$};\\
2. if \textsl{\{t$_{0}$\}$_{r_{0}}$$\leqslant$t'}, then for any index term \textsl{r} in \textsl{t} that \textsl{r$\leqslant$r$_{0}$}, \textsl{r$\leqslant$t$_{0}$} holds.\\
\end{prop}

These two propositions guarantee that the parser can easily infer the omitted \textsl{groupby} and \textsl{hid} suffixes from the necessary information in the construction pattern. The details of the inference is often trivial and thus is omitted.
Furthermore, in the extraction pattern the variable \textsl{x} of an element pattern \textsl{/n$\Rightarrow$x} or \textsl{//n$\Rightarrow$x} can also be omitted if it is not used in data construction. The query parser would automatically generate a variable for each abbreviated element pattern and parse its location in the \textsl{hid} suffix or \textsl{groupby} suffix in the construction pattern. 



For example, in the extraction pattern of the query Q2 the abbreviated element pattern ``\textsl{//book}'' would be treated as the pattern like ``\textsl{//book$\Rightarrow$\$b}'', and the extraction pattern has the matching term like T1 and is restructured to the matching term like T2 as the argument of the construction pattern. The construction pattern omits the suffix ``\textsl{hid (\$b,\$t,\$l,\$f)}'' in the inner most layer of the nested sets and the suffixes ``\textsl{groupby (\$b,\$t,((\$a$|$\$e),(\$l,\$f)))}'', ``\textsl{groupby \$t\%}'' and ``\textsl{groupby (\$l,\$f)\%}'' of the set patterns respectively. These suffixes can easily be inferred based on the properties listed above.

In inferring the omitted index terms in the construction patterns, there are often multiple choices for the index terms of a set \textsl{\{t\}} since the terms \textsl{\{t\}$_{r}$} and \textsl{\{t\}$_{r'}$} can be both valid when \textsl{r$\equiv$r'} but \textsl{r$\neq$r'}. XTQ adopts a specific ordering of the units in the index terms. The units are listed following the order of the occurrence in the pre-order traversal list of the original term. That is, for the original term \textsl{t} of the extraction pattern, the index units are listed as \textsl{(u$_{1}$,...,u$_{n}$)} iff: for \textsl{1$\leq$i$<$j$\leq$n}, there is \textsl{(t', t'')$\leqslant$t} that \textsl{u$_{i}$$\leqslant$t'} and \textsl{u$_{j}$$\leqslant$t''}. For example, in the above query the index term can also be ``\textsl{(\$b,(\$a$|$\$e),(\$f,\$l),\$t)}'' which indicates another restructuring route but results in the same bindings. However, when there is no \textsl{groupby} suffix, the parser would choose ``\textsl{(\$b,\$t,((\$a$|$\$e),(\$l,\$f)))}'' as the index term since the index units follow the occurrence order in the original term.

Since construction patterns are used for generating not only document fragments but also function arguments, there are two kinds of construction patterns which are designed specific for function invocation. One is the labeled patterns, i.e., the construction patterns attached with a label, e.g., ``\textsl{l1\&author$\Rightarrow$\$a}''. Such a pattern generates a labeled value of an option term for the function to differentiate its expected content type. Another kind of patterns is in the form \textsl{p$_{1}$::p$_{2}$}, which constructs a order set value like a list in conventional functional languages. We introduce this operation to seamlessly integrate the basic features of conventional functional languages into XTQ. These features would be further discussed in Section 5.

\subsubsection{Semantics of Data Construction}

Data construction is carried out by instantiating the construction pattern with a binding of its argument matching term, resulting in a value of a certain type. Before instantiation, the value should be preprocessed by an auxiliary mechanism to rearrange the locations of the elements. XTQ uses the ``\textsl{orderby}'' suffix and the \textsl{groupby} suffix to specify the explicit requests and the implicit guidelines of  ordering the set bindings.


The \textsl{orderby} suffix is used to rank the elements in a set binding. For a set binding \textsl{b}=\textsl{\{b$_{i}$ $|$ i$\in$\{1..m\}\}} with the suffix \textsl{orderby (o$_{1}$, ..., o$_{n}$)},  each binding \textsl{b$_{i}$} is associated with a value tuple\textsl{ v$_{i}$}=\textsl{(v$_{i_{1}}$, \dots,v$_{i_{n}}$)} by evaluating the order items \textsl{(o$_{1}$,\dots,o$_{n}$)} with the constants bound to the variables of \textsl{o$_{j}$} (\textsl{j}$\in$\{1..n\}). Then \textsl{b$_{i}$} are ranked in the set based on the lexicographical order on v$_{i}$. If there is no \textsl{orderby} suffix or there are multiple bindings which have the same ranking value tuple, the \textsl{groupby} suffix is used to decide the final order.

For a set binding \textsl{\{b$_{i}$ $|$ i$\in$\{1..m\}\}:\{t$_{e}$\}$_{r}$}  for data construction, each \textsl{b$_{i}$:t$_{e}$} (\textsl{i}$\in$\{1..m\}) is associated with the index binding \textsl{b$_{ind_{i}}$}=\textsl{(x$_{1}$$\mapsto$(v$_{i1}$,l$_{i1}$),\dots,x$_{n}$$\mapsto$ (v$_{in}$,l$_{in}$))} and  the location tuple \textsl{l$_{i}$} = \textsl{(l$_{i1}$,\dots,l$_{in}$)}. Then \textsl{b$_{i}$} are ranked in the set based on the lexicographical order on \textsl{l$_{i}$}. Especially, if two bindings \textsl{b$_{ind_{i}}$:t$_{i}$} and \textsl{b$_{ind_{j}}$:t$_{j}$} (\textsl{t$_{i}$$\neq$t$_{j}$}) have the same location tuple, that is, \textsl{t$_{i}$$|$t$_{j}$ $\ll$t$_{e}$}, then \textsl{b$_{i}$} ranks higher than \textsl{b$_{j}$} in the set if \textsl{t$_{i}$'$||$t$_{j}$'$<$t}, \textsl{t$_{i}$$\ll$t$_{i}$'} and \textsl{t$_{j}$$\ll$t$_{j}$'}. 
Especially, if \emph{r} is a content term  \emph{r'\%}, which means the location value of the index bindings are all set to \emph{l$_{\bot}$}, the processor can randomly choose an order for the elements in the folded set binding.


In instantiating the construction pattern, the location information in the transformed binding would be removed, i.e., any variable binding \textsl{\$x$\mapsto$(d,l)} be altered to \textsl{\$x$\mapsto$d}, since the hierarchy and the succession  would be reestablished in data construction. However, the location \textsl{l} can be used to instantiating a special construction pattern ``\textsl{loc(\$x)}''  as a special location value passed to the functions involving the location calculation.

XTQ uses the composite values, according to the composite structure of the construction patterns, as the result of data construction. The kinds of values and the data types they are associated with are specified in Figure 8 and Figure 9. Here we use \textsl{v} to range over the the values, use \textsl{t} to range over value types, use \textsl{c} to range over the constants of different types such as boolean, number, content data and location,  use \textsl{d} to range over document fragments, and use \textsl{lab} to range over the labels attached to a value in an option.

\begin{figure}[h]
\begin{center}
\begin{small}
\begin{tabular}{|p{0.13\textwidth} p{0.02\textwidth} p{0.78\textwidth}|}
\hline
$<$value$>$ &::=& $<$bool$>$ $|$ $<$num$>$ $|$ $<$cdata$>$ $|$ $<$loc$>$ $|$ $<$doc$>$   \\
&& `nil' $|$  $<$value$>$`::'$<$value$>$ $|$ `'$<$value$>$`,'$<$value$>$ $|$\\
 &&$<$value$>$`$||$' $<$value$>$) $|$ $<$label$>$`\&'$<$value$>$ $|$ `('$<$value$>$`)'\\
$<$valtype$>$ &::=& `bool' $|$ `num' $|$  `cdata' $|$ `loc' $|$ `doc' $|$ `*' $|$ `\{'$<$valtype$>$`\}' $|$ \\ &&$<$valtype$>$`,'$<$valtype$>$ $|$ $<$valtype$>$`$||$'$<$valtype$>$) $|$  \\
&&$<$labeltype$>$ $|$ `('$<$valtype$>$`)'\\
$<$labeltype$>$ &::=& $<$label$>$`\&'$<$valtype$>$`$|$'$<$label$>$`\&'$<$valtype$>$  $|$ \\ &&$<$label$>$`\&'$<$valtype$>$`$|$'$<$labeltype$>$\\
\hline
\end{tabular}
\end{small}
\caption{Values and types in XTQ}
\end{center}
\end{figure}

\begin{figure}
\begin{center}
\begin{small}
\begin{tabular}{|p{\textwidth}|}
\hline
c$_{bool}$ : bool (bool-type)\hspace{10pt} c$_{num}$ : num  (number-type) \hspace{10pt} c$_{cdata}$ : cdata (cdata-type)\\
c$_{loc}$ : loc (loc-type) \hspace{10pt}  d : doc (doc-type) \hspace{10pt} v : * (univ-type)
\\
 \textbf{nil}: \{t\} (empty-type)\hspace{10pt} $\dfrac{v : t, v' : \{t\}}{v::v' : \{t\}}$ (comp-set-type)\hspace{10pt}
$\dfrac{v : t, v' : t'}{(v,v') : (t, t')}$ (tpl-type)\\
$\dfrac{v : t, v' : t'}{(v||v') : (t' || t')}$ (dunion-type)\hspace{10pt}
$\dfrac{v:t}{lab\&v:(lab\&t | \ lab'\&t')}$ (opt-type)\\
\hline
\end{tabular}
\end{small}
\caption{Typing rules of XTQ}
\end{center}
\end{figure}
As the typing rules show, the correspondence between most kinds of the values and their types are quite straightforward. For the value of a set type, other than using the plain set of elements , we adopt the form of \textsl{v::v'}, i.e., the recursive definition of a list in functional language, because the elements of a set are always ordered as mentioned previously. Such a recursive definition also enables recursively processing the set elements. For convenience, we also denote the set \textsl{v$_{1}$::v$_{2}$::\dots::v$_{n}$::nil} as \textsl{\{v$_{1}$;v$_{2}$;\dots;v$_{n}$\}}.
Especially, we introduce the universal type denoted by ``*'' to represent the super-type of all the other types, that is, any value is an instance of the type ``*''.

\begin{figure}
\begin{center}
\begin{footnotesize}
\begin{tabular}{|p{\textwidth}|}
\hline
c[$\varepsilon$]$\rightarrow$c  (const) \hspace{10pt} \$x[\$x$\mapsto$(d,l)]$\rightarrow$ d (var-cont-inst)\hspace{10pt}
loc(\$x)[\$x$\mapsto$(d,l)]$\rightarrow$ l (var-loc-inst)\\
\vspace{1pt}
$\dfrac{p_{1}[b_{1}] \rightarrow n, p_{2}[b_{2}] \rightarrow v, \nu l\in L} { p_{1} \Rightarrow p_{2}[(b_{1},b_{2})] \rightarrow \{|n\Rightarrow_{\nu l} \delta(v)|\}}$ (elem-cons) $^{\dag}$
$\dfrac{p[b] \rightarrow v}{lab\&p[b] \rightarrow lab\&v}$\\
\vspace{1pt}
$\dfrac{p_{i}[b_{i}] \rightarrow v_{i}  \forall i\in\{1..n\}} { (p_{1},\dots, p_{n})[(b_{1},\dots,b_{n})] \rightarrow (v_{1}, ...,v_{n})}$ (tpl-cons)\hspace{10pt}
$\dfrac{p[b] \rightarrow v, b':t} { p\hspace{2pt} \textsl{hid}\hspace{2pt} t [(b,b')] \rightarrow v}$ (tpl-hid)\\
\vspace{1pt}
$\dfrac{p_{i}[b_{i}] \rightarrow v_{i}  \forall i\in\{1..n\}} { (p_{1} ||\dots || p_{n})[(b_{1},\dots,b_{n})] \rightarrow v_{1}||\dots||v_{n}}$ (dun-cons)\hspace{10pt} $\dfrac{p[b] \rightarrow v, b':t} { p\hspace{2pt} \textsl{elim}\hspace{2pt} t [(b||b')] \rightarrow v}$ (dun-elim)\\
\vspace{1pt}
$\dfrac{ p_{i}[b_{i}]\rightarrow v_{i} \exists i\in\{1..n\}} { (p_{1} | \dots | p_{n})[b_{i}] \rightarrow v_{i}}$ (option-cons) \hspace{10pt}
$\dfrac{ p[b_{i}] \rightarrow v_{i}  \forall i\in\{1..n\}} {\{p\}[\{b_{1};\dots; b_{n}\}] \rightarrow \{v_{1}; ...;v_{n}\}}$ (set-cons1)\\
\vspace{1pt}
$\dfrac{p[b] \rightarrow v_{1}, p'[b'] \rightarrow \{v_{2};\dots;v_{n}\}}{ p::p'[(b,b')] \rightarrow \{v_{1};v_{2};\dots;v_{n}\}}$ (set-cons2)\\
\vspace{1pt}
$\dag$ $\nu$l is a fresh location, and $\nu$l / l' holds for any n'$\Rightarrow_{l'}$d' $\in$ $\delta$(v).\\
\hline
\end{tabular}
\end{footnotesize}
\caption{Semantic rules of data construction}
\end{center}
\end{figure}

The semantics of data construction is specified in Figure 10. The semantic rules are in the form of \textsl{p[b]$\rightarrow$v}  where \textsl{p} is a construction pattern, \textsl{b} is an ordered binding and \textsl{v} is a composite value.  A composite value can be directly transformed to a SXD document fragment, denoted as  $\delta$(v), in a successive representation as below.

\begin{equation}
\delta(v) =
    \begin{cases}
        \nu d& \text{if $v=d$,}\\
        \{|\delta(v_{1});\dots;\delta(v_{n})|\}& \text{if $v=\{v_{1};\dots;v_{n}\}$,}\\
        \{|\delta(v_{1});\dots;\delta(v_{n})|\}& \text{if $v=(v_{1},\dots,v_{n})$,}\\
        \{|\delta(v_{1});\dots;\delta(v_{n})|\}& \text{if $v=v_{1}||\dots||v_{n}$.}\\
    \end{cases}
\end{equation}
In (1), the \textsl{$\nu$d} is a document fragment that \textsl{$\nu$d$\equiv$d} and the locations of \textsl{$\nu$d} are fresh. That means, in the transformation of \textsl{v} to \textsl{$\delta$(v)}, the document fragments maintains their content and the relations of the locations, meanwhile the new locations are introduced to indicate a new document hierarchy.

For example, given the binding sample-binding, the construction pattern of the query Q2 would generate the following XML fragments:

\begin{footnotesize}\begin{verbatim}
 results => {| ...;
    result => {|
      person => {|last=>"Artificial"; first=>"Data"|};
      book => {|
        title=>"Selected Poems And Four Plays of William Butler Yeats";
        role => "author" |} |};
    result => {|
      person => {|last=>"Aritifical"; first=>"Info."|};
      book => {|
        title=>"The Selected Poems of Paul Blackburn";
        role => "author" |} |};
    result => {|
      person => {|last=>"Blackburn"; first=>"Paul"|};
      book => {|
        title=>"The Selected Poems of Paul Blackburn";
        role => "author" |} |};
    result => {|
      person => {|last=>"Jarolim"; first=>"Edie"|};
      book => {|
        title=>"The Selected Poems of Paul Blackburn";
        role => "editor" |} |};
    result => {|
      person => {|last=>"Rosenthal"; first=>"M.L."|};
      book => {|
        title=>"Selected Poems And Four Plays of William Butler Yeats";
        role => "editor" |};
      book => {|
        title=> "The Selected Poems of Paul Blackburn";
        role => "editor" |}  |};
    result => {|
      person => {|last=>"Yeats"; first=>"Willam"|};
      book => {|
        title=>"Selected Poems And Four Plays of William Butler Yeats";
        role => "author" |} |};
    ... |}    
\end{verbatim}\end{footnotesize}



\section{Data Filtering}
\subsection{Syntax of the Conditions}

XTQ uses the conditions, including the predicate ones and the compound ones, in the \textsl{where} clause to filter out the unwanted data elements.
The syntax of the conditions are specified in Figure 11.

\begin{figure}[h]
\begin{center}
\begin{small}
\begin{tabular}{|p{0.20\textwidth} p{0.02\textwidth} p{0.70\textwidth}|}
  \hline
   $<$condition$>$ &::=& $<$predcond$>$ $|$ $<$compcond$>$ $|$ \textbf{(} $<$condition$>$ \textbf{)}\\
  $<$predcond$>$  &::=&  $<$comparison$>$ $|$ $<$quantification$>$ $|$ $<$funinvoke$>$ $|$ \\
               && \textbf{not} $<$predcond$>$  $|$ $<$predcond$>$ \textbf{and} $<$predcond$>$ $|$ \\
               && $<$predcond$>$ \textbf{or} $<$predcond$>$\\
  $<$comparison$>$ &::=&  $<$consptn$>$  $<$infix$>$ $<$consptn$>$\\
 $<$infix$>$ &::=& = $|$ $<$ $|$ $<=$ $|$  $\ll$     \\
 $<$quantification$>$ &::=& \textbf{(}$<$quantifier$>$ $<$term$>$ \textbf{in} \{$<$term$>$\}. $<$predcond$>$\textbf{)}\\
 $<$quantifier$>$ &::=& \textbf{foreach}  $|$ \textbf{forsome}\\
  $<$compcond$>$ &::=& $<$condition$>$ \textbf{par} $<$condition$>$ $|$ \\
  &&$<$condition$>$ \textbf{with} $<$predcond$>$\\
\hline
\end{tabular}
\caption{Syntax of conditions}
\end{small}
\end{center}
\end{figure}

Predicate conditions are essentially predicate function invocations to filter the values bound to the argument term.
A predicate condition can be an atomic one or a composite one combining atomic conditions with the boolean connectives ``\textsl{and}'', ``\textsl{or}'' and ``\textsl{not}''.

Atomic conditions include comparisons, canonical predicate function invocations and quantified conditions. A comparison is a binary predicate condition presented as an infix expression, e.g.,  ``\textsl{\$a1 = \$a2}''. Comparison operators include ``$=$'', ``$<$'' and``$<=$'' for value comparison  and ``$\ll$'' and ``$<$'' for location comparison. The values that can be compared include numerics and strings, which are the results of the construction patterns in the comparison. The location comparison can only be used to compare the location of the variables. Canonical predicate function invocations are often in the form of \textsl{fun(t)}, whereas the construction pattern which returns the boolean value can be directly used as the function invocation. For example, in the query Q3, the condition has the argument term ``\textsl{(\$t,(\$a$|$\$e))}'' and is specified as the construction pattern ``\textsl{loc(\$a)$\ll$loc(\$t) $|$ loc(\$t)$<$loc(\$t)}'', indicating that for each book we need the author preceding the title and the editors locating after the title.

\begin{footnotesize}\begin{verbatim}
Q3. query doc("bib.xml")//book(/title=>$t;(/author=>$a||/editor=>$e))
     where loc($a)<<loc($t) | loc($t)<loc($e)
     construct {book=>({author=>$a}||{editor=>$e})}
Q4. query doc("bib.xml")//book(/author=>$a||/editor=>$e)/email=>$m)
     where (foreach ($a|$e,{$m}).(foreach $m. endwith($m, "edu")))
     construct {book=>{author=>$a|editor=>$e}}
\end{verbatim}\end{footnotesize}

The quantified condition ``\textsl{foreach(forsome) t in \{t'\}$_{\$r}$. c}''  indicates that each (or some) of the binding(s) of the term t in the set binding of the term \textsl{\{t'\}$_{\$r}$} should satisfy the predicate condition \textsl{c}. The argument term of the quantified condition is the set term \textsl{\{t'\}$_{\$r}$} indicating the domain of the elements.
The quantified part ``\textsl{t in \{t'\}$_{r}$}'' can be abbreviated as \textsl{t} if \textsl{t=t'} and \textsl{r} can be omitted.  For example, in  the query Q4, the nested quantified condition gets the books whose authors and editors all have only email address ending with ``\textsl{edu}''.


XTQ uses boolean connectives ``\textsl{and}'' and ``\textsl{or}''  to combine two predicate conditions as a composite predicate condition. The argument term of a composite condition is the tuple consisting of the argument terms of the sub-conditions.  The connective  \textsl{not} is used to specify the negation of a predicate condition. For example, in the query Q5, the books would be conjunctively filtered by two conditions, the former condition restricting that none of the authors should have the last name ``Li'' and the latter one restricting that the editors should have the last name ``Liu'' or ``Li''.

\begin{footnotesize}\begin{verbatim}
Q5. query doc("bib.xml")//book(/author=>$a,/editor=>$e/last=>$l)
     where (foreach $a. not($a/last=>"Li")) and ($l="Liu" or $l="Li")
     construct {bookinfo=>({author=>$a},{editor=>$e})}
\end{verbatim}\end{footnotesize}

Since different predicate conditions are used to filter the data in different parts of the data structure, XTQ adopts compound conditions consisting of the predicate conditions with the connectives ``\textsl{par}'' and ``\textsl{with}'' to form a consistent global requirement of filtering the extracted data.

A compound condition composing two (compound) conditions \textsl{c$_{1}$} and \textsl{c$_{2}$} with the  connective ``\textsl{par}'' is a disjunctive condition where \textsl{c$_{1}$} and \textsl{c$_{2}$} are to filter values of different subterms of a d-union or option. Disjunctive conditions can effectively present requests of separately filtering data. For example, in the query Q6, the disjunctive condition filters the bindings of \textsl{\$a} and \textsl{\{\$e\}} respectively. Since \textsl{\$a$|$\{\$e\}} or \textsl{\$a$||$\{\$e\}} is not a valid term, the compound condition cannot be replaced with a predicate condition; in the query Q7, the title and the first two authors are listed for each book, and an empty "et-al" element is appended if the book has additional authors. In this query, the authors elements extracted by a disjunctive pattern is presented by the d-union \textsl{\{\$a\}$||$\{\$b\}}. The d-union is further filtered by a disjunctive condition so that the two groups are filtered separately and processed heterogeneously.

\begin{footnotesize}\begin{verbatim}
Q6. query doc("bib.xml")//book((/author=>$a||editor=>$e),/title=>$t)
     where $a/last=>"Li" par count({$e})>2
     construct {book=>({title=>$t}, {people=>$a|$e})}
Q7. query doc("bib.xml")//book((/author=>$a||/author=>$b),/title=>$t)
     where (count({$a})>0 and count({$a})<=2) par count({$b})>2
     construct {book=>({title=>$t}, ({author=>$a}||
                             ({author=>$b}.(1..2),et-al=>~)))}
\end{verbatim}\end{footnotesize}

The connective ``\textsl{with}'' is used to combine a (compound) condition \textsl{c$_{1}$} and a predicate \textsl{c$_{2}$} if \textsl{c$_{2}$} is to be used to filter the result data elements filtered by  \textsl{c$_{1}$}. The condition ``\textsl{c$_{1}$ with c$_{2}$}'' differs from ``\textsl{c$_{1}$ and c$_{2}$}'' in that it is processed sequentially rather than commutatively, which indicates a bottom-up order on filtering hierarchical data. For example, to find the books each of which has at least 3 authors who has the last name ``Li'' or has the email ending with ``edu'', the condition to filter the authors and the one to filter the groups of the authors work at different levels in the data hierarchy, and thus a bottom-up order to process the data filtering is required. In the query Q8, the first condition filters the bindings of \textsl{(\$l,\$m)}, and the second one filters the set binding of \textsl{\{(\$a, \{\$f\}$_{\$f}$,\{\$m\}$_{\$m}$)\}$_{\$a}$} of the qualified authors. If the connective ``\textsl{and}'' were used instead of ``\textsl{with}'', the predicate condition should have the argument term  ``\textsl{(\$l,\$m,\{(\$a,\{\$f\}$_{\$f}$,\{\$m\}$_{\$m}$)\}$_{\$a}$)}''. However, this term is not valid. In fact, such a condition is inconsistent. It should be syntactically equivalent to ``\textsl{(\$l =`Li' and count(\{\$a\})$>$2) or (endwith(\$m,`edu') and count(\{\$a\})$>$2)}'' following a simple boolean algebra, but the two conditions are not semantically equivalent  as we can consider a counter-example where there are two authors with the last name ``Li'' and other two with the email ending with ``edu''.

\begin{footnotesize}\begin{verbatim}
Q8.query doc("bib.xml")//book(/author=>$a(/last=>$l, /email=>$m),
                                /title=>$t)
    where ($l="Li" or endwith($m,"edu")) with count({$a})>2
    construct results=>{book=>({title=>$t},{authoremail=>$m})}
\end{verbatim}\end{footnotesize}

The connectives ``\textsl{par}'' and ``\textsl{with}'' can be used together to present complex data filtering request. In the query Q9, the values bound to \textsl{\$e} are separately filtered by the conditions ``\textsl{commondomain(\$ma,\$me)}'' and ``\textsl{\$b=\$e}'' in the two parts of disjunctive condition. The result set \textsl{\{\$e\}} in the first condition is further filtered by the condition ``\textsl{count(\{\$e\})$>$1}'' meanwhile the set in the second condition is filtered by the condition ``\textsl{count(\{\$e\})$>$2}''. After that, the set \textsl{\{\$e\}} gathering the values satisfying the two conditions respectively is filtered again by the condition ``\textsl{count(\{\$e\})$>$3}''.

\begin{footnotesize}\begin{verbatim}
Q9. query doc("bib.xml")//book((/author=>$a/email=>$ma||/author=>$b),
                                /editor=>$e/email=>$me)
     where ((commondomain($ma,$me) with count({$e} hid $a)>1)  par
             ($b=$e with count({$e} hid $b)>2)) with
           (count({$e} hid {$a|$b})>3)
     construct {book=>({authorA=>$a|authorB=>$b}, {editor=>$e})}
\end{verbatim}\end{footnotesize}

\subsection{Consistency and Completeness of the Conditions}

In an XTQ query,  a condition is to specify the semantic restrictions on the bindings of one or more portions of the original term.
In discussing the formal aspects of conditions, we often use a raw term to explicitly specify which parts of the overall binding are to be restricted by a condition.

\begin{definition}
A matching term \textsl{t} is a \textbf{raw term in} a matching term \textsl{t'} (or a \textsl{raw term} in short), iff:\\
1) \textsl{t $\leqslant$ t'},\\
2) \textsl{t} = \textsl{(t$_{1}$, \dots, t$_{n}$)} where each \textsl{t$_{i}$} (\textsl{i$\in$\{1..n\}}) is a raw term in \textsl{t'}, or\\
3) \textsl{t }= \textsl{(t$_{1}$ $|$ \dots $|$ t$_{n}$)} where each \textsl{t$_{i}$} (\textsl{i$\in$\{1..n\}}) is a raw term in \textsl{t'}.
\end{definition}
For example, the terms \textsl{\$a}, \textsl{(\$f,\$l)}, \textsl{(\{\$f\}$_{\$f}$,\$l)} and \textsl{\{(\$a, \{\$f\}$_{\$f}$,\{\$l\}$_{\$l}$)\}$_{\$a}$} are all raw terms in T1 whereas \textsl{\{\$a\}$_{\$a}$} is not. We say a term u is a \textbf{unit} term in a raw term \textsl{t}, denoted as \textsl{u$<_{u}$t}, iff: a) \textsl{u} is a variable, a set or a d-union; b) \textsl{u$\leqslant$t}; and c) for any \textsl{u'$\leqslant$t} that \textsl{u$<$u'}, \textsl{u'} is a tuple.
Apparently, a binding of a raw term is always a tuple of unit bindings.

For a condition \textsl{c} in a XTQ query, the \textbf{scope} of \textsl{c} is a raw term in the original term which indicates the portions of the original term to be restricted. Different conditions might have overlapped scopes which work on  different layers in the hierarchy of the original terms. For example, in query Q8 the atomic conditions respectively have the scopes \textsl{\$l},\textsl{ \$m} and \textsl{\{(\$a,\{\$l\}$_{\$l}$,\{\$m\}$_{\$m}$)\}$_{\$a}$}. As discussed previously, if we combine the atomic conditions conjunctively as a predicate condition, the composed condition would concurrently work on the different layers in the hierarchy, e.g.,
\textsl{\$l} and \textsl{\{\$l\}$_{\$l}$}, rather than sequentially, which leads to the inconsistency of data filtering. To formally specify the consistency, XTQ introduces the following definitions.

%
%
%
%
%
%
%

\begin{definition} For two raw terms \textsl{u} and \textsl{v} in \textsl{t}, \textsl{v} is \textbf{finer} than \textsl{u}, denoted by \textsl{t $\lessdot$ u}, iff for any term \textsl{v$_{0}$$<_{u}$v}, there is a term \textsl{u$_{0}$$<_{u}$u} that \textsl{v$_{0}$$<$u$_{0}$} or \textsl{$\exists$u$_{1}$.v$_{0}$$\lessdot$u$_{1}$$<$u$_{0}$}.
\end{definition}
For example, it is easy to see that \textsl{\$l$\lessdot$\{\$l\}$_{\$l}$}, \textsl{(\$a,\$e)$\lessdot$(\{\$a\}$_{\$a}$,\{\$e\}$_{\$e}$)} and \textsl{\{\$a\}$_{\$a}$$\lessdot$\{\$a\}$_{\$a}$$||$\{\$e\}$_{\$e}$}, however, \textsl{(\$a,\$e)$\lessdot$(\$a,\{\$e\}$_{\$e}$)} doesn't hold.

\begin{definition}
For a raw term \textsl{t} and a condition \textsl{c}, \textsl{c} is \textbf{consistent} to \textsl{t} iff:

1. if \textsl{c} is predicate condition whose argument is \textsl{u}, then u$\unlhd$t and there is a raw term r in t that r$\stackrel{*}{\leadsto}$u; The scope of \textsl{c} is \textsl{r};

2. if \textsl{c} = \textsl{c$_{1}$ par c$_{2}$} where \textsl{c$_{1}$} and \textsl{c$_{2}$} have the scopes \textsl{r$_{1}$} and \textsl{r$_{2}$} respectively, then there is a non-empty set \textsl{U} = \{\textsl{u} $|$ \textsl{u} is a raw term in t, \textsl{u$\unlhd$t}, \textsl{u=(u$_{1}$'$||$u$_{2}$')} or \textsl{u$\stackrel{*}{\leadsto}$(u$_{1}$'$|$u$_{2}$')} where  \textsl{u$_{1}$$\leqslant$ u$_{1}$'} and \textsl{u$_{2}$ $\leqslant$u$_{2}$'\}}, and there is \textsl{r$\in$U} such that \textsl{r$\lessdot$u} or \textsl{r$\leqslant$u} holds for any \textsl{u$\in$U}; The scope of \textsl{c} is \textsl{r};

3. if \textsl{c} = \textsl{c$_{1}$ with c$_{2}$} where \textsl{c$_{1}$} and \textsl{c$_{2}$} have the scopes \textsl{r$_{1}$} and \textsl{r$_{2}$} respectively, then \textsl{r$_{1}$$\lessdot$r$_{2}$}, or \textsl{c$_{1}$} = \textsl{c$_{1}$' par c$_{2}$'} and \textsl{r$_{1}$ $\equiv$ r$_{2}$};
The scope of \textsl{c} is \textsl{r$_{2}$}.
\end{definition}


Especially, a consistent predicate condition \textsl{c} whose argument term is \textsl{u} is also specified as the predicate \textsl{$\overline{c}$} whose argument is the scope of \textsl{c}. Here \textsl{c} having the scope \textsl{r} can be denoted as \textsl{c:r},  and \textsl{c[b$_{u}$]=$\overline{c}$[b$_{r}$]} for any binding \textsl{b$_{r}$$\stackrel{*}{\leadsto}$b$_{u}$}.

Processing data filtering in XTQ is to filter out the bindings not satisfying the conditions rather than find the ones satisfying them. These two issues are unified in conventional query where the elements are only conjunctively combined. However, they are different in XTQ since the disjunction is universally used in data extraction and construction.

For example, in the query Q2, the  condition ``\textsl{contains(\$t, (\$l hid \$a)) or contains(\$t, (\$f hid \$a))}'' has the scope  ``\textsl{(\$t,\$a,\$l,\$f)}'' which is a part of the option term ``\textsl{(\$t,\$a,\$l,\$f) $|$ (\$t,\$e,\$l,\$f)}''. This condition restricts the names of the authors, however, it doesn't impose the restriction on the editors' names  and thus all the editors' names would be maintained on default.
%
%
The conditions in a \textsl{where} clause often focus on restricting the bindings of certain raw terms and needn't cover the bindings of the disjunctive terms, like the editors' names in Q2. However, in processing an XTQ query, there should be a complete filtering request on all the disjunctive branches of the original binding.

\begin{definition}
A consistent condition \textsl{c} with the scope \textsl{r} is \textbf{parallelly complete} with respect to a raw term \textsl{t} iff: \\
1) there is not a term u$\unlhd$t that r$|$u$\unlhd$t or r$||$u$\unlhd$t,\\
2) if \textsl{c} = \textsl{c$_{1}$ with c$_{2}$}, then \textsl{c$_{1}$} is parallelly complete with respect to \textsl{r}.
\end{definition}

The conditions in a where clauses are often not complete, as shown in the previous examples. Before processing them, the parser would change them into a complete condition with a complementing mechanism. In this mechanism a special kind of wildcard predicates in the form of ``\textsl{any(u)}'' is used as the padding conditions. An \textsl{any(u)} is a condition  which has the scope \textsl{u}  and can be satisfied by any binding of \textsl{u} (even a null one).
A common condition can be complemented using the following \textsl{comp()} function.

\begin{definition}
For a condition \textsl{c} consistent to the raw term \textsl{t}, the complemented condition denoted as \textsl{comp(c,t)} is constructed as follows: \\
1) if \textsl{c} is parallel complete to \textsl{t} then \textsl{comp(c,t)=c}; otherwise\\
2) if c is a predicate or \textsl{c} = \textsl{c$_{1}$ par c$_{2}$} , then \textsl{comp(c,t)} = \textsl{c par any(u$_{1}$) par \dots par any(u$_{n}$)} where \\
\indent i) each \textsl{u$_{i}$$\leqslant$t},\\
\indent ii) \textsl{comp(c,t)} is consistent to \textsl{t},\\
\indent iii) there is not a term \textsl{u'} that \textsl{comp(c,t) par u'} is consistent to \textsl{t},\\
\indent iv) for any term \textsl{u$_{i}$'$\leqslant$t} that \textsl{c par any(u$_{1}$) par \dots par any(u$_{i-1}$) par any(u$_{i}$') par any(u$_{i+1}$) par \dots par any(u$_{n}$)} is consistent to \textsl{t}, \textsl{u$_{i}$'$\leqslant$u$_{i}$} holds;\\
3) if \textsl{c} = \textsl{c$_{1}$ with c$_{2}$}, then \textsl{comp(c,t)} = \textsl{(comp(c$_{1}$,t$_{1}$) with c$_{2}$) par any(u$_{1}$) par \dots par any(u$_{n}$)} where \textsl{t$_{1}$} is the scope of \textsl{c} and the conditions \textsl{any(u$_{1}$)}, \dots and \textsl{any(u$_{n}$)} satisfies \textsl{comp(c$_{2}$,t)} = \textsl{c$_{2}$ par any(u$_{1}$) par \dots par any(u$_{n}$)}.
\end{definition}
For example, the complemented condition of the query Q2 can be ``\textsl{contains(\$t, (\$l hid \$a)) or contains(\$t, (\$f hid \$a)) par any((\$e,\{\$l\},\{\$f\}))}'', and the scope of the complemented condition is ``\textsl{(\$t, \$a,\{\$l\}$_{\$l}$,\{\$f\}$_{\$f}$)$|$(\$t, \$e,\{\$l\}$_{\$l}$,\{\$f\}$_{\$f}$)}''.

\begin{prop}
For a condition \textsl{c} consistent to the term \textsl{t}, \textsl{comp(c,t)} is consistent to \textsl{c} and is parallelly complete with respect to \textsl{c}.
\end{prop}

\subsection{Semantics of Data Filtering}

%
%

The semantics of data filtering is built on a partial order on bindings and a special structure named filtering couple.

\begin{definition}
For the bindings \textsl{b:t} and \textsl{b':t}, \textsl{b} is \textbf{not larger than} \textsl{b'}, denoted as \textsl{b$\sqsubseteq$b'}, iff one of the following criteria is satisfied:\\
1. \textsl{b'} = \textsl{b};\\
2. if \textsl{b=\{b$_{i}$$|$i$\in$\{1.. n\}\}} and \textsl{b'=\{b$_{j}$$|$j$\in$ \{1..m\}\}}, then \textsl{$\forall$ i$\in$\{1.. n\}. $\exists$j$\in$\{1..m\}. b$_{i}$$\sqsubseteq$b$_{j}$'};\\
3. if \textsl{b=(b$_{1}$:t$_{1}$,\dots, b$_{n}$:t$_{n})$} and \textsl{b'=(b$_{1}$':t$_{1}$,\dots, b$_{n}$':t$_{n}$)}, then \textsl{b$_{i}$ $\sqsubseteq$b$_{i}$'} (i $\in$ \{1.. n\});\\
4. if \textsl{b=(b$_{1}$:t$_{1}$$||$\dots$||$ b$_{n}$:t$_{n})$} and \textsl{b'=(b$_{1}$':t$_{1}$$||$\dots$||$b$_{n}$':t$_{n}$)}, then \textsl{b$_{i}$=null or b$_{i}$ $\sqsubseteq$b$_{i}$'} (i $\in$ \{1.. n\}).
\end{definition}

It is easy to see that, for a binding \textsl{b}, all the bindings \textsl{b$_{i}$} that \textsl{b$_{i}$$\sqsubseteq$b} form a complete lattice. Naturally, the supremum  and the infimum of the bindings \textsl{b$_{i}$} and \textsl{b$_{j}$} that \textsl{b$_{i}$$\sqsubseteq$b} and \textsl{b$_{j}$$\sqsubseteq$b} are denoted by \textsl{b$_{i}$$\sqcup$b$_{j}$} and \textsl{b$_{i}$$\sqcap$b$_{j}$} respectively.

As mentioned previously, in the processing of data query, data filtering occurs between data restructuring and data construction. It is the restructured binding of the construction pattern's argument term rather than the original binding to be filtered by the conditions. Especially, in the discussion of data filtering, we don't consider filtering the folded set binding involving content terms. That means, all the content term \textsl{\{t\%,t'\}$_{t\%}$} would be transformed back to the term \textsl{t'} for processing the filtering, and then the \textsl{set-fold} rule would be applied to the filtered result to generate the bindings for instantiating the construction patterns. For example, in processing the query Q2, it is the binding \emph{restr-binding} that is to be filtered by the condition.
Since the conditions specify the restrictions on the bindings of the raw terms, we often resort to the raw form of a transformed binding or a filtered binding to facilitate specifying the semantics of data filtering.

\begin{definition}
For an original binding \textsl{b:t} and the bindings \textsl{b':t'},\textsl{b'':t'} and \textsl{b$_{u}$:u} that \textsl{b$\stackrel{*}{\leadsto}$b'}, \textsl{b''$\sqsubseteq$b'} and \textsl{b$_{u}$$\leqslant$b''}, the \textbf{projection of b$_{u}$ in b} is a binding, denoted as \textsl{$\overline{b_{u}}$:$\overline{u}$} where \textsl{$\overline{u}$} is a raw term in \textsl{b}, that \textsl{$\overline{b_{u}}$$\stackrel{*}{\leadsto}$b$_{u}$':u} and \textsl{b$_{u}$$\sqsubseteq$b$_{u}$'}; and for any binding \textsl{b$_{\overline{u}}$:$\overline{u}$} that \textsl{b$_{\overline{u}}$$\stackrel{*}{\leadsto}$b$_{\overline{u}}$':u} and \textsl{b$_{u}$$\sqsubseteq$b$_{\overline{u}}$'}, \textsl{$\overline{b_{u}}$$\sqsubseteq$b$_{\overline{u}}$} holds.
\end{definition}
For example, for a binding \textsl{b=\{(b$_{1}$:\$a,b$_{2}$:\$b), (b$_{3}$:\$a,b$_{4}$:\$b)\}:\{(\$a,\$b)\}$_{(\$a,\$b)}$}, the projection \textsl{$\overline{b}$}=\textsl{(\{b$_{1}$,b$_{3}$\},\{b$_{2}$,b$_{4}$\})}. In the rest of this subsection, we assume the existence of an original binding \textsl{b$_{O}$:t$_{O}$} and the binding \textsl{$\overline{b}$:$\overline{t}$} is the projection of the binding \textsl{b:t} in \textsl{b$_{O}$}.

\begin{prop}
For the binding \textsl{b$_{u}$:u} and \textsl{b:t} that \textsl{b$_{u}$$\leqslant$b}, \textsl{$\overline{b_{u}}$$\unlhd$$\overline{b}$}.
\end{prop}


\begin{definition}
A couple \textsl{(b,s)} is a \textbf{filtering couple} where\\
1) \textsl{b:t} is a binding, and \\
2) \textsl{s} is a set of bindings, named restriction set, where there is a raw term \textsl{r} in \textsl{$\overline{t}$} that \textsl{b$_{r}$:r} holds for each \textsl{b$_{r}$$\in$s} and there is a \textsl{b':r$\unlhd$$\overline{b}$} that \textsl{b$_{r}$$\subseteq$b'}.
\end{definition}
A filtering couple \textsl{(b, s)} is to specify the status of data filtering. Here the restriction set is assumed to contain the elements which satisfying the condition in filtering and thus \textsl{r} is assumed to be the scope of the condition.


We also use \textsl{s:\{r\}} to denote a restriction set with its term as a set without an index. To facilitate specifying the semantics of data filtering, we introduce the following operations on the restriction sets.

\begin{definition}

1. The filtering couple \textsl{(b, s':\{r'\})} is the expanded couple of the filtering couple \textsl{(b, s:\{r\})}, denoted as \textsl{(b, s)$\hookrightarrow$(b, s')}, where

1) \textsl{r$\lessdot$r'};


2) for any pair of the bindings \textsl{b$_{r}$:r$\in$s} and \textsl{b$_{u}$:u$\leqslant$$\overline{b}$} that \textsl{(b$_{r}$,b$_{u}$)$\unlhd$$\overline{b}$}, if \textsl{(r,u)$\unlhd$r'}, then there is a binding \textsl{b$_{r}$'$\in$s'} that \textsl{(b$_{r}$,b$_{u}$)$\unlhd$b$_{r}$'};

3) if \textsl{r$_{1}||$r$_{2}$$\leqslant$r'}, and for any term \textsl{u$\leqslant$r$_{2}$}, \textsl{u$\leqslant$r} does not hold, then for any \textsl{b$_{r}$'$\in$s'} that b$_{r}$'$\sqsubseteq$b$_{r0}$$\unlhd$$\overline{b}$,
if \textsl{b$_{r2}$:r$_{2}$$\leqslant$b$_{r}$'} then \textsl{b$_{r2}$}$\leqslant$b$_{r0}$; and

4) for each \textsl{b$_{r}$'$\in$s'} and for any \textsl{b$_{u}$:u$\leqslant$b$_{r}$'} where \textsl{u$\leqslant$r}, there is a binding \textsl{b$_{r}$$\in$s} that \textsl{b$_{u}$$\leqslant$b$_{r}$} and \textsl{b$_{r}$$\unlhd$b$_{r}$'}.

\noindent 2. The filtering couple \textsl{(b,s)} is the combined couple of the couples \textsl{(b$_{1}$,s$_{1}$)} and \textsl{(b$_{2}$,s$_{2}$)}, denoted as \textsl{(b,s)} = \textsl{(b$_{1}$,s$_{1}$) $\sqcup$ (b$_{2}$,s$_{2}$)} where

1) \textsl{b} = \textsl{b$_{1}$ $\sqcup$ b$_{2}$};

2) \textsl{s$_{1}$:\{r\}}, \textsl{s$_{2}$:\{r\}} and \textsl{s:\{r\}};

3) a binding \textsl{b$_{r}$} = \textsl{(b$_{u_{1}}$,\dots, b$_{u_{n}}$)$\in$s} where \textsl{b$_{u_{i}}$:u$_{i}$} and \textsl{u$_{i}$$<_{u}$r} iff:

\indent\indent    i) \textsl{b$_{r}$$\sqsubseteq$b$_{r}$':r$\unlhd$b} where \textsl{b$_{r}$'} = \textsl{(b$_{u_{1}}$',\dots, b$_{u_{n}}$')};

\indent\indent    ii) \textsl{$\exists$b$_{r}$''. b$_{r}$''=(b$_{u_{1}}$'',\dots, b$_{u_{n}}$'')$\in$s$_{1}$$\cup$s$_{2}$} and \textsl{$\forall$i$\in$\{1..n\}.b$_{u_{i}}$''$\sqsubseteq$b$_{u_{i}}$'};

\indent\indent    iii) b\textsl{$_{u_{i}}$} =$\bigsqcup$\{\textsl{b$_{v}$:u$_{i}$} $|$ \textsl{$\exists$b$_{r0}$:r$\in$s$_{1}$$\cup$s$_{2}$. b$_{v}$$\in$b$_{r0}$} and \textsl{b$_{v}$$\sqsubseteq$b$_{u_{i}}$'}\}.
%
%
%
%
%

%
%

\noindent 3. The bound set of a restriction set \textsl{s:\{r\}} on a binding \textsl{b:u} where \textsl{u$<_{u}$r}, denoted as \textsl{s$\downarrow_{b}$}, is a new restriction set that \textsl{s$\downarrow_{b}$} = \textsl{\{b$_{r'}$ $|$ (b,b$_{r'}$)$\in$s\}}.
\end{definition}
The expanding operation is to extend the restriction set \textsl{s:\{r\}} to a new restriction set \textsl{s':\{r'\}} by including the elements in \textsl{b} which are not in the scope \textsl{r} but in the scope \textsl{r'}. For example,  for a set \textsl{s:\{\$a\}=\{\$a$\mapsto$(d$_{a11}$,l$_{a11}$)\}}, it can be expanded to \textsl{s':\{(\{\$a\},\$t)\}=\{(\{\$a$\mapsto$(d$_{a11}$,l$_{a11}$)\},\$t$\mapsto$(d$_{t1}$,l$_{t1}$))\}}. The combination operation is to combine two filtering couples \textsl{(b$_{1}$,s$_{1}$)} and \textsl{(b$_{2}$,s$_{2}$)} which are to restrict the bindings  with parallel conditions. For example, \textsl{s$_{1}$=\{\{\$a$\mapsto$(d$_{a11}$,l$_{a11}$)\}$||$\{\$e$\mapsto$null\}\}},
\textsl{s$_{2}$=\{\{\$e$\mapsto$(d$_{e11}$,l$_{e11}$)\}$||$\{\$a$\mapsto$null\}\}} and \textsl{s} = \textsl{\{\{\$a$\mapsto$(d$_{a11}$,l$_{a11}$)\}$||$\{\$e$\mapsto$(d$_{e11}$,l$_{e11}$)\}\}}.
The bound set is to narrow the restriction set by fixing a unit binding in it.
Especially, \textsl{s$\downarrow_{b}$=\{$\varepsilon$\}} if \textsl{r=t} and \textsl{b$\in$s}. We also use \textsl{r$\downarrow$u} to denote the term of the new bindings in the bound set.

\begin{figure}
\begin{center}
\begin{footnotesize}
\begin{tabular}{|p{\textwidth}|}
\hline
\textbf{Predicate Condition Filtering}\\
Assumption: ${b:t, \hspace{5pt} s:\{r\}, \hspace{5pt} c:r', \hspace{5pt} r \lessdot r'}$\\
\vspace{1pt}
$\dfrac{ (b,s)\hookrightarrow(b,s_{0}:\{r'\}), s'=\{b_{r'}\hspace{2pt} | b_{r'}\in s_{0} \hspace{2pt} \wedge \hspace{2pt}\overline{c}[b_{r'}]\rightarrow true\}, \hspace{5pt} b[s'] \rightarrow b'}
{(b, s) |_{f} c \rightarrow (b', s')}$  (rset-expand)\\
\vspace{1pt}
$\dfrac{\forall u\leqslant \overline{t}. \neg u\leqslant r, (\overline{t},r)\unlhd t_{O}}{b[s]\rightarrow b}$ (conj-unrestr) \hspace{5pt}
$\dfrac{\overline{t}|r \unlhd t_{O} \vee \overline{t}||r \unlhd t_{O}}{b[s]\rightarrow \bot}$(disj-unrestr)\\
\vspace{1pt}
$\dfrac{s = \emptyset}{b[s] \rightarrow \bot}$ (filter-out)
\hspace{5pt}
$\dfrac{ u<_{u}\overline{t}, u<_{u}r, \overline{b}\equiv (b_{u}:u, b_{0}), b[s\downarrow b_{u}]\rightarrow b'}{b[s] \rightarrow b' }$ (conj-restr)\\
\vspace{1pt}
$\dfrac{\exists u\leqslant\overline{t}.u\leqslant r, \neg (\exists u<_{u}\overline{t}.u<_{u}r), b=(b_{1},\dots, b_{n}), b_{i}[s]\rightarrow b_{i}' (i\in\{1..n\})}{b[s] \rightarrow !(b_{1}', \dots, b_{n}') ^{\dag}}$ (tuple-comp)\\
\vspace{1pt}
$\dfrac{\exists u\leqslant\overline{t}.u\leqslant r, \neg (\exists u<_{u}\overline{t}.u<_{u}r), b=\{b_{1},\dots, b_{n}\}, b_{i}[s]\rightarrow b_{i}' (i\in\{1..n\})}
{b[s] \rightarrow !\{b_{1}',\dots, b_{n}'\} ^{\ddag}}$ (set-comp)\\
\vspace{1pt}
$\dfrac{\exists u\leqslant\overline{t}.u\leqslant r, \neg (\exists u<_{u}\overline{t}.u<_{u}r), b=b_{1}||\dots||b_{n}, b_{i}[s]\rightarrow b_{i}' (i\in\{1..n\})}{b[s] \rightarrow !(b_{1}'|| \dots|| b_{n}')^{\sharp}}$(dunion-comp)\\
\vspace{1pt}
$\dag$ $\forall$i$\in$\{1..n\}.b$_{i}$$\neq$$\bot$ $\Rightarrow$ !(b$_{1}$,\dots,b$_{n}$)=(b$_{1}$,\dots,b$_{n}$), otherwise !(b$_{1}$,\dots,b$_{n}$)=$\bot$\\
$\ddag$ $\exists$ b$\in$S. b$\neq\bot \Rightarrow$ !S=\{b$_{i}$$\in$S $|$ b$_{i}$$\neq\bot$\}, otherwise !S=$\bot$\\
$\sharp$ if b$_{i}$=$\bot$(i$\in$\{1..n\}) then !(b$_{1}$$||\dots||$b$_{n}$)=$\bot$ , otherwise !(b$_{1}$$||\dots||$b$_{n}$) = (b$_{1}$'$||\dots||$b$_{n}$') where b$_{i}$'=null if b$_{i}$=$\bot$ otherwise b$_{i}$'=b$_{i}$.\\
\hline
\textbf{Compound Condition Filtering}\\
${(b,s)|_{f}c_{1} \rightarrow (b_{1},s_{1}), \hspace{5pt}(b,s)|_{f}c_{2} \rightarrow (b_{2},s_{2}), \hspace{5pt}c_{1}\hspace{2pt} par\hspace{2pt} c_{2}:r, }$\\ $\dfrac{\hspace{5pt}(b_{1},s_{1})\hookrightarrow(b_{1},s_{1}':\{r\}), (b_{2},s_{2})\hookrightarrow(b_{2},s_{2}':\{r\})\hspace{5pt} }
{\hspace{5pt}(b,s) |_{f} (c_{1}\hspace{2pt} par \hspace{2pt}c_{2}) \rightarrow (b_{1},s_{1}')\sqcup (b_{2},s_{2}')\hspace{5pt}}$ (par-comp)\\
\vspace{1pt}
$\dfrac{(b,s)|_{f}c_{1} \rightarrow (b_{1},s_{1}),  (b_{1},s_{1})|_{f}c_{2} \rightarrow (b_{2},s_{2})}{(b,s)|_{f} (c_{1}\hspace{2pt} with\hspace{2pt} c_{2}) \rightarrow (b_{2},s_{2})}$ (with-comp)\\
\hline
\end{tabular}
\end{footnotesize}
\caption{Semantic rules of data filtering}
\end{center}
\end{figure}

The semantics of data filtering is defined in Figure 12. The semantic rules are specified as either \textsl{(b,s)$|_{f}$c$\rightarrow$(b',s')}, indicating that filtering a couple \textsl{(b,s)} by the condition \textsl{c} should result in a couple \textsl{(b',s')}, or \textsl{b[s]$\rightarrow$b'}, indicating that filtering \textsl{b} by the restriction set \textsl{s} should result in a binding \textsl{b'}. Here we especially introduce a symbol ``$\bot$'' to denote the ``false'' binding indicating the filtering result of unsatisfying the conditions. The predicate condition filtering rules show how a filtering result of a predicate condition is formed by decomposing the binding to the sub-bindings and narrowing the restriction set to a certain bound set. 

For example, a condition ``\textsl{contains(\$t, (\$l hid \$a)) or contains(\$t, (\$f hid \$a))}'' might have the following restriction set s=
\{(b$_{\$t,t1}$, b$_{\$a,a11}$, \$l$\mapsto$("Yeats",l$_{l11}$), \$f$\mapsto$("Willam" ,l$_{f11}$)),
(b$_{\$t,t2}$, b$_{\$a,a21}$, \$l$\mapsto$("Blackburn",l$_{l21}$), \$f$\mapsto$("Paul",l$_{f21}$))\}
and the following filtering steps illustrate how the filtering rules are used in carrying out the filtering of predicate conditions on the binding restr-binding in Section 3.4.

\begin{small}
\textsl{b$_{r1}$ [\{$\varepsilon$\}] $\rightarrow$ b$_{r1}$ $\vdash$ b$_{r1}$[\{b$_{\$t,t1}$\}] $\rightarrow$ b$_{r1}$ $\vdash$ \dots \\
 \indent \indent \indent $\vdash$ b$_{r1}$ [s$\downarrow$b$_{\$a,a11}$] $\rightarrow$ b$_{r1}$ $\vdash$ b$_{r1}$[s]$\rightarrow$ b$_{r1}$, \\
\indent b$_{r2}$ [$\emptyset$] $\rightarrow$ $\bot$ $\vdash$ b$_{r2}$ [s$\downarrow$b$_{\$a,a12}$] $\rightarrow$ $\bot$ $\vdash$  b$_{r2}$[s] $\rightarrow$ $\bot$, ...\\
\indent b$_{restr-binding}$ [s] $\rightarrow$ \{..., b$_{r1}$, b$_{r4}$, ...\}\\
}\end{small}

The \textsl{par-comp} rule indicates that the filtering of a parallel condition be the supremum of the bindings and the union of the restriction sets. The \textsl{with-comp} rule indicates that the filtering of a sequential condition is carried out by filtering the conditions sequentially. For example, in processing the query Q2, the complemented condition ``\textsl{contains(\$t, (\$l hid \$a)) or contains(\$t, (\$f hid \$a)) par any((\$e,\{\$l\},\{\$f\}))}'' is used and the filtering result of restr-binding is  \{..., b$_{r1}$, b$_{r3}$, b$_{r4}$, b$_{r6}$, b$_{r7}$,...\}. Therefore, the final construction result of the query is like:

\begin{footnotesize}\begin{verbatim}
 results => {| ...;
    result => {|
      person => {|last=>"Blackburn"; first=>"Paul"|};
      book => {|
        title=>"The Selected Poems of Paul Blackburn";
        role => "author" |} |};
    result => {|
      person => {|last=>"Jarolim"; first=>"Edie"|};
      book => {|
        title=>"The Selected Poems of Paul Blackburn";
        role => "editor" |} |};
    result => {|
      person => {|last=>"Rosenthal"; first=>"M.L."|};
      book => {|
        title=>"Selected Poems And Four Plays of William Butler Yeats";
        role => "editor" |};
      book => {|
        title=> "The Selected Poems of Paul Blackburn";
        role => "editor" |}  |};
    result => {|
      person => {|last=>"Yeats"; first=>"Willam"|};
      book => {|
        title=>"Selected Poems And Four Plays of William Butler Yeats";
        role => "author" |} |};
    ... |}
\end{verbatim}\end{footnotesize}

\begin{definition} For the binding \textsl{b:t} and the condition \textsl{c} with the scope \textsl{r}, a binding \textsl{b$_{0}$:u$\leqslant$b} where \textsl{$\overline{u}$$\leqslant$r} \textbf{is effective for} \textsl{c} in \textsl{b} iff:\\
1) if \textsl{c} is a predicate condition with the scope \textsl{r}, then  there is a binding \textsl{b$_{r}$:r}  that \textsl{$\overline{b_{0}}$$\leqslant$b$_{r}$}, \textsl{b$_{r}$$\in$$\overline{b}$} and \textsl{$\overline{c}$[b$_{r}$] $\rightarrow$ true};\\
2) if \textsl{c} =\textsl{ c$_{1}$ par c$_{2}$} where \textsl{c$_{1}$} and \textsl{c$_{2}$} respectively have the scopes \textsl{r$_{1}$} and \textsl{r$_{2}$}, then there are the bindings \textsl{b$_{1}$:t}, \textsl{b$_{2}$:t},
\textsl{b$_{1}$':u} and \textsl{b$_{2}$':u} that \\
\indent i) \textsl{b} = \textsl{b$_{1}$$\sqcup$b$_{2}$} and \textsl{b$_{0}$} = \textsl{b$_{1}$'$\sqcup$b$_{2}$'},\\
\indent ii) for any \textsl{b$_{u1}$:u$_{1}$$\leqslant$b$_{1}$} where \textsl{$\overline{u_{1}}$$\leqslant$r$_{1}$}, \textsl{b$_{u1}$} is effective for \textsl{c$_{1}$} in \textsl{b$_{1}$}, and\\
\indent iii)  for any \textsl{b$_{u2}$:u$_{2}$$\leqslant$b$_{2}$} where \textsl{$\overline{u_{2}}$$\leqslant$r$_{2}$}, \textsl{b$_{u2}$} is effective for \textsl{c$_{2}$} in \textsl{b$_{2}$}.\\
3) if \textsl{c} = \textsl{c$_{1}$ with c$_{2}$}, then \\
\indent i) \textsl{b$_{0}$}  is effective for \textsl{c$_{2}$} in \textsl{b}, and \\
\indent ii) for any \textsl{b$_{1}$:u$_{1}$$\leqslant$b$_{0}$} and \textsl{$\overline{u_{1}}$$\leqslant$r$_{1}$} where \textsl{r$_{1}$} is the scope of \textsl{c$_{1}$}, \textsl{b$_{1}$} is effective for \textsl{c$_{1}$} in \textsl{b}.
\end{definition}

The effectiveness of the bindings indicates that all the  bindings in the filtering result should contribute to the satisfaction of the condition in the global binding. It is an interesting property which makes the bindings satisfying the conditions in a ``as much as possible'' way. For example, the following query shows how the effectiveness is embodied in XTQ.

\begin{footnotesize}\begin{verbatim}
Q10. query doc("bib.xml")//book=>$b(/author=>$a, editor=>$e)
     where $a/last = $e/last
     construct result=>(pairs=>{{(author=>$a, editor=>$e)}} groupby $b,
                        grp1=>{{(author=>$a, {editor=>$e})}} groupby $b,
                        grp2=>{({author=>$a}, {editor=>$e})} groupby $b)
\end{verbatim}\end{footnotesize}

In this query, the pairs of \textsl{(\$a,\$e)} in each book are presented in three ways. Assume that in a certain book the bindings \textsl{(\$a$\mapsto$(d$_{a1}$,l$_{a1}$), \$e$\mapsto$(d$_{e1}$,l$_{e1}$))}, \textsl{(\$a$\mapsto$(d$_{a1}$,l$_{a1}$), \$e$\mapsto$(d$_{e2}$,l$_{e2}$))}, \textsl{(\$a$\mapsto$(d$_{a2}$,l$_{a2}$), \$e$\mapsto$(d$_{e3}$,l$_{e3}$))} satisfy the condition. Then for the construction pattern, the \textsl{pairs} branch would generate the ordered set value  \textsl{\{(a1,e1);(a1,e2);(a2,e3)\}}, the \textsl{grp1} branch would generate the value  \textsl{\{(a1,\{e1;e2\});(a2,\{e3\})\}}, and the \textsl{grp2} branch would generate the value \textsl{\{(\{a1,a2\},\{e1,e2,e3\})\}}.

\begin{definition} A binding \textsl{b} satisfies a condition, denoted by \textsl{b} $\models$ \textsl{c}, iff:
 for any binding \textsl{b$_{0}$}:\textsl{u}  that \textsl{b$_{0}$$\leqslant$b} and \textsl{$\overline{u}$$\leqslant$r} where \textsl{r} is the scope of \textsl{c}, \textsl{b$_{0}$} is effective for \textsl{c} in \textsl{b}.
\end{definition}

Based on the above definitions, we have the following fact indicating the soundness of data filtering.
\begin{prop}
For a binding \textsl{b:t} and a condition \textsl{c} complete with respect to \textsl{t}, if \textsl{(b,\{$\epsilon$\})$|_{f}$c$\rightarrow$(b',s')}, then \textsl{b'$\models$c}.
\end{prop}

\section{Function in XTQ}
\subsection{Enhanced Data Extraction in Functions}



The function in XTQ is a QWC statement which uses an input value as its data source and constructs an output value. The input value is represented by the keyword ``\#input'' in the query clause. As the extraction patterns indicate, the input value can be parsed and extracted in two ways: one is to use document extraction patterns by treating the value as a common document fragment; the other is to use the value extraction patterns to parse the composite structure of the value and extract the components of interest. The former one has been introduced in Section 3.3, and the latter one is specified in Figure 13.

\begin{figure}
\begin{center}
\begin{small}
\begin{tabular}{|p{0.14\textwidth} p{0.02\textwidth} p{0.73\textwidth}|}
\hline
$<$valexptn$>$ &::=& $<$atomvptn$>$ $|$ $<$compvptn$>$\\
$<$atomvptn$>$ &::=& `\textbf{*}'  $|$ $<$value$>$ $|$ $<$var$>$ $|$ $<$var$>$ $<$extrptn$>$  \\
$<$compvptn$>$ &::=& `\{'$<$atomvptn$>$`\}'  $|$  $<$valexptn$>$`::'$<$valexptn$>$ $|$ \\
&&`('$<$valexptn$>$`)' $|$ $<$valexptn$>$`,'$<$valexptn$>$ $|$ \\
&& $<$valexptn$>$`$||$'$<$valexptn$>$ $|$ $<$optvptn$>$\\
$<$optvptn$>$ &::=& $<$labelptn$>$ `$|$' $<$labelptn$>$  $|$ $<$labelptn$>$ `$|$' $<$optvptn$>$  \\
$<$labelptn$>$ &::=& $<$label$>$`\&'$<$valexptn$>$ \\
\hline
\end{tabular}
\caption{Value extraction pattern for function}
\end{small}
\end{center}
\end{figure}

As Figure 13 shows, the value extraction patterns are specified according to the composite structure of each types, and thus matching the pattern with the values of the compatible structure would result in a binding of a similar structure.
The semantics of matching the value extraction pattern is listed in Figure 14. Here we use \textsl{p:t} to denote that the value extraction pattern \textsl{p} corresponds to the type \textsl{t}, use \textsl{p$_{d}$} to denote a document extraction pattern, and use \textsl{p$_{o}$} to denote an option extraction pattern. We also use the symbol ``$\bot$'' to denote ``false'' result, i.e., the result of a mismatching of value, and the ``!'' function for maintaining the valid bindings as in Figure 12. In Figure 14, we introduce two special functions \textsl{$\delta$'(v)} and \textsl{$\lambda$(v)} which respectively generate a document fragment and a valid location from the value \textsl{v} so as to make the binding sound. As the \textsl{doc-extr} and the \textsl{var-match} rules show, the \textsl{$\delta$'(v)} function works like the \textsl{$\delta$(v)} function, but the former maintains the existing locations of the elements in \textsl{v} meanwhile specifying the successive relationships among the elements when necessary. The \textsl{$\lambda$(v)} function returns the location of \textsl{v} or generates a new location \textsl{l$_{v}$} for the value \textsl{v} (if it has not one) and the \textsl{l$_{v}$} is in the proper successive relations with the locations of preceding or succeeding values.

\begin{figure}[h]
\begin{center}
\begin{footnotesize}
\begin{tabular}{|p{\textwidth}|}
  \hline
${v \hspace{2pt}|\hspace{2pt} * \rightarrow \varepsilon}$ (univ-match) \hspace{10pt}  ${v\hspace{2pt} | \hspace{2pt}v \rightarrow \varepsilon}$ (value-match)\\
$\dfrac{v \neq v'}{v\hspace{2pt} | \hspace{2pt}v' \rightarrow \bot }$(value-mismatch)
\hspace{10pt}
$\dfrac{v:t, p:t', \hspace{2pt}t\neq t'}{ v \hspace{2pt}|\hspace{2pt} t \rightarrow \bot}$ (type-mismatch)\\
$\dfrac{\delta'(v)\hspace{2pt} |\hspace{2pt} p_{d} \rightarrow b} {v\hspace{2pt} |\hspace{2pt} p_{d} \rightarrow b}$ (doc-extr)\hspace{10pt}
$v\hspace{2pt} | \hspace{2pt}x \rightarrow x\mapsto(\delta'(v), \lambda(v))$ (var-match)\\
$\dfrac{v \hspace{2pt}|\hspace{2pt} x \rightarrow b, v\hspace{2pt} |\hspace{2pt} p \rightarrow b'}{ v \hspace{2pt}|\hspace{2pt} x \hspace{2pt} p \rightarrow !(b,b')}$ (var-ptn-match)\hspace{10pt}
$\dfrac{v \hspace{2pt}| \hspace{2pt}p \rightarrow b, v'\hspace{2pt} |\hspace{2pt} p' \rightarrow b'}{(v,v')\hspace{2pt} |\hspace{2pt} (p,p') \rightarrow !(b,b')}$ (tpl-match)\\
$\dfrac{v_{i}\hspace{2pt} | \hspace{2pt}p \rightarrow b_{i} (i\in\{1..n\})}{ \{v_{1};...;v_{n}\} \hspace{2pt}|\hspace{2pt} \{p\} \rightarrow !\{b_{1},...,b_{n}\}}$ (set-match) \hspace{2pt}
$\dfrac{v_{1} \hspace{2pt}| \hspace{2pt}p \rightarrow b, \hspace{2pt}\{v_{2};\dots;v_{n}\}'\hspace{2pt} | \hspace{2pt}p' \rightarrow b'} {\{v_{1};\dots;v_{n}\} \hspace{2pt}|\hspace{2pt} p::p' \rightarrow !(b,b')}$ (rec-match)\\
$\dfrac{v\hspace{2pt} | \hspace{2pt}p \rightarrow b, v' \hspace{2pt}|\hspace{2pt} p' \rightarrow b'}{(v || v') \hspace{2pt}|\hspace{2pt} (p || p') \rightarrow !(b || b')}$ (dunion-match) \hspace{10pt}
$\dfrac{v\hspace{2pt} | \hspace{2pt}p \rightarrow b}{l\&v \hspace{2pt}| \hspace{2pt}l\&p|p_{o} \rightarrow b}$ (opt-match)\\
\hline
\end{tabular}
\end{footnotesize}
\caption{Semantic rules of term extraction patterns}
\end{center}
\end{figure}


For example, the query Q11 gathers the information of the books from three bookstores, i.e., ``bsa.xml'', ``bsb.xml'' and ``bsc.xml'' to find the minimal price of each book and the corresponding bookstore and book balance.  In this query the function \textsl{mapget} is defined to fetch  the contents associated with the specified key in a group of \textsl{(key, content)} pairs.
In the invocation, the minimal price of each book is passed as the key, and the group containing the pairs of the price and the bookstore information is passed as the content.
In the function body, the input value  would be parsed and bound to the matching term ``\textsl{(\$key,\{\$m,(\$k, \$c)\}$_{\$m}$}'', and then the set of values of \textsl{\$c} satisfying the condition ``\textsl{\$k=\$key}'' would be returned.

\begin{footnotesize}\begin{verbatim}
Q11. declare mapget(*, {(*,*)}):{*} as (
      query #input[($key, {$m[($k, $c)]})]
      where $k = $key
      construct {$c} )
query (doc("bsa.xml")||doc("bsb.xml")||doc("bsc.xml"))
        (/storename=>$n,//book(/title=>$t,/price=>$p,/balance=>$b)
construct  {bookinfo=>(title=>$t%, mapget(min({$p}),
                {($p%, (price=>$p%,{(store=>$n, balance=>$b)}))}))}
\end{verbatim}\end{footnotesize}


A trick in designing the set pattern is that there is often a variable to match the elements of the set, which plays the role of the index variable like \textsl{\$m} in Q11.  Thus the bindings from the term extraction pattern and the ones from the document extraction pattern have the same form and can be used seamlessly.
Furthermore, these variables can also be omitted in practice, following the same way as omitting the variable \textsl{x} in the element pattern  \textsl{/n$\Rightarrow$x} as previously mentioned.

XTQ functions often use ``*'' to denote a generic value type and adopt a loose type checking for the values, and the value matching the ``*'' can be dynamically parsed by finer value extraction patterns. The runtime typing error would be treated as a special query error as if a value in the function image is missing, which was common for semi-structured data.

%

XTQ allows recursive query functions. For example, in the query Q12, a function \textsl{successive()} is defined to find in a set of elements the first subset whose original locations are successive. In this function, the argument set is analyzed by checking whether the first item and the second item are successive and the first item is appended with the recursive invocation result of the remainder of the set. In Q12, the ``bib2.xml'' combines the books information in a plain element ``books'', and the book information is in a regular expression structure  \textsl{(title, author+)+}. In this scenario, the query can easily parse each book's title and authors using the \textsl{successive()} function. Even though, to parse complex regular expression structured documents, specific functions and the auxiliary mechanisms are required.

\begin{footnotesize}\begin{verbatim}
Q12. declare successive({(doc,loc)}):{(doc,loc)} as
      query #input([($i1, $l1)::nil] ||
                   [($i2,$l2)::$t[($i3,$l3)::*]]||
                   [($i4,$l4)::[($i5,$l5)::*]])
      where $l2<<$l3 par not($l4<<$l5)
      construct {($i1,$l1)}||($i2,$l2)::successive($t)||{($i4,$l4)}
query doc("bib2.xml)//books(/title=>$t;/author=>$a)
where loc($t)<loc($a)
construct books=>{book=>(title=>$t, successive({(author=>$a, loc($a))}))}
\end{verbatim}\end{footnotesize}

\subsection{Function Invocation using Values}
As a functional query language, XTQ adopts a uniform view for the queries and the functions.
However, unlike many other languages like XPath/XQuery which treat queries as special functions, XTQ treats a function as a special query which extracts the output value from the image of the function.
In XTQ, a special virtual value denoted by \textsl{\#func:\{(cdata, \{(*, *)\})\}} is used to contain the images of the functions which are in the form of the tuples ``(fun\_name, \{(input\_value, output\_value)\})''. That means, a function invocation ``\textsl{fun(v)}'' can be represented as the query like ``\textsl{\#func[\{(``fun'', \{(v,\$out)\})\}]}''.

The query style of function invocation is used in the query clause where only the fresh variables are allowed to occur in the extraction patterns.
We adopt this restriction because the bindings from the extraction patterns are often to be filtered by the conditions and thus cannot be used directly as function arguments. Further, using the query style of function invocation can also allow the bindings both from the document extraction patterns and from the function invocations to be used seamlessly in data restructuring.

For example, in the query Q13 which categorizes the book titles by the number of authors who have the email ending with ``edu'', the \textsl{count()} function is invoked by join the input element \textsl{\$in} with the the set \textsl{\{\$a\}}, thus the \textsl{\$out} associated with the \textsl{\$in} would be bound to the number of the set accordingly and be used in the set folding. The function ``\textsl{count(\{\$a\})}'' is invoked actually after the authors of a book are filtered by the condition  ``\textsl{endwith(\$m,`edu')}''. However, if the function were directly invoked in the query clause, such a ``deferred'' invocation cannot be easily presented.

\begin{footnotesize}\begin{verbatim}
Q13.query doc("bib.xml")//book(/author=>$a/email=>$m, /title=>$t),
          #func[{("count",{($in,$out)})}]
    where endwith($m,"edu") with $in = {$a}
    construct book=>{(author_number=>$out%, {title=>$t})}
\end{verbatim}\end{footnotesize}

An important issue about the function invocation query is that the predicate conditions involving the output of the function can only have the scope of the whole output rather than the finer values. For example, for the value extraction pattern of a output value like ``\textsl{\$out[\{\$a\}]}'', the predicate conditions can not use \emph{\$a} as its argument but use a quantified condition like ``\textsl{for \$a in \{\$a\}.c}''. That is because filtering out an element, e.g. a value bound to \emph{\$a}, in the output would make the whole output value invalid and thus also filter out the input value.

%
%
%


Another interesting and useful feature of virtual function value is that it provides something of a ``reflection'' of functions that enables the runtime allocation and invocation of functions, which can even be used to implement higher-order functions, i.e., using functions as arguments. Such a function invocation is encapsulated as an image document fragment like

\begin{footnotesize}\begin{verbatim}
funinvoke=>{|
   funname=>"eq";
   arg=>{|tuple=>{|
            item=>{| funinvoke => {|
                        funname=>"headnode";
                        arg=>{|tuple=>{|item=>{|extern=>"member"|};
                                        item=>{|value=>"year"|}|}|}|}|};
            item=>{| value=>"2012" |} |} |} |}
\end{verbatim}\end{footnotesize}

A function invocation image contains the function name and the image of the argument values passed to the function. As shown above, there are three kinds of argument values in the image: the first one is the ``value'' element which directly encloses the argument value to be passed, e.g.,``\textsl{value$\Rightarrow$`year'}''; the second one is the ``\textsl{extern}'' element which would import a value at runtime, e.g. ``\textsl{extern$\Rightarrow$`member'}''; and the third one is another ``\textsl{funinvoke}'' element, i.e., a nested function invocation. The above image can be generated by certain functions from the function invocation string, e.g., ``\textsl{condimg(`eq(headnode((extern:member),year),2012)')}''.

Under the support of the function invocation image and the auxiliary mechanisms, the two functions \textsl{applyinvoke()}  and \textsl{resolvearg()} are mutually and recursively defined to invoke a function with the specified import context, as shown in Q14. In the \textsl{applyinvoke()} function, the invocation image is joined with the virtual function value by the argument value which is the value result of the \textsl{resolvearg()} function resolving the argument image by recursively enumerating and substituting the elements with the concrete values.

\begin{footnotesize}\begin{verbatim}
Q14.declare applyinvoke({(*,*)}, doc):* as (
      query #input[
            ($import,
             $funinvoke/funinvoke(/funname=>$fn, /arg=>$a))],
            #func[{($fun,{($in, $out)})}]
      where $fn=$fun and $in = toval(resolvearg($import,$a)
      construct {$out} )

    declare resolvearg(({(*,*)}, doc):doc as (
      query #input[
            ($import[{($iname, $ival)}],
             $arg(/$vtag=>$vcont || /~=>$cons || /value=>$argv ||
                  /extern=>$ename || /funinvoke=>$f))]
      where ($vtag != "value" and $vtag!="extern" and $vtag!="funinvoke"
             and $tag!="~")  par $iname=$ename
      construct {$cons | $argv | ($ival hid $ename) |
                 applyinvoke($import, $f) |
                 $vtag=>resolvearg($import, $vcont)}
\end{verbatim}\end{footnotesize}

The \textsl{applyinvoke()} function is  powerful since it can carry out the invocation at runtime. Here we use an example to show its expressiveness.  The \textsl{twindow()} function in Q15 is an simplified emulation of the tumbling window clause in XQuery 3.0. In this function, certain non-overlapped subsequences named \textsl{windows} are extracted from a sequence of items, satisfying the start condition and the end condition which are encoded as a function invocation image. The query program invoking the \textsl{twindow()} function can extract the sequences of the books which are published in 2012. It  passes the \textsl{twindow} function the image of the start condition ``\textsl{eq(headnode((extern:member),year),2012)}'' and the image of the end condition ``\textsl{neq(headnode((extern:member),year), 2012)}''.

\begin{footnotesize}\begin{verbatim}
Q15.declare twindow({*}, doc, doc):{(*,*,*)} as
      query #input[
             ($seq([{$start}],[{$end}]),
              $startcond, $endcond)]
      where loc($start) < loc($end) and
            applyinvoke({("member", $start)}, $startcond) and
            applyinvoke({("member", $end)}, $endcond)
      construct ((subSeq($seq, pos(head({$start})), pos(head({$end}))),
                  $start, $end)::
                 twindow(endSeq($seq,pos($end)),$startcond,$endcond))

query doc("bib.xml")//book=>$b,
      #func[{("twindow",{($in,$out[{($w,$s,$e)}])})}]
where $in = ({$b}, condimg("eq(headnode((extern:member),year),2012)"),
               condimg("neq(headnode((extern:member),year),2012)"))
construct windows=>{window=>(content=>$w,start=>$s,end=>$e)}
\end{verbatim}\end{footnotesize}

The above program shows that the new feature introduced in the forthcoming version of XQuery can be encoded in common function invocations of XTQ. The window feature is very expressive and interesting \cite{window} in querying streaming XML data. It can also be used to process the sequential data elements, e.g., the well known regular expression structured XML data, just by specifying the proper starting and ending conditions and combining and comparing the window fragments. As the full encoding of the window features is quite technical and complicated, we omit the discussion of the details and leave it for further reports.

\section{Conclusion}
Declarativeness is always a goal of designing a query language. Existing XML query languages have tried to achieve it in several ways:  adopting the program form where data extraction, data filtering and data construction are separately presented;  using pattern-based approaches to intuitively present data extraction and data construction;  specifying compositional semantics to be coherent with the syntax structure. Nevertheless, as previously mentioned, there is still a limitation of declarativeness in them to present complex queries such as restructuring data hierarchy and handling heterogeneous data, since they seldom concern presenting and utilizing  hierarchical and disjunctive relationships among data elements.

In this paper, we introduced XTQ, a new pattern-based functional query language, to overcome the limitation. In comparison with existing languages, XTQ has the advantages in four aspects.

Firstly, XTQ  uses expressive patterns to explicitly specify the conjunctive, disjunctive and hierarchical composition of data elements with matching terms, and thus can present complex data operations declaratively. Existing languages, such as TQL, XDuce, Xcerpt and XTreeQuery, may recognize the importance of the disjunctive or hierarchical composition and use certain mechanisms to indicate them, but as far as we know there is no language which deploys an explicit and coherent approach to present and manipulate them as XTQ does. Therefore, XTQ is more expressive in presenting complex queries.

Secondly, XTQ uses a simple but powerful rewriting system for restructuring matching term to coherently present data transformation and deductively infer its process.¡¡Although data transformation is a central task of XML query, existing languages seldom address this issue since they do not present complex data structure in single query but leave building data hierarchy to layered subqueries. Some logic languages like XTreeQuery and Xcerpt can carry out simple transformation through specifying the new relationships between data elements, just like the distribution rules in XTQ, but they lack a systematic mechanism to manipulate nested sets. Further, XTQ also adopts many mechanisms, such as the folding extraction pattern, the folded set term  and the ordering of the set elements, to meet practical data transformation requirements. These mechanisms work together to enable XTQ declaratively present complex data transformation and construction requests.

Thirdly, XTQ adopts a consistent data filtering mechanism on composite data structures.  As hierarchical and disjunctive structure cannot be presented in existing languages, the conditions in these languages only concerns the restrictions on tuples of data elements, and thus filtering a composite structure need to be carried out in several clauses or subqueries. For example, a simple request like the query Q8 cannot be easily specified in a single query. XTQ introduces the connectives \emph{par} and \emph{with} with a consistent mechanism and a sound semantics for filtering disjunctive and hierarchial data structure, which enables data filtering on complex semantic structure be presented structurally and consistently.

Finally, although borrowing the basic features from conventional functional languages, XTQ seamlessly embed them into the pattern mechanisms so as to make the language more suitable for presenting data query than existing functional languages such as XDuce and CDuce. The virtual function document value also enables XTQ to present high-order function in a style consistent to document query.

The core of XTQ has been implemented based on the operational semantics. Now we are improving the prototype with
more efficient mechanisms in combining data extraction with condition evaluation. Further, since label-keyword query is a promising topic in querying large XML documents \cite{xmlkeyword}, we are working to improve XTQ to contain features of presenting label-keyword queries. The flexible structure manipulation in XTQ program is promising in specifying uniform views for
heterogeneous data in XML integration, and the composition and
deduction of the views based on extraction pattern and restructuring
mechanism are practical and interesting problems to study. We would also consider introducing the schema-based patterns into the language, which would make it as expressive as the typed languages like CQL in parsing the schema-based documents.
Additionally, fragmenting and distributing big XML documents becomes
a practical issue in XML data processing. How to adapt XTQ to fit
for manipulating various kinds of distributed document fragments is
also an interesting problem we concern.

\appendix
\section{Proofs of Propositions}

\newdefinition{prf}{Proof of Proposition}

\noindent\textbf{Proof of Proposition 1}. \\
Assume that \textsl{d$|$p$\rightarrow$b} or \textsl{e$|$p$\rightarrow$b}, we prove the proposition by induction on the structure of \textsl{p}. \\
1. If \textsl{p} is an atomic pattern, It is trivial to prove that the hypothesis holds.\\
2. If \textsl{p} =\textsl{/p'} or \textsl{//p'} or \textsl{@p'}, then\\
\indent 1) if \textsl{d} is a singleton set, i.e., \textsl{d}=\textsl{\{e'\}}, and \textsl{e'$|$p'$\rightarrow$b'}, then \textsl{b':mt(p')}, and thus \textsl{b=\{b'\}:\{mt(p')\}$_{mt(p')}$=mt(p)} and the hypothesis holds;\\
\indent 2) if \textsl{d} = \textsl{\{e$_{i}$$|$i$\in$\{1..n\}\}}, then following the \textsl{ep-comp} rule, either \textsl{b} = \textsl{\{b$_{j}$$|$j$\in$\{1..m\}\}} (\textsl{b$_{j}$ (j$\in$\{1..m\}}) is not null and is distinct in the set, or \textsl{b} = \textsl{\{b'\}} where \textsl{b'} is null. For the both cases,
\textsl{b:\{mt(p')\}$_{mt(p')}$=mt(p)}.\\
3. If \textsl{p} = \textsl{/p$_{1}$(p$_{2}$)}, then \\
\indent 1) if \textsl{d} is a singleton set, i.e., \textsl{d} = \textsl{\{e'\}}, and \textsl{e'$|$p$_{1}$$\rightarrow$b$_{1}$}, then \textsl{b$_{1}$:mt(p$_{1}$)};
\indent \indent i) if \textsl{b$_{1}$} is not null, i.e., \textsl{e'} = \textsl{n$\Rightarrow_{l}$d'} matches \textsl{p$_{1}$}, then according to the hypothesis if \textsl{d'$|$p$_{2}$$\rightarrow$b$_{2}$} then \textsl{b$_{2}$:mt(p$_{2}$)}, and thus \textsl{b:\{(mt(p$_{1}$),mt(p$_{2}$))\}=mt(p)} and the hypothesis holds;\\
\indent \indent ii) if \textsl{b$_{1}$} is null, then \textsl{b$_{1}$:(mt(p$_{1}$),mt(p$_{2}$))} as the part 6) in the Definition 3 indicates,  and thus \textsl{b:mt(p)} and the hypothesis holds;\\
\indent 2) if \textsl{d} = \textsl{\{e$_{i}$$|$i$\in$\{1..n\}\}}, then following the \textsl{tree-comp} rule and 2.2) it is easy to know the hypothesis holds.\\
4. If \textsl{p} = \textsl{//p$_{1}$(p$_{2}$)}, similar to 3. We only need to consider the singleton set case, i.e., \textsl{d=\{e'\}}. By induction on the structure of \textsl{e}, it is easy to prove that the hypothesis holds and so does for the case of the multi-element set.\\
5. If \textsl{p} = \textsl{/p$_{1}$;p$_{2}$}, it is easy to know from the \textsl{fold-comp} rule that the hypothesis holds for when the bindings are not null, and if the binding is null then \textsl{b:\{mt(p$_{1}$)\}}$_{mt(p_{1})}$ and thus \textsl{b:\{(mt(p$_{1}$),mt(p$_{2}$))\}}$_{mt(p_{1})}$ and the hypothesis holds.\\
6. If \textsl{p} = \textsl{(p$_{1}$,p$_{2}$)} or \textsl{(p$_{1}$$||$p$_{2}$)}, it is trivial to prove that the hypothesis holds.\\

This completes the induction. $\Box$\\

\noindent\textbf{Proof of Proposition 2}. \\
It is trivial by induction on the rules involving set restructuring.$\Box$\\

\newtheorem{lemma}{Lemma}


\begin{definition}
1. For an index term \textsl{r}, \textsl{r} is \textsl{in normal form} (or \textsl{is normal} in short) iff:

1) \textsl{r} = \textsl{x};

2) \textsl{r} = \textsl{r$_{1}$ $|$ \dots $|$ r$_{n}$} (called \textsl{normal option}) where each \textsl{r$_{i}$} (\textsl{i$\in$\{1..n\}}) is in normal form, or

3) \textsl{r} = \textsl{((r$_{1}$), r$_{2}$)} (called \textsl{normal tuple}) where \textsl{r$_{1}$} is in normal form and \textsl{r$_{2}$} is an \textsl{index unit} which is either a variable or a normal option.\\
2. For an index term \textsl{r}, the \textsl{normalization} of \textsl{r} denoted by \textsl{$\overline{r}$} is defined as :

1) if \textsl{r} = \textsl{x}, then \textsl{$\overline{r}$} = \textsl{x};

2) if \textsl{r} = \textsl{r$_{1}$ $|$ \dots $|$ r$_{n}$}, then \textsl{$\overline{r}$} = \textsl{$\overline{r_{1}}$ $|$ \dots $|$ $\overline{r_{n}}$};

3) if \textsl{r} = \textsl{(r$_{1}$, r$_{2}$)}, then \textsl{$\overline{r}$} = \textsl{$\overline{r_{1}}\circ\overline{r_{2}}$} where \textsl{$\overline{r_{1}}\circ\overline{r_{2}}$} is defined as \textsl{(\dots(($\overline{r_{1}}$,u$_{1}$), u$_{2}$),\dots,u$_{m}$)} if \textsl{$\overline{r_{2}}$} = \textsl{(\dots(u$_{1}$, u$_{2}$),\dots,u$_{m}$)}.
\end{definition}

\begin{lemma}
For the terms \textsl{t}, \textsl{t'} and \textsl{r} where each set term \textsl{\{u\}$_{v}$$\leqslant$t} has the normal index term \textsl{v} and \textsl{t'} contains no content term,  \textsl{t$\stackrel{*}{\leadsto}$\{t'\}$_{r}$} iff \textsl{t$\stackrel{*}{\leadsto}$\{t'\}$_{\overline{r}}$}.
\end{lemma}
\noindent\textbf{Proof.} $\Rightarrow$. \\
The lemma can be proved by the induction on the structure of \textsl{r}.\\
1.  \textsl{r} = \textsl{x}. Trivial.\\
2. \textsl{r} = \textsl{(r$_{1}$,r$_{2}$)}. Then there are the terms \textsl{t$_{1}$} and \textsl{t$_{2}$} that \textsl{t $\stackrel{*}{\leadsto}$ \{t$_{1}$\}$_{r_{1}}$ $\stackrel{*}{\leadsto}$ \{\{t$_{2}$\}$_{r_{2}}$\}$_{r_{1}}$ $\leadsto$ \{t$_{2}$\}$_{(r_{1},r_{2})}$ $\stackrel{*}{\leadsto}$ \{t'\}$_{r}$}. By the hypothesis, \textsl{t $\stackrel{*}{\leadsto}$ \{t$_{1}$\}$_{\overline{r_{1}}}$} and  \textsl{t$_{1}$ $\stackrel{*}{\leadsto}$ \{t$_{2}$\}$_{\overline{r_{2}}}$}, and thus \textsl{t $\stackrel{*}{\leadsto}$ \{\{t$_{2}$\}$_{\overline{r_{2}}}$\}$_{\overline{r_{1}}}$ $\leadsto$ \{t$_{2}$\}$_{(\overline{r_{1}},\overline{r_{2}})}$}. Assume that \textsl{$\overline{r_{2}}$} = \textsl{(...(u$_{1}$,u2),...,u$_{m}$)}, obviously \textsl{t$\stackrel{*}{\leadsto}$\{\{t$_{2}$\}$_{\overline{r_{2}}}$\}$_{\overline{r_{1}}}$} iff \textsl{t$\stackrel{*}{\leadsto}$\{\{...\{t$_{2}$\}$_{u_{m}}...$\}$_{u_{1}}$\}$_{\overline{r_{1}}}$$\stackrel{*}{\leadsto}$ \{t$_{2}$\}$_{\overline{r_{1}}\circ\overline{r_{2}}}$}. The hypothesis holds.\\
3. \textsl{r} = \textsl{r$_{1}$ $|$ ... $|$ r$_{n}$}. Trivial.\\
This completes the induction.\\
$\Leftarrow$. By induction on the structure of \textsl{r}, and it is easy to prove the hypothesis since the above induction steps are reversible. \\
$\Box$

\begin{lemma}
For the \textsl{t} where each set term \textsl{\{u\}$_{v}$$\leqslant$t} has the normal index term \textsl{v}, and the term \textsl{r} which is not a content term, there exists the term \textsl{t$_{0}$} that if \textsl{t $\stackrel{*}{\leadsto}$ \{t$_{0}$\}$_{\overline{r}}$}, then for any \textsl{t'} that contains no content term and \textsl{t $\stackrel{*}{\leadsto}$\{t'\}$_{r}$, t$_{0}$ $\stackrel{*}{\leadsto}$ t'} holds.
\end{lemma}
\noindent\textbf{Proof.}  We prove the lemma by the induction on the structure of \textsl{r}.\\
1. \textsl{r} = \textsl{x}. It is easy to know that \textsl{t $\equiv$ (t$_{1}$,t$_{2}$)} (where t$_{2}$ can be $\epsilon$) that either \textsl{t$_{1}$=\{t$_{1}$'\}$_{x}$} or \textsl{t$_{1}$=o$_{1}$$||$...$||$o$_{m}$} or \textsl{o$_{1}$$|$...$|$o$_{m}$} where \textsl{o$_{i}$$\stackrel{*}{\leadsto}$\{o$_{i}$'\}$_{x}$(i$\in$\{1..m\})}. In the former case, let \textsl{t$_{0}$=(t$_{1}$',t$_{2}$)}, and for any \textsl{t'} that \textsl{t$\stackrel{*}{\leadsto}$\{t'\}$_{x}$}, \textsl{t$_{0}$$\stackrel{*}{\leadsto}$t'} holds. In the latter case, by the hypothesis, there is \textsl{o$_{i}$$\stackrel{*}{\leadsto}$\{u$_{i}$\}$_{x}$} that \textsl{u$_{i}$$\stackrel{*}{\leadsto}$o$_{i}$'}, and similarly \textsl{t$_{0}$=((u$_{1}$$|$...$|$u$_{m}$),t$_{2}$)} is the term required.\\
2. \textsl{r} = \textsl{r$_{1}$ $|$ ... $|$ r$_{n}$}. Then \textsl{t $\equiv$ (t$_{1}$,t$_{2}$)} that \textsl{t$_{1}$=\{t$_{1}$'\}$_{r}$} or
\textsl{t$_{1}$ $\equiv$ o$_{1}$ $||$ \dots $||$ o$_{n}$} that \textsl{o$_{i}$$\stackrel{*}{\leadsto}$\{o$_{i}$'\}$_{r_{i}}$} and \textsl{t'$\equiv$ (o$_{1}$'$|$\dots$|$o$_{n}$', t$_{2}$}. In the former case, let \textsl{t$_{0}$=(t$_{1}$', t$_{2}$)}. For the latter case, by the hypothesis, \textsl{o$_{i}$$\stackrel{*}{\leadsto}$\{u$_{i}$\}$_{r_{i}}$} and \textsl{u$_{i}$$\stackrel{*}{\leadsto}$t$_{i}$'}, thus \textsl{t$_{0}$ = (u$_{1}$$|$\dots$|$u$_{n}$, t$_{2}$)}.\\
3. \textsl{$\overline{r}$} = \textsl{u$\circ$r'} where \textsl{u} is variable or normal option. Then from the above two cases there is \textsl{t$_{1}$} that \textsl{t$\stackrel{*}{\leadsto}$\{t$_{1}$\}$_{u}$} and \textsl{t$_{1}$$\stackrel{*}{\leadsto}$\{t'\}$_{r'}$}, by the induction hypothesis and Lemma 1 the \textsl{t$_{0}$} exists.\\
This completes the induction.$\Box$\\

\begin{lemma}
\end{lemma}
\noindent\textbf{Proof of Proposition 3}.\\
We prove a variant of the proposition as follows:\\
For two matching terms \textsl{t} and \textsl{t'} where each set term \{u\}$_{v}$ $\leqslant$t has the normal index term v, whether \textsl{t$\stackrel{*}{\leadsto}$t'} is decidable.\\
We firstly prove the proposition by restricting t' to contain no content term.\\
The proposition can be proved by the induction on the structure of \textsl{t} and \textsl{t'}. We use \textsl{x} to denote the variables, use \textsl{t}, \textsl{r}, \textsl{o} and \textsl{u} to denote terms.\\
1. \textsl{t} = \textsl{x}. Trivial. \\
2. \textsl{t} = \textsl{\{t$_{1}$\}$_{r_{1}}$}. Here \textsl{t'} can only be a set or a tuple.\\
\indent 1) if \textsl{t'}=\textsl{\{t$_{1}$'\}$_{r_{1}'}$},  from Lemma 1 we know \textsl{t$\stackrel{*}{\leadsto}$t'} iff \textsl{\{t$_{1}$\}$_{r_{1}}$$\stackrel{*}{\leadsto}$\{t$_{1}$'\}$_{\overline{r_{1}'}}$}. Since \textsl{r$_{1}$} is normal, it is easy to know that \textsl{$\overline{r_{1}'}$=r$_{1}$$\circ$ r$_{2}$'}, then \textsl{\{t$_{1}$\}$_{r_{1}}$$\stackrel{*}{\leadsto}$\{t$_{1}$'\}$_{\overline{r_{1}'}}$} iff \textsl{t$_{1}$$\stackrel{*}{\leadsto}$\{t$_{1}$'\}$_{r_{2}'}$} which is decidable from the induction hypothesis.\\
\indent 2) if \textsl{t'=(t$_{1}$,\dots, t$_{n}$)}, from the restructuring rules it is easy to know that there must be a tpl-dupl rule being applied to \textsl{t} and thus \textsl{t$\stackrel{*}{\leadsto}$t'} iff \textsl{t$\stackrel{*}{\leadsto}$t$_{i}$} (\textsl{i$\in$\{1..n\}}) which is decidable from the induction hypothesis.\\
3. \textsl{t} = (t$_{1}$,\dots, t$_{n}$). \\
\indent 1) If \textsl{t'} = \textsl{x} or \textsl{t'=t$_{1}$'$||$\dots$||$t$_{m}$'}, then \textsl{t$\stackrel{*}{\leadsto}$t'} doesn't hold.\\
\indent 2) If \textsl{t' = \{t$_{a}$\}$_{r}$}, then from Lemma 2 there is \textsl{t$_{0}$ that t$\stackrel{*}{\leadsto}$\{t$_{0}$\}$_{\overline{r}}$} and \textsl{t$_{0}$$\stackrel{*}{\leadsto}$t$_{a}$}, Lemma 2 gives the approach of inferring \textsl{t$_{0}$} from \textsl{t} and \textsl{r}, and \textsl{t$_{0}$$\stackrel{*}{\leadsto}$t$_{a}$} is decidable from the induction hypothesis.\\
\indent 3) If t' = (t$_{1}$', \dots, t$_{m}$'), then by induction on m it is easy to know t $\equiv$ (t$_{1}$*, \dots, t$_{m}$*) that t$_{i}$* has the same variables as t$_{i}$' (i$\in$\{1..m\}). It is also easy to know that t$\stackrel{*}{\leadsto}$t' iff t$_{i}$*$\stackrel{*}{\leadsto}$t$_{i}$'(i$\in$\{1..m\}) which are decidable from the induction hypothesis.\\
4. \textsl{t} = \textsl{t$_{1}$ $||$ \dots $||$ t$_{n}$}. Here \textsl{t'} can only be a d-union or a set.\\
\indent 1) If \textsl{t'} = \textsl{t$_{1}$'$||$\dots$||$t$_{m}$'}, then by induction on \textsl{m} \textsl{t$\equiv$t$_{1}$*$||$\dots$||$t$_{m}$*} that \textsl{t$_{i}$*} has the same variables as \textsl{t$_{i}$' (i$\in$\{1..m\})}. It is also easy to know that \textsl{t$\stackrel{*}{\leadsto}$t'} iff \textsl{t$_{i}$*$\stackrel{*}{\leadsto}$t$_{i}$'(i$\in$\{1..m\})} which are decidable from the hypothesis.\\
\indent 2) If \textsl{t'} = \textsl{\{t$_{1}$'\}$_{r}$},  then from Lemma 2 there is \textsl{t$_{0}$} that \textsl{t$\stackrel{*}{\leadsto}$\{t$_{0}$\}$_{\overline{r}}$ }and \textsl{t$_{0}$$\stackrel{*}{\leadsto}$t$_{1}$'} which is decidable from the induction hypothesis.\\
5. \textsl{t} = \textsl{t$_{1}$ $|$ \dots $|$ t$_{n}$}. Here \textsl{t'} can be an option, a tuple or a set.\\
\indent 1) If \textsl{t'} = \textsl{t$_{1}$' $|$ \dots $|$ t$_{m}$'}, then by induction on \textsl{m} \textsl{t$\equiv$t$_{1}$*$|$\dots$|$t$_{m}$*} that \textsl{t$_{i}$*} has the same variables as \textsl{t$_{i}$'} (\textsl{i$\in$\{1..m\}}). It is also easy to know that \textsl{t$\stackrel{*}{\leadsto}$t'} iff t$_{i}$*$\stackrel{*}{\leadsto}$t$_{i}$'(\textsl{i$\in$\{1..m\}}) which are decidable from the induction hypothesis.\\
\indent 2) If \textsl{t'} = \textsl{\{t$_{1}$'\}$_{r}$}, then from Lemma 3 there is \textsl{t$_{0}$} that \textsl{t$\stackrel{*}{\leadsto}$\{t$_{0}$\}$_{\overline{r}}$} and \textsl{t$_{0}$$\stackrel{*}{\leadsto}$t$_{1}$'} which is decidable from the induction hypothesis.\\
This completes the induction.

When \textsl{t'} contains content term, it is easy to know that there is \textsl{t''} which contains no content term that \textsl{t$\stackrel{*}{\leadsto}$t'} iff \textsl{t$\stackrel{*}{\leadsto}$t''}. Therefore Proposition 3 holds.
$\Box$\\


\begin{definition}
For the bindings \textsl{b$_{1}$}, \textsl{b$_{2}$} and \textsl{b} that \textsl{b$_{1}$$\unlhd$b} and \textsl{b$_{2}$$\unlhd$b}, \textsl{b$_{1}$} and \textsl{b$_{2}$} are \textbf{conjunctively related} in \textsl{b}, denoted as \textsl{b$_{1}$*b$_{2}$}, if there are two bindings \textsl{b$_{1}$'} and \textsl{b$_{2}$'} that \textsl{b$_{1}$$\unlhd$b$_{1}$'}, \textsl{b$_{2}$$\unlhd$b$_{2}$'} and \textsl{(b$_{1}$',b$_{2}$')$\leqslant$b}.
\end{definition}


\begin{lemma}
For the bindings and terms \textsl{b:t}, \textsl{b':t'}, \textsl{b$_{1}$:t$_{1}$} and \textsl{b$_{2}$:t$_{2}$} that \textsl{b$\stackrel{*}{\leadsto}$b'},  \textsl{b$_{1}$$\leqslant$b'},  \textsl{b$_{2}$$\leqslant$b'} and \textsl{(t$_{1}$,t$_{2}$)$\leqslant$t'}, if \textsl{b$_{1}$*b$_{2}$} in \textsl{b}, then \textsl{(b$_{1}$,b$_{2}$)$\leqslant$b'}.
\end{lemma}
Proof. Assume that there are the bindings b$_{1}$':t$_{1}$' and b$_{2}$':t$_{2}$' that b$_{1}$$\unlhd$b$_{1}$', b$_{2}$$\unlhd$b$_{2}$' and (b$_{1}$',b$_{2}$')$\leqslant$b. It is easy to know by induction on the restructuring rules that (t$_{1}$,t$_{2}$)$\unlhd$(t$_{1}$',t$_{2}$'), and there must be a restructuring route u$_{0}$= (t$_{1}$',t$_{2}$')$\stackrel{1}{\leadsto}$u$_{1}$$\stackrel{1}{\leadsto}$\dots$\stackrel{1}{\leadsto}$u$_{n}$ where (t$_{1}$,t$_{2}$)$\leqslant$u$_{n}$. We prove the hypothesis by induction on n. If n=0, then (t$_{1}$,t$_{2}$)$\leqslant$(t$_{1}$',t$_{2}$')$\leqslant$t, and thus (b$_{1}$,b$_{2}$)$\leqslant$b' since (b$_{1}$,b$_{2}$) would be invariant during restructuring. Assume that when n$\leqslant$k the hypothesis holds. If n=k+1, then there should be b'':t'' that b$\stackrel{*}{\leadsto}$b''$\stackrel{*}{\leadsto}$b', u$_{k}$$\leqslant$t''. It is easy to prove that there should be b$_{1}$'':t$_{1}$'' and b$_{2}$'':t$_{2}$'' that (t$_{1}$'',t$_{2}$'')$\leqslant$u$_{k}$, b$_{1}$''$\unlhd$b$_{1}$', b$_{2}$''$\unlhd$b$_{2}$',  t$_{1}$$\leqslant$t$_{1}$'' and t$_{2}$$\leqslant$t$_{2}$''. According to the induction assumption, (b$_{1}$'',b$_{2}$'')$\leqslant$b'' holds. The hypothesis can be proved trivially by induction on the restructuring rules applying to (t$_{1}$'',t$_{2}$'')$\stackrel{1}{\leadsto}$u$_{k+1}$' where (t$_{1}$,t$_{2}$)$\leqslant$u$_{k+1}$'. $\Box$\\

\noindent\textbf{Proof of Proposition 4.}\\
It is enough to prove that for each \textsl{b:t} and \textsl{t$\leadsto$t'}, there is \textsl{b':t'} that \textsl{b$\curvearrowright$b'}.
The hypothesis is proved by the induction on the restructuring rules. The discussion on the rules incurring no set transformation is trivial, so we only list the induction on the following rules involving set transformation.\\
1. set-distr. If \textsl{t=(t$_{1}$, \{t$_{2}$\}$_{r}$)}, \textsl{t'=\{(t$_{1}$, t$_{2}$)\}$_{r}$} and \textsl{b:t=(b$_{0}$,\{b$_{i}$$|$ i$\in$\{1..n\}\})}, let \textsl{b'=\{(b$_{0}$,b$_{i}$)$|$ i$\in$\{1..n\}\}}. It is easy to see that \textsl{(b$_{0}$,b$_{i}$):(t$_{1}$, t$_{2}$)}. Since for each \textsl{b$_{i}$} there is the index sub-binding \textsl{b$_{r_{i}}$:r$\unlhd$b$_{i}$}, it is easy to know that \textsl{b$_{r_{i}}$} is also the sub-binding of \textsl{(b$_{0}$,b$_{i}$)} in \textsl{b'}. Thus \textsl{b':t'}.\\
2. set-flatten. If \textsl{t=\{\{t$_{0}$\}$_{r_{1}}$\}$_{r_{2}}$}, \textsl{t'=\{t$_{0}$\}$_{(r_{2},r_{1})}$} and \textsl{b:t=\{c$_{i}$=\{b$_{i,j}$ $|$ j$\in$\{1..n$_{i}$\}\} $|$ i$\in$\{1..n\}\}}, let \textsl{b'=\{b$_{i,j}$ $|$ i$\in$\{1..n\}, j$\in$\{1..n$_{i}$\}\}}. For each \textsl{b$_{i,j}$} there is the index sub-binding \textsl{b'$_{i,j}$:r$_{2}$$\unlhd$b$_{i,j}$}, and for each \textsl{c$_{i}$} there is also a sub-binding \textsl{c'$_{i}$:r$_{1}$$\unlhd$c$_{i}$}  and obviously \textsl{c'$_{i}$$\unlhd$b$_{i,j}$}. By induction on the transformation steps it is easy to know that \textsl{c'$_{i}$*b'$_{i,j}$} in b$_{i,j}$, therefore, \textsl{(c'$_{i}$,b'$_{i,j}$):(r$_{2}$,r$_{1}$)} is unique for \textsl{b$_{i,j}$} in \textsl{b'} and thus is the index sub-binding. Thus \textsl{b':t'}.\\
3. set-merge. It is trivial.\\
This completes the induction. $\Box$\\

\noindent\textbf{Proof of Proposition 5.}\\
We prove that \textsl{b$_{0}$$\leqslant$b' $\Leftrightarrow$ b$_{0}$$\leqslant$b''} for any \textsl{b$_{0}$:t$_{0}$},  by induction on the structure of \textsl{t$_{0}$}.\\
1. \textsl{t$_{0}$$\leqslant$t}. We prove that \textsl{b$_{0}$$\leqslant$b'} iff \textsl{b$_{0}$$\leqslant$b} by induction on the transformation steps. It is trivial to know that the hypothesis holds for \textsl{b$\stackrel{1}{\curvearrowright}$b'} by induction on the transformation rules. Assume that the hypothesis holds for \textsl{b$\stackrel{1}{\curvearrowright}$b$_{1}$$\stackrel{1}{\curvearrowright}$\dots b$\stackrel{1}{\curvearrowright}$b$_{k}$}, for \textsl{b$_{k}$$\stackrel{1}{\curvearrowright}$b$_{k+1}$}, the hypothesis also holds by induction on the transformation rules. Therefore, \textsl{b$_{0}$$\leqslant$b' $\Leftrightarrow$ b$_{0}$$\leqslant$b''}.\\
2.  \textsl{t$_{0}$=\{t$_{1}$\}$_{\{r_{1}\}}$}. According to the assumption, for any \textsl{b$_{1}$:t$_{1}$}, \textsl{b$_{1}$$\leqslant$b' $\Leftrightarrow$ b$_{1}$$\leqslant$b''}. If \textsl{t'=t$_{0}$}, there is only one set binding of \textsl{t$_{0}$} and it is easy to know that the hypothesis holds. Otherwise, assume that \textsl{b$_{0}$$\leqslant$b'}, it is easy to know that \textsl{t$\stackrel{*}{\leadsto}$t'$\stackrel{*}{\leadsto}$\{(t$_{0}$,u$_{0}$)\}$_{r0}$} where \textsl{u$_{0}$} contains no set terms, and \textsl{b'$\curvearrowright$b'$_{0}$} and \textsl{b''$\curvearrowright$b''$_{0}$} where the bindings of \textsl{t$_{0}$}, e.g., \textsl{b$_{0}$}, are not transformed. As the definition of set binding shows, in \textsl{b$_{0}$'} and \textsl{b$_{0}$''} each \textsl{(b$_{n}$,c$_{n}$):(t$_{0}$,u$_{0}$)} has the unique index binding \textsl{d$_{n}$:r$_{0}$} . According to the induction assumption, for the tuple \textsl{(b$_{0}$,c$_{0}$)$\in$b$_{0}$'} which has the index binding \textsl{d$_{0}$}, \textsl{d$_{0}$$\leqslant$b'$_{0}$$\Leftrightarrow$d$_{0}$$\leqslant$b''$_{0}$}. Assume that \textsl{b$_{0}$=\{b$_{11}$,\dots,b$_{1n}$\}} where \textsl{b$_{1i}$:t$_{1}$(i$\in$\{1..n\})}, \textsl{d$_{0}$*b$_{1i}$} in \textsl{b'$_{0}$} and thus \textsl{d$_{0}$*b$_{1i}$} in \textsl{b''$_{0}$}, since \textsl{d$_{0}$} is also the index binding of a tuple \textsl{(b$_{m}$,c$_{m}$)} in \textsl{b''}, and thus\textsl{ b$_{1i}$$\in$b$_{m}$} (\textsl{i$\in$\{1..n\}}), that is \textsl{b$_{0}$$\subseteq$b$_{m}$}, and similarly \textsl{b$_{m}$$\subseteq$b$_{0}$} in the reverse way. That is, \textsl{b$_{0}$=b$_{m}$$\leqslant$b$_{0}$''}. The hypothesis holds.\\
3. \textsl{t$_{0}$=(t$_{1}$,\dots,t$_{n}$)}. Assume that \textsl{b$_{0}$=(b$_{1}$,\dots,b$_{n}$)} where \textsl{b$_{1}$:t$_{1}$,\dots, b$_{n}$:t$_{n}$}. According to the induction assumption, \textsl{b$_{1}$$\leqslant$b''},\dots, and \textsl{b$_{n}$$\leqslant$b''}. According to the Lemma, it is easy to prove \textsl{(b$_{1}$,\dots,b$_{n}$)$\leqslant$b''} by induction on \textsl{n}.\\
4. \textsl{t$_{0}$=t$_{1}$$||$\dots$||$t$_{n}$}.  If \textsl{b$_{0}$=b$_{1}$$||$\dots$||$b$_{n}$}, then the hypothesis holds since there is no restructuring rule except set-merge applying to an enumeration.\\
5. \textsl{t$_{0}$=(t$_{1}$$|$\dots$|$t$_{n}$)}. According to the transformation rules, \textsl{b$_{0}$=b$_{1}$} where \textsl{b$_{1}$:t$_{1}$},\dots, or \textsl{b$_{0}$=b$_{n}$} where \textsl{b$_{n}$:t$_{n}$}, and according to the induction assumption,  \textsl{b$_{1}$$\leqslant$b''},\dots, or \textsl{b$_{n}$$\leqslant$b''}. It is easy to know that in applying the merging rules, any \textsl{b$_{i}$:t$_{i}$}  should be in \textsl{b$_{0}$':t$_{0}$} in \textsl{b''}, therefore \textsl{b$_{0}$$\leqslant$b''}.\\
6. \textsl{t$_{0}$= \{(\{(t$_{2}$)\}$_{\{r_{2}\}}$,t$_{1}$)\%\}$_{t_{1}\%}$}.
Assume that for \textsl{b'} the transformation follows the restructuring route \textsl{t$_{a}$=\{(t$_{1}$,t$_{2_{a}}$)\}$_{r_{a}}$$\leadsto$t$_{a}$'=\{(\{(t$_{1}$,t$_{2_{a}}$)\}$_{r_{1}}$,t$_{1}$\%)\}$_{t_{1}\%}$$\stackrel{*}{\leadsto}$t$_{0}$}.
and for \textsl{b''} the route is \textsl{t$_{b}$=\{(t$_{1}$,t$_{2_{b}}$)\}$_{r_{b}}$$\leadsto$t$_{b}$'=\{(\{(t$_{1}$,t$_{2_{b}}$)\}$_{r_{1}}$,t$_{1}$\%)\}$_{t_{1}\%}$$\stackrel{*}{\leadsto}$t$_{0}$}.
Thus \textsl{\{(t$_{1}$,t$_{2 a}$)\}$_{ra}$$\stackrel{*}{\leadsto}$\{(t$_{2}$)\}$_{\{r_{2}\}}$} and \textsl{\{(t$_{1}$,t$_{2_{b}}$)\}$_{r_{b}}$$\stackrel{*}{\leadsto}$\{(t$_{2}$)\}$_{\{r_{2}\}}$}, and each \textsl{b$_{a}$:(t$_{1}$,t$_{2_{a}}$)} or \textsl{b$_{b}$:(t$_{1}$,t$_{2_{b}}$)} in the set binding \textsl{b$_{s_{a}}$:\{(t$_{1}$,t$_{2_{a}}$)\}$_{r_{a}}$} or \textsl{b$_{s_{b}}$:\{(t$_{1}$,t$_{2_{b}}$)\}$_{r_{b}}$} corresponds to a binding of \textsl{t$_{1}$}. If we alter the restructuring routes as \textsl{t$_{a}$$\stackrel{*}{\leadsto}$\{(t$_{2}$)\}$_{\{r_{2}\}}$} and \textsl{t$_{b}$$\stackrel{*}{\leadsto}$\{(t$_{2}$)\}$_{\{r_{2}\}}$} bypassing the set-folding step, each set binding \textsl{b$_{s_{a}}$} or \textsl{b$_{s_{b}}$} would be transformed to the binding \textsl{b$_{s_{a}}$':\{(t$_{2}$)\}$_{\{r_{2}\}}$ and bsb':\{(t$_{2}$)\}$_{\{r_{2}\}}$}, and in the transformation each tuple \textsl{(t$_{1}$,t$_{2_{a}}$)} or (t$_{1}$,t$_{2_{b}}$) would be transformed to one or more bindings of \textsl{t$_{2}$}. Assume that the binding \textsl{b'} and \textsl{b''} are accordingly altered to \textsl{c'} and \textsl{c''}, By the induction assumption, each binding \textsl{b$_{s}$:\{(t$_{2}$)\}$_{\{r_{2}\}}$$\leqslant$c'$\Leftrightarrow$b$_{s}$$\leqslant$c''} and \textsl{b$_{s}$} can either be transformed from a certain \textsl{b$_{s_{a}}$:\{(t$_{1}$,t$_{2_{a}}$)\}$_{r_{a}}$} or a certain \textsl{b$_{s_{b}}$:\{(t$_{1}$,t$_{2_{b}}$)\}$_{r_{b}}$}. In the transformation of the step \textsl{t$_{a}$$\leadsto$t$_{a}$'} or \textsl{t$_{b}$$\leadsto$t$_{b}$'}, the elements in \textsl{b$_{s_{a}}$} or \textsl{b$_{s_{b}}$} would be classified by the content term \textsl{t$_{1}$\%}, and then be transformed to\textsl{ b$_{s}$':t$_{0}$} which actually corresponds to the result of classifying the elements of \textsl{b$_{s}$} by the content term \textsl{t$_{1}$\%}. According to the transformation rule for set-folding, classifying the elements of \textsl{b$_{s}$} by the content term \textsl{t$_{1}$\%} only depends on the variable bindings in bs involving the variables in \textsl{t$_{1}$}, that is, \textsl{b$_{s_{a}}$} or \textsl{b$_{s_{b}}$} would lead to the same binding \textsl{b$_{s}$'}=\textsl{b$_{0}$}. The hypothesis holds.\\
This completes the induction.\\
Therefore, the hypothesis holds for any \textsl{b$_{0}$}, especially \textsl{b$_{0}$=b'} then \textsl{b$_{0}$=b''}. That is, \textsl{b'=b''}. $\Box$\\

\noindent\textbf{Proof of Proposition 6}\\
\noindent It is trivial to prove the proposition by induction on the structure of r and the steps of restructuring involving application of set-distr rules.\\

\noindent\textbf{Proof of Proposition 7}\\
We prove it by induction on the structure of the condition \textsl{c}.\\
1. \textsl{c} is predicate. According to the requirements listed in i),ii) and iv), it is easy to know that for each \textsl{u$_{i}$(i$\in$\{1..n\})}, there exists \textsl{u'} that \textsl{u$_{i}$$||$u'$\leqslant$t}. That is, each \textsl{u$_{i}$} is a branch of a d-union term in \textsl{t}. According to iii) the \textsl{u$_{i}$(i$\in$\{1..n\})} cover all the disjunctive branches of \textsl{t}. Therefore, \textsl{comp(c,t)} is parallelly complete  with respect to \textsl{t}.\\
2. \textsl{c} = \textsl{c$_{1}$ par c$_{2}$}. The discussion is the same as in 1.\\
3. \textsl{c} = \textsl{c$_{1}$ with c$_{2}$}. Similar to 1,  \textsl{comp(c,t)} is parallelly complete with respect to \textsl{t}. Then we only need to prove the consistency. Since the each condition in \textsl{comp(c$_{1}$,t$_{1}$)}, i.e., \textsl{c$_{1}$} or \textsl{any(u)} has the scope \textsl{r$\leqslant$t$_{1}$} or \textsl{r$\lessdot$t$_{1}$}, that is, \textsl{comp(c$_{1}$,t$_{1}$)} also has the scope \textsl{t$_{1}$},  and thus \textsl{comp(c,t)} has the same scope as \textsl{comp(c$_{2}$,t)} and is consistent to \textsl{t}. \\\
This completes the induction. $\Box$.\\

\noindent\textbf{Proof of Proposition 8}\\
We prove the proposition by induction on the structure of \textsl{b}.\\
1. \textsl{b} is a variable binding. Trivial.\\
2. \textsl{b} = \textsl{(b$_{1}$,\dots, b$_{n}$)}. If \textsl{b$_{u}$$\leqslant$b$_{i}$}, then by the induction assumption, \textsl{$\overline{b_{u}}$$\unlhd$$\overline{b_{i}}$$\unlhd$$\overline{b}$}. Otherwise, \textsl{b$_{u}$=(b$_{1}$,\dots,b$_{m}$)}, and \textsl{$\overline{b_{u}}$$\equiv$($\overline{b_{1}}$,\dots,$\overline{b_{m}}$) $\unlhd$$\overline{b}$}.\\
3. \textsl{b} = \textsl{\{b$_{i}$ $|$ i$\in$\{1..n\}\}}. If \textsl{b$_{u}$$\leqslant$b$_{i}$}, then by the induction assumption the hypothesis holds. Otherwise \textsl{b$_{u}$}=\textsl{b} and the hypothesis holds.\\
4. \textsl{b} = \textsl{b$_{1}$$||$\dots$||$b$_{n}$}. Similar to 3.\\
This completes the induction. $\Box$\\

\begin{lemma}
1. For a binding \textsl{b} and a restriction set \textsl{s:\{r\}} that \textsl{b[s]$\rightarrow$b'}, if \textsl{$\overline{b}$$\equiv$(b$_{u}$:u, b$_{v}$)} and \textsl{u$<_{u}$r},
then for any \textsl{b$_{u1}$:u$_{1}$$\leqslant$b'} where \textsl{$\overline{u_{1}}$$\unlhd$u}, \textsl{$\overline{b_{u1}}$$\unlhd$b$_{u}$} holds.

2. For the bindings \textsl{b:t}, \textsl{b':t} and the restriction set \textsl{s:\{r\}} that \textsl{b[s]$\rightarrow$b'}, if \textsl{b$_{0}$:u$\leqslant$b'} and \textsl{$\overline{u}$$\unlhd$r}, then \textsl{$\exists$b$_{r}$$\in$s.$\overline{b_{0}}$$\unlhd$b$_{r}$}.
\end{lemma}

\noindent Proof.\\
1. We can prove the lemma by induction on the deduction of \textsl{b[s]$\rightarrow$b'}. Since \textsl{$\overline{b}$$\equiv$(b$_{u}$:u, b$_{v}$)} and \textsl{u$<_{u}$r}, the only filtering rule can be applied to infer \textsl{b[s]$\rightarrow$b'} is conj-restr. Assume that \textsl{$\overline{b}$$\equiv$(b$_{u'}$,b$_{v}$')} and \textsl{b[s$\downarrow$b$_{u'}$]$\rightarrow$b'}, if \textsl{u} and \textsl{u'} are different, the hypothesis holds by the induction assumption. Otherwise, \textsl{b[s$\downarrow$b$_{u}$]$\rightarrow$b'}, since \textsl{$\forall$u$_{0}$$\leqslant$u. $\neg$u$\unlhd$r$\downarrow$u}, therefore, in inferring the \textsl{b$_{u1}$}, only the rule conj-unrestr can be applied, and thus \textsl{b$_{u1}$$\leqslant$b}. By Proposition 8 \textsl{$\overline{b_{u1}}$$\unlhd$(b$_{u}$:u, b$_{v}$)}, and since \textsl{u$_{1}$$\unlhd$u}, \textsl{$\overline{b_{u1}}$ $\unlhd$ b$_{u}$} holds.\\
This completes the induction.$\Box$\\

\noindent 2. We prove this lemma by induction on the final step of the deduction of b[s]$\rightarrow$b'.\\
1) conj-unrestr. Since \textsl{$\forall$u$\leqslant$$\overline{t}$.$\neg$u$\leqslant$r}, it is trivial.\\
2) disj-unrestr and filter-out. Since \textsl{b'=$\bot$}, it is trivial.\\
3) conj-restr. There should be \textsl{$\overline{b}$$\equiv$(b$_{u'}$:u',b$_{v}$)} and \textsl{b[s$\downarrow$b$_{u'}$]$\rightarrow$b'}. Assume that \textsl{$\overline{u}$$\equiv$(u$_{1}$,...,u$_{n}$)} and \textsl{$\overline{b_{0}}$$\equiv$(b$_{u1}$:u$_{1}$,..., b$_{un}$:u$_{n}$)} where \textsl{u$_{i}$$\unlhd$u$_{i}$'} and \textsl{u$_{i}$'$<_{u}$r (i$\in$\{1..n\})}. If \textsl{$\forall$i$\in$\{1..n\}.u'$\neq$u$_{i}$'}, then \textsl{$\overline{u}$$\unlhd$r$\downarrow$u'}, and by induction assumption, there is \textsl{b$_{r}$'$\in$s$\downarrow$b$_{u}'$} that \textsl{$\overline{b_{0}}$$\unlhd$b$_{r}$'}, and thus \textsl{$\overline{b_{0}}$$\unlhd$(b$_{r}$',b$_{u'}$)}=\textsl{b$_{r}$$\in$s}. The hypothesis holds. If \textsl{u'=u$_{i}$'} for certain \textsl{i$\in$\{1..n\}} (without loss of generality we assume \textsl{u'=u$_{1}$'}), then by induction assumption,
there is \textsl{$\overline{b_{0}}$'=(b$_{u2}$,\dots,b$_{un}$) $\unlhd$ b$_{r}$'$\in$s$\downarrow$b$_{u1}$'} for any \textsl{b$_{0}$'$\unlhd$b$_{0}$}. From the previous lemma we know that \textsl{b$_{u1}$$\unlhd$b$_{u1}$'}, therefore, \textsl{$\overline{b_{0}}$$\unlhd$(b$_{u1}$,b$_{r}$')=b$_{r}$$\in$s}. The hypothesis holds.\\
4) tuple-comp. Assume that \textsl{b'=(b$_{1}$',\dots,b$_{n}$')}, and since b\textsl{$_{0}$$\leqslant$b'}, either \textsl{$\exists$k.b$_{0}$$\leqslant$b$_{k}$'} or \textsl{b'=(b$_{1}$,\dots,b$_{m}$')}. For the former case, the hypothesis holds by the induction assumption. For the latter case, there should exists \textsl{u'$\leqslant$$\overline{u}$} and \textsl{u'$<_{u}$r}, which contradicts the precondition of the rule tuple-comp.\\
5) set-comp. Similar to 4) there should be \textsl{b$_{n}$$\in$b} that \textsl{b$_{n}$[s]$\rightarrow$b$_{n}$'} and \textsl{b$_{0}$$\leqslant$b$_{n}$'}, and the hypothesis holds by the induction assumption.\\
6) dunion-comp. Similar to 4) and 5).\\
This completes the induction. $\Box$\\

\begin{lemma}
For the filtering couples \textsl{(b:t,s:\{r\})} and \textsl{(b':t,s':\{r'\})} and the condition \textsl{c} = \textsl{c$_{1}$ with c$_{2}$} where \textsl{c$_{1}$:r$_{1}$} and \textsl{(b,s)$|_{f}$c$\rightarrow$(b',s')}, if there are a binding \textsl{b$_{0}$:u$\leqslant$b'}, a binding \textsl{b''$\sqsubseteq$b'} and a binding \textsl{b$_{r}$$\in$s'} that \textsl{$\overline{b_{0}}$$\leqslant$b$_{r}$} and \textsl{b$_{r}$$\unlhd$$\overline{b''}$}, then for any \textsl{b$_{u1}$:u$_{1}$$\leqslant$b$_{0}$} and \textsl{$\overline{u_{1}}$$\leqslant$r$_{1}$}, \textsl{b$_{u1}$} is effective for \textsl{c$_{1}$} in \textsl{b''}.
\end{lemma}
\noindent \textbf{Proof.} Assume that \textsl{(b,s)$|_{f}$c$_{1}$$\rightarrow$(b$_{1}$,s$_{1}$)}. The lemma is proved by induction on the structure of \textsl{c$_{1}$}.\\
1. \textsl{c$_{1}$} is a predicate.  Since  \textsl{b$_{r}$$\in$s'}, by the predicate filtering rules, \textsl{(b$_{1}$,s$_{1}$:\{r$_{1}$\}) $\hookrightarrow$(b$_{1}$,s$_{1}$':\{r'\})} and \textsl{s'$\subseteq$s$_{1}$'}, and thus \textsl{b$_{r}$$\in$s$_{1}$'}. By Definition 20.1, there is a binding \textsl{b$_{r1}$$\in$s$_{1}$} that \textsl{$\overline{b_{u1}}$$\leqslant$b$_{r1}$} and \textsl{b$_{r1}$ $\unlhd$ b$_{r}$ $\unlhd$$\overline{b''}$}. The hypothesis holds.\\
2. \textsl{c$_{1}$} = \textsl{c$_{a}$ with c$_{b}$}. Assume that \textsl{(b,s)$|_{f}$c$_{a}$$\rightarrow$(b$_{a}$,s$_{a}$)}, then \textsl{(b$_{a}$,s$_{a}$)$|_{f}$(c$_{b}$ with c$_{2}$) $\rightarrow$(b',s')}. By induction assumption, \textsl{b$_{u1}$} is effective for \textsl{c$_{b}$} in \textsl{b''}, and thus there is \textsl{b$_{rb}$$\in$s$_{b}$} that \textsl{$\overline{b_{u1}}$$\leqslant$b$_{rb}$$\unlhd$$\overline{b''}$}. For any \textsl{b$_{ua}$:u$_{a}$$\leqslant$b$_{u1}$} that \textsl{$\overline{u_{a}}$$\leqslant$r$_{a}$}, by induction assumption, \textsl{b$_{ua}$} is effective for \textsl{c$_{a}$} in \textsl{b''}. Therefore, \textsl{b$_{u1}$} is effective for \textsl{c$_{1}$} in \textsl{b''}. The hypothesis holds.\\
3. \textsl{c$_{1}$} = \textsl{c$_{a}$ par c$_{b}$}. Assume that \textsl{(b,s)$|_{f}$c$_{a}$$\rightarrow$(b$_{a}$,s$_{a}$)}, \textsl{(b,s)$|_{f}$c$_{b}$$\rightarrow$(b$_{b}$,s$_{b}$)}, \textsl{(b$_{a}$,s$_{a}$:\{r$_{a}$\}) $\hookrightarrow$(b$_{a}$,s$_{a}$':\{r$_{1}$\})},  \textsl{(b$_{a}$,s$_{b}$:\{r$_{b}$\}) $\hookrightarrow$(b$_{b}$,s$_{b}$':\{r$_{1}$\})},
\textsl{b$_{a}$'=b$_{a}$$\sqcap$b''}, \textsl{b$_{b}$'=b$_{b}$$\sqcap$b''}. Then \textsl{b$_{a}$$\sqcup$b$_{b}$=b$_{1}$} and \textsl{$\bigsqcup$(s$_{a}$'$\cup$s$_{b}$') = s$_{1}$}. By Definition 20.1, there is a binding \textsl{b$_{r1}$$\in$s$_{1}$} that \textsl{$\overline{b_{u1}}$$\leqslant$b$_{r1}$} and \textsl{b$_{r1}$ $\unlhd$ b$_{r}$ $\unlhd$$\overline{b''}$}. Assume that \textsl{b$_{u1}$=b$_{ua}$$\sqcup$b$_{ub}$} where \textsl{b$_{ua}$$\leqslant$b$_{a}$'} and \textsl{b$_{ub}$$\leqslant$b$_{b}$'}. Since the effectiveness of \textsl{b$_{u1}$} on \textsl{c$_{a}$} and \textsl{c$_{b}$} are symmetrically defined, we only need to discuss the case of \textsl{c$_{a}$}. If \textsl{b$_{ua}$} is not null, then there should be a binding \textsl{b$_{ra}$'$\in$s$_{a}$'} that \textsl{b$_{ra}$'$\sqsubseteq$b$_{r1}$} and \textsl{$\overline{b_{ua}}$$\leqslant$b$_{ra}$'}. According to Definition 20, there is a binding \textsl{b$_{ra}$$\in$s$_{a}$} that \textsl{$\overline{b_{ua}}$$\leqslant$b$_{ra}$} and \textsl{b$_{ra}$$\unlhd$b$_{ra}$'$\unlhd$b$_{a}$'}. If \textsl{c$_{a}$} is a predicate,  \textsl{b$_{ua}$} is effective for \textsl{c$_{a}$} in \textsl{b$_{a}$'}. If \textsl{c$_{a}$ = c$_{a1}$ with c$_{a2}$}, by induction assumption, \textsl{b$_{ua}$} is effective for \textsl{c$_{a}$} in \textsl{b$_{a}$'}. If \textsl{c$_{a}$} = \textsl{c$_{a1}$ par c$_{a2}$}, the condition \textsl{c$_{1}$} is the same as \textsl{c$_{a1}$ par c$_{a2}$ par c$_{b}$}, and the discussion is similar. Therefore, the hypothesis holds.\\
This completes the induction. $\Box$ \\

\begin{lemma}
For the filtering couple \textsl{(b,s:\{r\})} and \textsl{(b',s':\{r'\})} and a condition \textsl{c} that \textsl{(b,s)$|_{f}$c$\rightarrow$(b',s')},
if a) \textsl{b$_{r}$$\unlhd$$\overline{b}$} for any \textsl{b$_{r}$$\in$s} and b) for each \textsl{b$_{u}$:u$\leqslant$$\overline{b}$(u$\leqslant$r)} there is \textsl{b$_{r}$$\in$s} that \textsl{b$_{u}$$\leqslant$b$_{r}$}, then  \textsl{b$_{r}$'$\unlhd$$\overline{b'}$} for any \textsl{b$_{r}$'$\in$s'}, and for each \textsl{b$_{u}$:u$\leqslant$$\overline{b'}$(u$\leqslant$r')} there is \textsl{b$_{r}$'$\in$s'} that \textsl{b$_{u}$$\leqslant$b$_{r}$'}.
\end{lemma}
\noindent Proof.\\
We prove the lemma by induction on the structure of \textsl{c}.\\
1. \textsl{c} is a predicate. We firstly prove \textsl{b$_{r}$'$\leqslant$b'}. It is enough to prove that if \textsl{(b,s)$\hookrightarrow$(b,s$_{0}$:\{r'\})} then \textsl{b$_{r0}$$\unlhd$$\overline{b}$} for any \textsl{b$_{r0}$$\in$s$_{0}$}. Since there is \textsl{b$_{r1}$$\unlhd$$\overline{b}$} that \textsl{b$_{r0}$$\subseteq$b$_{r1}$}, we prove that \textsl{b$_{r0}$=b$_{r1}$}. If not, there should be \textsl{b$_{u}$} that \textsl{b$_{u}$$\leqslant$b$_{r1}$} and \textsl{$\neg$b$_{u}$$\leqslant$b$_{r0}$}. It is easy to see that there should be \textsl{b$_{u0}$:u$_{0}$$\leqslant$b$_{u}$} that \textsl{u$_{0}$$\leqslant$r}, or \textsl{(r,u$_{0}$)$\leqslant$r}, or \textsl{r$||$r$_{1}$$\leqslant$r'} and \textsl{u$_{0}$$\leqslant$r$_{1}$}. According to the Definition 20.1.2-4), all these three cases would lead to \textsl{b$_{u0}$$\leqslant$b$_{r1}$}, that means, \textsl{b$_{r1}$=b$_{r0}$}.

We prove the existence of \textsl{b$_{r}$'}  using reduction to absurdity. Since \textsl{b$_{u}$$\leqslant$$\overline{b'}$} and \textsl{u$\leqslant$r'}, it is easy to know that there is \textsl{b$_{u}$':u'$\leqslant$$\overline{b}$} that \textsl{b$_{u}$$\leqslant$b$_{u}$'} and \textsl{u'$<_{u}$r'}. If the \textsl{b$_{r}$'} doesn't exist, then \textsl{s'$\downarrow$b$_{u}$'} is empty, and thus for any \textsl{b$_{0}$$\leqslant$b} that \textsl{$\overline{b_{0}}$$\equiv$(b$_{u}$',b$_{v}$)}, \textsl{b$_{0}$[s']$\rightarrow$$\bot$}. As the result there is no \textsl{b$_{u}$$\leqslant$$\overline{b'}$} and it contradicts the premise.

The hypothesis holds.\\
2. \textsl{c} = \textsl{c$_{1}$ par c$_{2}$}. Assume that \textsl{(b,s)$|_{f}$c$_{1}$$\rightarrow$(b$_{1}$,s$_{1}$)} and \textsl{(b,s)$|_{f}$c$_{2}$$\rightarrow$(b$_{2}$,s$_{2}$)}, and \textsl{(b$_{1}$,s$_{1}$)$\hookrightarrow$(b$_{1}$,s$_{1}$':\{r'\})} and \textsl{(b$_{2}$,s$_{2}$)$\hookrightarrow$(b$_{2}$,s$_{2}$':\{r'\})}. According to the Definition 20.2.3), for any \textsl{b$_{r}$'=(b$_{u1}$,\dots,b$_{un}$)$\in$s'} where \textsl{b$_{ui}$:u$_{i}$} and \textsl{u$_{i}$$<_{u}$r'}, \textsl{b$_{ui}$=$\bigsqcup$U$_{i}$} where \textsl{U$_{i}$}=\{\textsl{b$_{v}$:u$_{i}$} $|$ \textsl{$\exists$b$_{r0}$:r$\in$s$_{1}$$\cup$s$_{2}$. b$_{v}$$\in$b$_{r0}$} and \textsl{b$_{v}$$\sqsubseteq$b$_{u_{i}}$'}\}. On the other hand, according to the proposition and the hypothesis, there is \textsl{b$_{v1}$$\leqslant$$\overline{b_{1}}$} and \textsl{b$_{v2}$$\leqslant$$\overline{b_{2}}$} that \textsl{b$_{v1}$,b$_{v2}$$\in$U$_{i}$}, thus \textsl{b$_{v1}$$\sqcup$b$_{v2}$$\leqslant$$\overline{b_{1}}$$\sqcup$$\overline{b_{2}}$=$\overline{b'}$}, that means, \textsl{b$_{ui}$$\leqslant$$\overline{b'}$}, and thus \textsl{b$_{r}$'$\unlhd$$\overline{b'}$} holds.

For any \textsl{b$_{u}$$\leqslant$$\overline{b'}$} (\textsl{u$\leqslant$r'}), since \textsl{b$_{u}$=b$_{u1}$$\sqsubseteq$b$_{u2}$} where \textsl{b$_{u1}$$\leqslant$$\overline{b1}$} and \textsl{b$_{u2}$$\leqslant$$\overline{b2}$}, and there is \textsl{b$_{r1}$$\in$s$_{1}$'} and \textsl{b$_{r2}$$\in$s$_{2}$'} that \textsl{b$_{u1}$$\in$b$_{r1}$} and \textsl{b$_{u2}$$\in$b$_{r2}$}, according to Definition 20.2.3) there is \textsl{b$_{r}$'$\in$s'} that \textsl{b$_{u}$'$\leqslant$b$_{r}$'} and \textsl{b$_{u1}$$\sqcup$b$_{u2}$$\subseteq$b$_{u}$'}, and from the above discussion we know that \textsl{b$_{u}$'=b$_{u}$=b$_{u1}$$\sqcup$b$_{u2}$}. That is, \textsl{b$_{u}$$\leqslant$b$_{r}$'}.

The hypothesis holds.\\
3. \textsl{c} = \textsl{c$_{1}$ with c$_{2}$}. The hypothesis holds by the induction assumption.\\
This completes the induction. $\Box$\\

\noindent \textbf{Proof of Proposition 9}

We prove \textsl{b'$\models$c} by induction on the structure of \textsl{c}. We assume that there is a binding \textsl{b$_{0}$:u$\leqslant$b'} and \textsl{$\overline{u}$$\leqslant$r} where \textsl{r} is the scope of \textsl{c}.

1. \textsl{c} is a predicate condition. According to the previous lemma, there is a \textsl{b$_{r}$$\in$s} that \textsl{$\overline{c}$[b$_{r}$]$\rightarrow$true} and \textsl{$\overline{b_{0}}$$\unlhd$b$_{r}$}. Since \textsl{$\overline{u}$$\leqslant$r}, it is easy to know that \textsl{$\overline{b_{0}}$$\leqslant$b$_{r}$}.

2. \textsl{c} = \textsl{c$_{1}$ par c$_{2}$}. Assume that \textsl{c$_{1}$} and \textsl{c$_{2}$} have the scopes \textsl{r$_{1}$} and \textsl{r$_{2}$} respectively, and \textsl{(b,\{$\epsilon$\})$|_{f}$c$_{1}$$\rightarrow$(b$_{1}$,s$_{1}$)} and \textsl{(b,\{$\epsilon$\})$|_{f}$c$_{2}$$\rightarrow$(b$_{2}$,s$_{2}$)}. Since  \textsl{b$_{0}$$\leqslant$ b'=b$_{1}$$\sqcup$b$_{2}$}, there should be \textsl{b$_{1}$':u$\leqslant$b$_{1}$} and \textsl{b$_{2}$':u$\leqslant$b$_{2}$} that \textsl{b$_{1}$'$\sqsubseteq$b$_{0}$} and \textsl{b$_{2}$'$\sqsubseteq$b$_{0}$}, and it is easy to know that \textsl{b$_{0}$=b$_{1}$'$\sqcup$b$_{2}$'}. By the induction assumption, for a binding \textsl{b$_{u1}$:u$_{1}$$\leqslant$b$_{1}$'} where \textsl{u$_{1}$$\leqslant$r$_{1}$}, \textsl{b$_{u1}$} is effective for \textsl{c$_{1}$} in \textsl{b$_{1}$}, and so does for the case of \textsl{c$_{2}$}. The hypothesis holds.

3. \textsl{c} = \textsl{c$_{1}$ with c$_{2}$}. According to the Lemma 6 and Lemma 7, the hypothesis is a special case where \textsl{b''=b'}, and thus it holds.

\end{document}